\documentclass[apjl,twocolumn]{openjournal}
\usepackage{natbib} 
\usepackage[dvipsnames]{xcolor}
\usepackage{aas_macros} 
\usepackage{bm}
\usepackage{amssymb}
\usepackage{amsmath}
\usepackage{threeparttable}
\usepackage[title]{appendix}
\usepackage{hyperref}	
\hypersetup{colorlinks=true,linkcolor=blue,citecolor=blue,filecolor=blue,urlcolor=blue}
\usepackage[caption=false]{subfig}
\usepackage{soul}                          

\makeatletter
\newcommand{\HI}{{\rm H\,{\scriptstyle I}}}

\newcommand{\HeI}{{\rm He\,{\scriptstyle I}}}

\newcommand{\CIV}{{\rm C\,{\scriptstyle IV}}}

\newcommand{\OII}{{\rm [O\,{\scriptstyle II}]}}
\newcommand{\OIII}{{\rm [O\,{\scriptstyle III}]}}
\newcommand{\OVI}{{\rm O\,{\scriptstyle VI}}}

\newcommand{\Rmnum}[1]{\expandafter\@slowromancap\romannumeral #1@}

\newcommand{\xHI}{x_{\rm HI}}

\newcommand{\br}{\boldsymbol{r}}
\newcommand{\A}{\mbox{\AA}}
\newcommand{\TIGM}{T_{\rm IGM}}

\newcommand{\Mh}{M_{\rm h}}

\begin{document}
\title[JWST galaxies $\times$ IGM]{\vspace{-0.5cm}JWST ASPIRE: How Did Galaxies Complete Reionization? \\ Evidence for Excess IGM Transmission around $\OIII$ Emitters during Reionization\vspace{-15mm}}

\author{Koki Kakiichi,$^{1,2}$\thanks{E-mail: \href{mailto:koki.kakiichi@nbi.ku.dk}{koki.kakiichi@nbi.ku.dk} (KK)}
Xiangyu Jin,$^{3}$
Feige Wang,$^{4,3}$
Romain A. Meyer,$^{5}$
Enrico Garaldi,$^{6,7}$
Sarah E.~I.~Bosman,$^{8,9}$ 
Frederick B.~Davies,$^{9}$
Xiaohui Fan,$^{3}$
Maxime Trebitsch,$^{10}$ 
Jinyi Yang,$^{4,3}$ 
Eduardo Ba\~{n}ados,$^{9}$
Jaclyn B. Champagne,$^{3}$
Anna-Christina Eilers,$^{11}$
Joseph F. Hennawi,$^{12,13}$
Fengwu Sun,$^{14,3}$ 
Yunjing Wu,$^{15}$
Siwei Zou,$^{16,15}$ 
Rahul Kannan,$^{17}$ 
Aaron Smith,$^{18}$
George D. Becker,$^{19}$ 
Valentina D'Odorico,$^{20,21,22}$
Thomas Connor,$^{23}$
Zihao Li,$^{1,2}$
Weizhe Liu,$^{3}$
Klaudia Protu\v{s}ov\'{a},$^{8}$
Fabian Walter$^{9}$,
Huanian Zhang$^{24}$
}

\affiliation{$^{1}$Cosmic Dawn Center (DAWN)}
\affiliation{$^{2}$Niels Bohr Institute, University of Copenhagen, Jagtvej 128, DK-2200 Copenhagen N, Denmark}
\affiliation{$^{3}$Steward Observatory, University of Arizona, 933 N. Cherry Ave., Tucson, AZ 85719, USA}
\affiliation{$^{4}$Department of Astronomy, University of Michigan, 1085 S. University Ave., Ann Arbor, MI 48109, USA}
\affiliation{$^{5}$Department of Astronomy, University of Geneva, Chemin Pegasi 51, 1290 Versoix, Switzerland}
\affiliation{$^{6}$Kavli Institute for the Physics and Mathematics of the Universe, The University of Tokyo, 5-1-5 Kashiwanoha, 277-8583, Japan}
\affiliation{$^{7}$Institute for Fundamental Physics of the Universe, via Beirut 2, 34151 Trieste, Italy}
\affiliation{$^{8}$Institute for Theoretical Physics, Heidelberg University, Philosophenweg 12, D–69120, Heidelberg, Germany}
\affiliation{$^{9}$Max Planck Institut f\"{u}r Astronomie, K\"{o}nigstuhl 17, D-69117, Heidelberg, Germany}
\affiliation{$^{10}$
LUX, Observatoire de Paris, Universit\'{e} PSL, Sorbonne Universit\'{e}, CNRS, 75014 Paris, France}
\affiliation{$^{11}$MIT Kavli Institute for Astrophysics and Space Research, Massachusetts Institute of Technology, Cambridge, MA 02139, USA}
\affiliation{$^{12}$Department of Physics, Broida Hall, University of California, Santa Barbara, CA 93106-9530, USA}
\affiliation{$^{13}$Leiden Observatory, Leiden University, P.O. Box 9513, NL-2300 RA Leiden, The Netherlands}
\affiliation{$^{14}$Center for Astrophysics $|$ Harvard \& Smithsonian, 60 Garden St., Cambridge MA 02138 USA}
\affiliation{$^{15}$Department of Astronomy, Tsinghua University, Beijing 100084, China}
\affiliation{$^{16}$Chinese Academy of Sciences South America Center for Astronomy, National Astronomical Observatories, CAS, Beijing 100101, China}
\affiliation{$^{17}$Department of Physics and Astronomy, York University, 4700 Keele Street, Toronto, ON M3J 1P3, Canada}
\affiliation{$^{18}$Department of Physics, The University of Texas at Dallas, 800 W Campbell Rd, Richardson, TX 75080, USA}
\affiliation{$^{19}$Department of Physics \& Astronomy, University of California, Riverside, CA 92521, USA}
\affiliation{$^{20}$INAF - Osservatorio Astronomico, via G.B. Tiepolo, 11, I-34143 Trieste, Italy}
\affiliation{$^{21}$Scuola Normale Superiore, Piazza dei Cavalieri, I-56126 Pisa, Italy}
\affiliation{$^{22}$IFPU - Institute for Fundamental Physics of the Universe, via Beirut 2, I-34151 Trieste, Italy}
\affiliation{$^{23}$Center for Astrophysics $\vert$\ Harvard\ \&\ Smithsonian, 60 Garden St., Cambridge, MA 02138, USA}
\affiliation{$^{24}$Department of Astronomy, Huazhong University of Science and Technology, Wuhan, Hubei 430074, People’s Republic of China}

\begin{abstract}
The spatial correlation between galaxies and the Ly$\alpha$ forest of the intergalactic medium (IGM) provides insights into how galaxies reionized the Universe. Here, we present initial results on the spatial cross-correlation between $\OIII$ emitters and Ly$\alpha$ forest transmission at $5.4<z<6.5$ from the JWST ASPIRE NIRCam/F356W Grism Spectroscopic Survey in $z>6.5$ QSO fields. Using data from five QSO fields, we find $2\sigma$ evidence for excess Ly$\alpha$ forest transmission at $\sim20-40\,\rm cMpc$ around $\OIII$ emitters at $\langle z\rangle\simeq5.86$, indicating that $\OIII$ emitters reside within a highly ionized IGM. At smaller scales, the Ly$\alpha$ forest is preferentially absorbed, suggesting gas overdensities around $\OIII$ emitters. Comparing with models, including THESAN cosmological radiation hydrodynamic simulations, we interpret the observed cross-correlation as evidence for significant large-scale fluctuations of the IGM and the late end of reionization at $z<6$, characterized by ionized bubbles over $50\rm\,cMpc$ around $\OIII$ emitters. The required UV background necessitates an unseen population of faint galaxies around the $\OIII$ emitters with average LyC leakage of $\log_{10} \langle f_{\text{esc}} \xi_{\text{ion}} \rangle / [{\text{erg}^{-1} \text{Hz}}] \simeq 24.5$ down to $M_{\text{UV}} = -10$. Furthermore, we find that the number of observed $\OIII$ emitters near individual transmission spikes is insufficient to sustain reionization in their surroundings, even assuming all $\OIII$ emitters harbour AGN with $100\,\%$ LyC escape fractions. Despite broad agreement, a careful analysis of ASPIRE and THESAN, using the observed host halo mass from the clustering of $\OIII$ emitters, suggests that the simulations underpredict the observed excess IGM transmission around $\OIII$ emitters, challenging our model of reionization. Potential solutions include larger ionized bubbles at $z<6$, further enhancement of large-scale UV background or temperature fluctuations of the IGM, and possibly a patchy early onset of reionization at $z>10$. Current observational errors are dominated by cosmic variance, meaning future analyses of more QSO fields from JWST will improve the results.

\end{abstract}


\section{Introduction}
Understanding what drove cosmic reionization is one of the key problems in modern cosmology. Observations of the cosmic microwave background fluctuations have established that the mid-point of reionization is at $z\simeq7.64\pm0.74$ \citep{Planck2020}. However, important details about what drove cosmic reionization and how fast it proceeded still remain unsolved. Recent observations of the Ly$\alpha$ forest towards background quasars show mounting evidence for reionization ending as late as $z\simeq5.3$ \citep{Becker2015, Bosman2018, Bosman2020, Eilers2018, Yang2020}. The concordance model of reionization assumes that this process is driven by an abundant, faint population of galaxies, with high Lyman continuum (LyC) escape fractions of $f_{\rm esc}\simeq10-20\,\%$. While large efforts have been put into charting the demographics of galaxies out to $z\sim15$ \citep[e.g.,][]{Donnan2023, Harikane2023a, McLeod2024}, due to the lack of knowledge about the ionizing power of galaxies, whether galaxies indeed drove reionization still remains unclear. While the LyC leakage from individual galaxies can now be indirectly estimated with JWST based on the spectroscopic properties such as UV continuum slope \citep{Chisholm2022}, $\OIII/\OII$ line ratio \citep{Izotov2018,Nakajima2020,Flury2022}, rest-optical nebular emission line strength \citep{Zackrisson2017,Topping2022}, and a combination thereof \citep{Choustikov2023,Saxena2023b,Jaskot2024}, they are still limited to a handful of bright enough objects. Furthermore, some luminous galaxies at intermediate redshifts show evidence for significant ionizing leakage both through direct LyC detection \citep{Marques-Chaves2021,Marques-Chaves2022} and through the Ly$\alpha$ line profile \citep{Matthee2022,Naidu2022}. Such luminous systems may contribute significantly to the total ionizing budget, at least in the reionization of their local environment. The surprisingly abundant population of faint active galactic nuclei (AGN) recently discovered by JWST \citep[e.g.,][]{Kocevski2023,Harikane2023b,Matthee2023b,Kokorev2023} could also contribute to reionization (\citealt{Madau2015,Madau2024,Dayal2024}, but see also \citealt{Kulkarni2019b,Shen2020}). The role of galaxies and AGN in reionization thus still remains an open question.

The formation of ionized bubbles around galaxies and the accompanying fluctuations in the physical state of the intergalactic medium (IGM) represent universal predictions of all cosmological reionization simulations \citep[e.g.][]{Gnedin2014,OShea2015,Pawlik2017,Ocvirk2020,Rosdahl2022,Kannan2022}. Different simulations vary in their predictions of the detailed reionization morphology and the extent of the spatial fluctuations in the ionizing background, temperature, and self-shielded gas in the IGM, depending on the ionizing source models and numerical resolutions.
However, the consensus of all theoretical works on reionization is that galaxies must be surrounded by large-scale ionized regions in the IGM. While this picture is widely accepted, we have not yet directly seen a three-dimensional map of galaxies and the IGM during reionization. Such visualizations would represent the most striking evidence of the reionization process and underscore the potential of 21-cm tomography and the Square Kilometre Array (SKA) \citep[e.g.][]{Furlanetto2006,Mellema2013}, enabling the direct mapping of galaxies in ionized bubbles \citep{Zackrisson2020}. Establishing the direct spatial connection between galaxies and the ionized IGM should thus be an important milestone in our understanding of how and whether galaxies drove cosmic reionization.

The spatial correlation between galaxies and the Ly$\alpha$ forest transmission of the IGM provides a way forward for testing this picture observationally. Since the Ly$\alpha$ forest transmission is sensitive to the amount of neutral hydrogen in the IGM, spatially correlating galaxies with the Ly$\alpha$ forest enables us to directly probe the ionization state of the intergalactic hydrogen around galaxies. Furthermore, as the spatial fluctuations of the Ly$\alpha$ forest optical depths depend on the fluctuations of gas overdensities, the UV background \citep{Becker2018}, thermal structures \citep{DAloisio2015}, and self-shielding absorbers \citep{Davies2016}, the spatial correlation between galaxies and Ly$\alpha$ forest transmission presents a powerful test to examine the physical processes shaping the IGM at the tail end of reionization. 

Dedicated spectroscopic surveys conducted in the foreground of bright background quasars, where exquisite Ly$\alpha$ forest spectra are available, as well as spectroscopic IGM tomographic surveys, have measured the spatial correlation between galaxies and the Ly$\alpha$ forest transmission both at cosmic noon ($z\sim2-3$) \citep[e.g.][]{Adelberger2003,Turner2014,Rudie2012,Bielby2017,Chen2020,Newman2024} and at the tail end of the reionization epoch ($z\sim5-6$) \citep{Kakiichi2018,Meyer2019,Meyer2020,Kashino2023}. These studies have shown that galaxies in the post-reionized universe are predominantly surrounded by large-scale gas overdensities up to several tens of comoving Mpc \citep[e.g.][]{Newman2024}, which is indicated by the {\it excess absorption} of Ly$\alpha$ forest in the vicinity of galaxies. This reflects the fact that galaxy formation takes place in the overdense regions of the large-scale cosmic web \citep{Turner2017,Nagamine2021,Newman2024}. The spatial correlation between galaxies and the Ly$\alpha$ forest becomes more complex towards higher redshifts. At these redshifts, reionization is expected to leave additional imprints on the spatial correlation between galaxies and the Ly$\alpha$ forest transmission \citep{Davies2018,Keating2020,Nasir2020}. Cosmological radiation hydrodynamic simulations \citep{Garaldi2022} predict that the additional impact of reionization, such as the ionized bubbles and the fluctuations in the UV background around galaxies, produces large-scale {\it excess transmission} in the Ly$\alpha$ forest around galaxies during the final stages of reionization. 

As the galaxy-Ly$\alpha$ forest cross-correlation relates to the collective properties of galaxies and the photoionization of the IGM, the measurement can be used to estimate the population-averaged LyC leakage and the relative contribution of galaxies to reionization \citep{Kakiichi2018}. \citet{Meyer2020} have measured the cross-correlation between Ly$\alpha$ emitters (LAEs) and Ly$\alpha$ forest using the MUSE observation of eight quasar fields, and inferred that a population-averaged LyC leakage of $\langle f_{\rm esc}\rangle\simeq0.14$ at $z\sim5.8$ is required to explain the observed signal. 
As the mean Ly$\alpha$ forest transmission is sensitive to the collective meta-galactic UV background including all ionizing galaxies in the same volume, it allows us to estimate the average LyC leakage from all galaxies including the faint population that are not individually detected \citep{Inoue2006,Kuhlen2012,Becker2013}. This presents a complementary measure to the indirect estimates of the LyC leakage from individual galaxies using their spectroscopic properties \citep[e.g.][]{Saxena2023b,Jaskot2024}. If our understanding of LyC leakage and how galaxies drove reionization is correct, the observed galaxy-Ly$\alpha$ forest cross-correlation signal should be explainable using the model of reionization with LyC leakage consistent with that inferred from individual galaxies and its expected extrapolation to fainter systems.


The ground-based effort to measure the cross-correlation between galaxies and the Ly$\alpha$ forest along multiple quasar sightlines \citep{Meyer2020} has highlighted the need to significantly increase both the number of galaxies and the number of surveyed quasar fields for more accurate measurements. To address this, we have designed the observational strategy for the JWST ASPIRE spectroscopic redshift survey of quasar fields at $z=6.5-6.8$ \citep{Wang2023} to enable robust measurements of the spatial correlation between galaxies and the Ly$\alpha$ forest. ASPIRE targets a total of 25 quasar fields with the NIRCam Wide-Field-Slitless Spectroscopy (WFSS) mode using the F356W filter \citep{Greene2017}, enabling us to homogeneously survey galaxies at $5.3<z<7$ using the $\OIII4960,5008$ doublet emission lines. The ASPIRE survey has already uncovered a large number of $\OIII$ emitters in quasar fields at $z>6$ \citep{Wang2023,Wu2023,Zou2024,Jin2024,Champagne2024a,Champagne2024b}. Thanks also to the wide field of view of NIRCam, this enables one order of magnitude increase in the sample size compared to the previously-largest ground-based spectroscopic galaxy survey in $z>6$ quasar fields \citep{Meyer2020}. This enables us to robustly perform the spatial correlation analysis between galaxies and the Ly$\alpha$ forest transmission during the final stages of reionization (see \citealt{Garaldi2024b}). In a separate paper \citep{Jin2024}, we also have presented the effective optical depth analysis of the Ly$\alpha$ forest transmission around $\OIII$ emitters in ASPIRE quasar fields. 

In this paper, we present the analysis of statistical galaxy-Ly$\alpha$ forest cross-correlation in the initial 5 quasar fields (out of 25) from the ASPIRE survey. We first describe the observations and the data reduction of both JWST data and quasar absorption spectroscopy in Section \ref{sec:obs}. We then present the spatial correlation between $\OIII$ emitters and the Ly$\alpha$ forest transmission in Section \ref{sec:map}. We highlight the individual associations between $\OIII$ emitter overdensities and the Ly$\alpha$ forest transmission spikes, arguing that the observed $\OIII$ emitters only make a minor contribution to the total ionizing budget. Section \ref{sec:xcorr} presents the statistical cross-correlation analysis between $\OIII$ emitters and Ly$\alpha$ forest transmission at $5.4<z<6.5$. Section \ref{sec:error} presents the analysis of the error budget and compares it with the theoretical covariance matrix. In Section \ref{sec:interpretation}, we discuss the physical interpretation of the cross-correlation signal using models based on an analytic radiative transfer/halo model-based framework as a guideline. In Section \ref{sec:thesan}, we show the comparison of our ASPIRE result with the THESAN cosmological radiation-hydrodynamic simulations and argue that the late end of reionization at $z<6$ and the large-scale IGM fluctuations inside ionized bubbles around $\OIII$ emitters are likely required to explain the observed cross-correlation. A reader interested in the physical implications of the observed galaxy-Ly$\alpha$ forest cross-correlation may jump to this section. Finally, we summarize our results in Section \ref{sec:conclusion}. 

Throughout this paper we assume cosmological parameters $(\Omega_m, \Omega_\Lambda, \Omega_b, h, \sigma_8,n_s)=(0.3089,0.6911,0.0486,0.6774,0.8159,0.9667)$ \citep{Planck2020}. We use cMpc (pMpc) to indicate distances in comoving (proper) units. All magnitudes in this paper are quoted in the AB system \citep{Oke1983}.

\section{Observations and Data}\label{sec:obs}

\subsection{NIRCam WFSS data}

We use the JWST/NIRCam WFSS data from A SPectroscopic survey of biased halos In the Reionization Era (ASPIRE) (GO 1: 2078, P.I.: Wang). The programme targets 25 quasars between $z=6.5$ and $6.8$ in total. This paper utilizes the data from 5 quasar fields where high signal-to-noise quasar spectra are available, sufficient to identify individual Ly$\alpha$ forest transmission spikes. The quasar fields included in our analysis are listed in Table \ref{tab:summary}.

\begin{table}
    \centering
    \caption{Summary of the quasar fields analysed in this paper. }
    \label{tab:summary}
    \begin{tabular}{llllll}
    \hline
    Quasar     & Redshift & $N_{\rm OIII}^{\dagger}$  & Instrument      & Exp. time$^\star$ & Ref. \\
    \hline
    J1104+2134 & 6.7662   & 6               & LRIS     & 2.0 hrs    &  [1]   \\
    J2002-3013 & 6.6876   & 8               & GMOS   & 2.3 hrs    &  [1]    \\
    J1526-2050 & 6.5869   & 15              & X-Shooter & 12.2 hrs   &  [2]    \\
    J0226+0302 & 6.5405   & 8               & X-Shooter & 6.5 hrs   &   [2]   \\
    J0224-4711 & 6.5222   & 12              & X-Shooter & 8.6 hrs   &   [2]   \\
    \hline
    \multicolumn{6}{l}{$^{\dagger}$ Number of $\OIII$ emitters in the Ly$\alpha$ forest region.} \\
    \multicolumn{6}{l}{$^{\star}$ Exposure time of the quasar spectrum.} \\
    \multicolumn{6}{l}{[1] \citet{Yang2020}, [2] \citet{Dodorico2023}}
\end{tabular}
\end{table}

The WFSS observation is obtained in F356W together with direct imaging in F115W, F200W, and F356W. For all fields, the on-source grism exposure time is $2834\rm\,s$. The direct imaging in the F115W, F200W, and F356W filters is obtained with exposure times of $472\rm\,s$, $2800\rm\,s$, and $472\rm\,s$, respectively.  The quasar is placed at a position $(X_{\rm offset}=-60.5\arcsec, Y_{\rm offset} = 7.5\arcsec)$ in module A to allow sufficient area around the quasars to be covered by the WFSS footprint. While this provides asymmetric spatial coverage around the quasar sightline, it will not affect our results as we examine the statistical spatial correlation between $\OIII$ emitters and the Ly$\alpha$ forest. The data were reduced using the combination of the standard JWST pipeline (\texttt{CALWEBB}; version 1.8.3, \citealt{Bushouse2022}) and some custom scripts as detailed in \citet{Wang2023} and \citet{Yang2023}. We use the calibration reference files (\texttt{jwst\_1015.pmap}) from version 11.16.15 of the standard Calibration Reference Data System (CRDS). We refer readers to \citet{Wang2023} and \citet{Yang2023} for a more detailed description of the process.

In order to extract spectra from the WFSS observations, we constructed the spectral tracing models using the spectral traces of point sources observed in the Large Magellanic Cloud (LMC) field (PID 1076) \citep{Sun2022a,Sun2023}. Then we extract both 2D and 1D spectra of all sources detected in the F356W direct imaging. The 2D spectrum of each source is extracted from each individual exposure and the exposures are then stacked to make a 2D spectrum after resampling them to a common wavelength and spatial grids following the histogram2D technique in the \texttt{PypeIt} software \citep{Prochaska2020}. We then extracted 1D spectra from the stacked 2D spectra using optimal extraction algorithms. 

To search for $\OIII$ emitters in the ASPIRE quasar fields, we used a set of scripts to automatically search for line emitters from both the extracted 1D spectra and the coadded 2D spectra. The line emitter searching algorithm based on 1D spectra is detailed in \citet{Wang2023}. To reduce the visual inspection efforts, we also introduced a line emitter searching algorithm based on the coadded 2D spectra (Wang et al. in prep). Briefly, we used the {\tt Photutils} (version 1.13.0, \citealt{Bradley2024}) for searching for bright blobs on the coadded 2D spectra with at least three connected pixels having a $S/N>0.8$ and the integrated line emission at $>2\sigma$ significance. The blob searching was done for all pixels within $\pm2$ pixels from the dispersion trace center pixels. To identify potential $\OIII$ emitters, we first assume all identified lines with $S/N>5$ (if exists) as the $\OIII$ $\lambda5008$ line and then ask if a corresponding $\OIII$ $\lambda4960$ or H$\beta$ line exists. If one of such case (i.e., $\OIII$ $\lambda5008$ with $S/N>5$ and $\OIII$ $\lambda4960$ or H$\beta$ with $S/N>2$) was identified in the coadded 2D spectra of a given object, we treat it as a $\OIII$ emitter. We found that such algorithm can recover all $\OIII$ emitters except for the faintest one (ASPIRE-J0305M31-O3-023) in \citet{Wang2023}. Since the combination of the 1D and 2D line emitter searching algorithms can reduce the required visual inspection effort by more than a factor of five, we decided to only visually inspect objects that are classified as $\OIII$ emitters in both algorithms. More details and the full $\OIII$ emitter catalogue will be presented in Wang et al. (in prep) and the numbers of $\OIII$ emitters used in this paper are listed in Table \ref{tab:summary}.

\subsubsection{UV magnitudes and $\OIII$ luminosities}

\begin{figure}
    \centering
	\includegraphics[width=0.9\columnwidth]{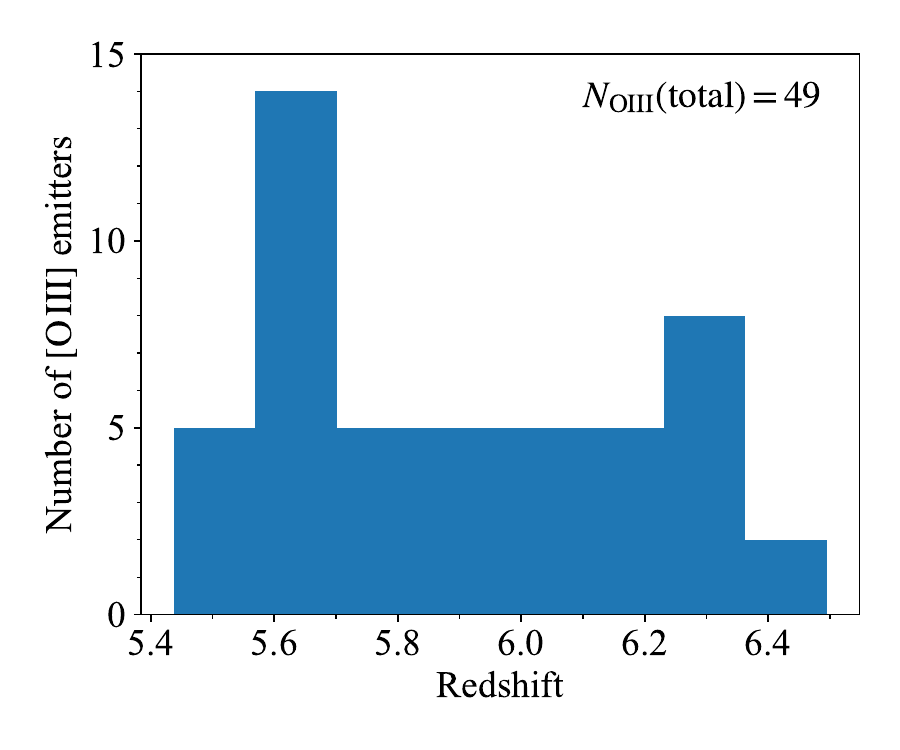}
 \vspace{-0.4cm}
    \caption{The redshift distribution of $\OIII$ emitters in the Ly$\alpha$ forest regions in the five ASPIRE quasar fields.}
    \label{fig:hist}

    \centering
	\includegraphics[width=0.96\columnwidth]{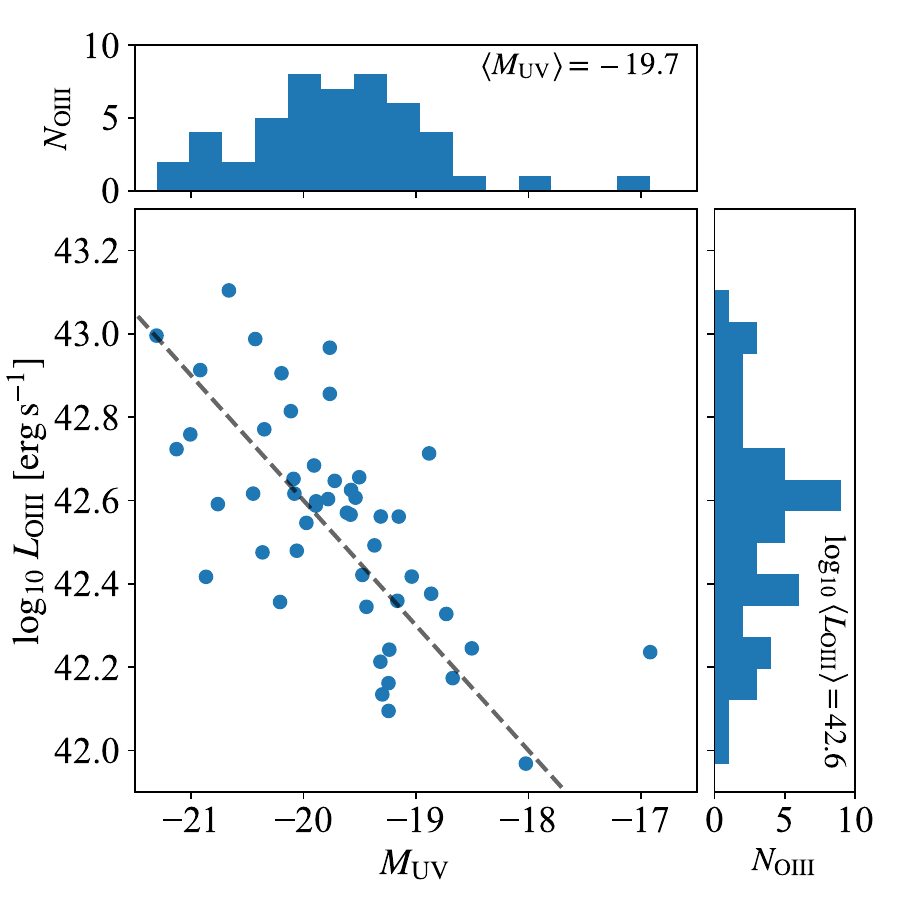}
 \vspace{-0.2cm}
    \caption{The relation between the $\OIII$ and UV luminosities for our sample of $\OIII$ emitters. The top and right histograms show the number of $\OIII$ emitters in each bin. The dashed line shows the linear fit from \citet{Matthee2023b}.}
    \label{fig:UV_OIII}
\end{figure}

The redshift distribution of $\OIII$ emitters in the Ly$\alpha$ forest regions of our five ASPIRE quasar fields is shown in Figure \ref{fig:hist}. We find 49 $\OIII$ emitters in the Ly$\alpha$ forest redshift range of the background quasars appropriate for galaxy-Ly$\alpha$ forest cross-correlation analysis. The median (mean) redshift of the sample is $\langle z_{\rm OIII}\rangle=5.861$ (5.895). 

The UV magnitudes of the $\OIII$ emitters are measured from their F115W magnitudes using $M_{\rm UV}=m_{\rm F115W}+2.5\log_{10}(1+z)-5\log_{10}(D_{\rm L}(z_{\rm OIII})/10\rm\,pc)$ where $D_{\rm L}$ is the luminosity distance. We assume flat UV continua. The $\OIII$ line luminosities are measured from the F356W WFSS spectra. The relation between the UV magnitudes and $\OIII$ luminosities is shown in Figure \ref{fig:UV_OIII}. The average UV magnitude of the sample is $\langle M_{\rm UV}\rangle=-19.7$, which is approximately one magnitude fainter than the typical $L_{\rm UV}^*$ of Lyman-break galaxies at $z\sim6$ (\citealt{Bouwens2021}, $M_{\rm UV}^\ast=-20.93\pm0.09$). We compare our $L_{\rm OIII}-M_{\rm UV}$ relation with \citet{Matthee2023b} and find that our sample is consistent and typical of $\OIII$ emitters found in the literature. 

\subsection{Quasar spectra}

We use archival ground-based optical spectroscopy of five quasars we targeted. We use VLT/X-Shooter spectra of J1526$-$2050, J0226$+$0302, and J0224$-$4711 from XQR-30 and E-XQR-30 sample available from the public repository\footnote{\url{https://github.com/XQR-30/Spectra}} \citep{Dodorico2023}. We use the Keck/LRIS spectrum of J1104$-$2134 and the Gemini/GMOS spectrum of J2002$-$3013 from \citet{Yang2020}. The latter spectra were reduced using \texttt{PypeIt} \citep{Prochaska2020zndo,Prochaska2020}, and the details of data reduction can be found in \citet{Yang2020}. 

Following \citet{Yang2020}, we perform a power-law continuum fitting on the quasar spectrum to reconstruct the intrinsic quasar continuum flux. We assume a broken power-law with a spectral index $\alpha_{\rm \lambda}$ of $-1.5$ and a break at 1000\AA\ \citep{Shull2012ApJ} and adopt the wavelength ranges of 1245$-$1285~\AA\ and 1310$-$1380~\AA\ in the quasar rest-frame when performing the power-law fitting. Following \citet{Jin2023ApJ}, we mask spectral pixels which are likely contaminated by strong sky emission lines. The best-fit quasar continuum flux is then used to normalise the Ly$\alpha$ forest flux to derive the IGM transmission. 



We define the usable regions of the Ly$\alpha$ forest towards the background quasars. The minimum usable redshift is set at the Ly$\beta$ line of the background quasars, $z_{\rm min}=\lambda_{\beta}/\lambda_\alpha(1+z_Q)-1$. We have tested the impact of the minimum redshift on our $\OIII$ emitter-Ly$\alpha$ forest cross-correlation measurement (Section \ref{sec:xcorr}). We find that setting the minimum redshift of the Ly$\alpha$ forest to the rest-frame 1040\,\AA~to avoid the intrinsic Ly$\beta + \OVI$ emission from the quasar results in $\sim8\,\%$ difference in the measured mean Ly$\alpha$ forest transmission around $\OIII$ emitters compared to the default choice. This difference is much smaller than our current error budget and thus does not affect our conclusions.

The maximum redshift is determined by the near-zone size for each quasar. We measure the near-zone size of each quasar by smoothing the continuum normalised spectra with a top-hat filter with 10\,\AA~width and find the near-zone redshift $z_{\rm NZ}$ where the flux first drops below $10\%$. To make sure that the analysed Ly$\alpha$ forest regions are not influenced by the quasar's radiation field, we additionally remove a $\Delta z=0.05$ region bluewards of the near-zone redshift ($\approx21\,\rm cMpc$ at $z=6$). The maximum redshift of Ly$\alpha$ forest is thus set to be $z_{\rm max}=z_{\rm NZ}-\Delta z$. 

\subsubsection{Identifying the transmission spikes}

We identify the transmission spikes in the Ly$\alpha$ and Ly$\beta$ forests using the Gaussian-matched filter method (e.g.~\citealt{Barnett2017}). We use Gaussian kernels with $\sigma=[10,15,20]\rm\,km\,s^{-1}$ and convolve them with the spectrum. We then record the signal-to-noise ratio (SNR) of the matched filter search for each kernel width and keep the maximum SNR at each pixel. We select the local peaks with $\rm SNR>5$ in the matched-filter search. We additionally require that the peak transmission at the spike in the original continuum-normalised flux is $>3\sigma$ and the Ly$\alpha$ forest transmission is $>0.02$ (corresponding to $\tau_{\rm eff}\lesssim4$) to ensure the significance of the transmission spike is above the noise.

\begin{figure*}
	\includegraphics[width=\textwidth]{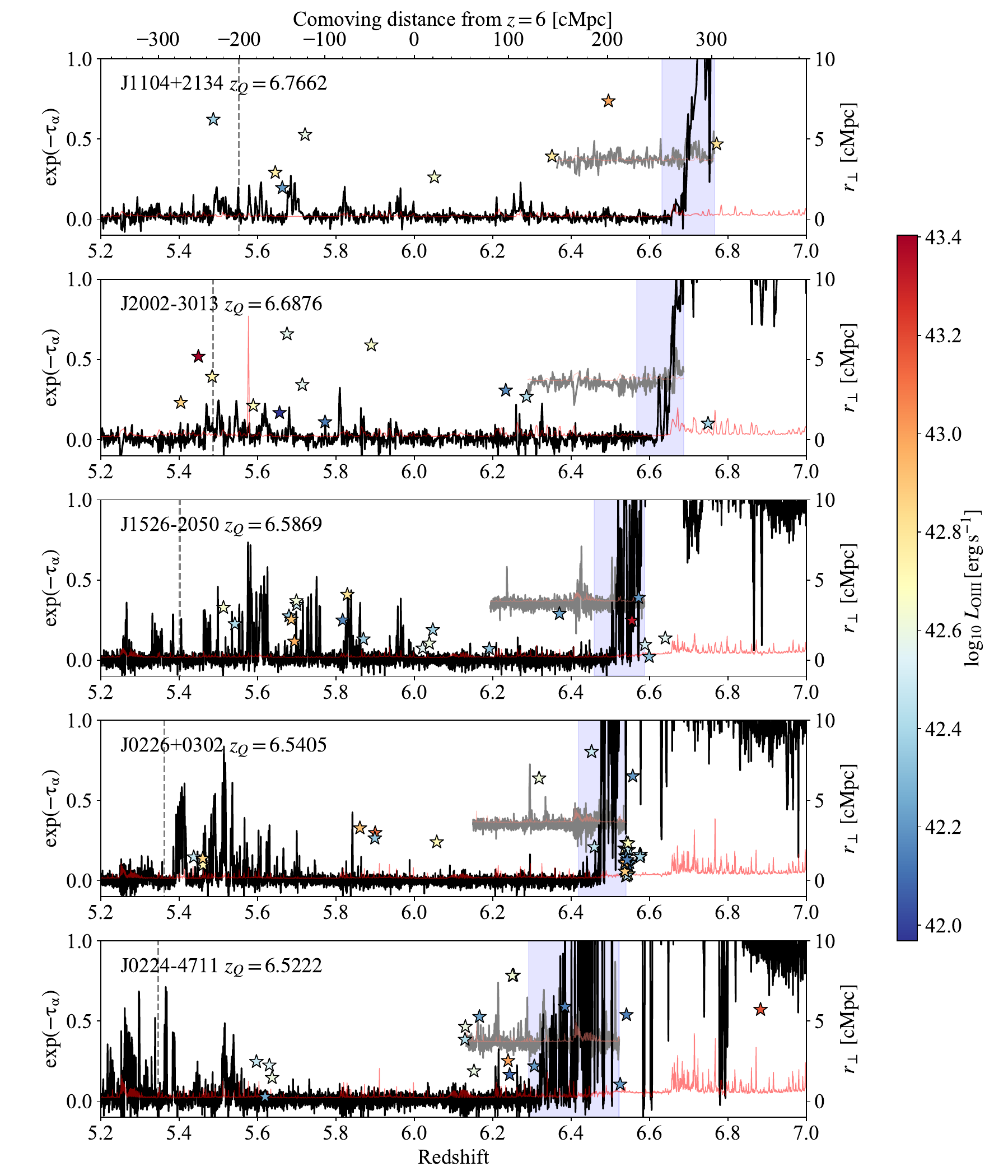}
    \vspace{-0.3cm}
    \caption{Ly$\alpha$ forest transmission $e^{-\tau_\alpha}$ (black) in the five quasar fields alongside the spectroscopic redshifts of the $\OIII$ emitters (star symbols). The top $x$-axis indicates the comoving line-of-sight distance relative to $z=6$ and the right $y$-axis refers to the angular comoving separation $r_\perp$ between $\OIII$ emitters and the quasar sightline. The noise in the quasar spectra is shown in red. The region of the quasar spectra covering the Ly$\beta$ forest is shown in grey and offset vertically for convenience. The $\OIII$ luminosities of the $\OIII$ emitters are indicated by the colour bar. The proximity zones (including $\Delta z=0.05$ offset, see main text) of the background quasars are marked by the blue shaded region. The vertical dashed line indicates the Ly$\beta$ redshift.}
    \label{fig:map}
\end{figure*}

\section{Spatial correlation between galaxies and Ly$\alpha$ forest}\label{sec:map}

In Figure \ref{fig:map} we show the overview of the spatial correlation between $\OIII$ emitters and Ly$\alpha$ forest transmission $e^{-\tau_\alpha}$ along five quasar fields from the ASPIRE survey. We find $\simeq6-15$ $\OIII$ emitters in the Ly$\alpha$ forest region of each quasar field, as shown in Table \ref{tab:summary}, with a total of 49 objects in all five fields. The $\OIII$ emitters are generally found in the vicinity of Ly$\alpha$ transmission spikes. While there is some small offset of $\sim10\rm\,cMpc$ between the $\OIII$ redshift and the transmission spikes along the line of sight, they are typically located within $\sim50\rm\,cMpc$ distance around the transmission spikes. We measure the statistical cross-correlation in Section \ref{sec:xcorr}. As we will discuss below, we interpret this large-scale correlation between $\OIII$ emitters and Ly$\alpha$ forest transmission spikes as evidence that star-forming galaxies reside in the region of highly ionized IGM at the tail end of reionization. We discuss the required physical state of the IGM around $\OIII$ emitters with the help of theoretical models (Section \ref{sec:interpretation}) and the result in the context of full radiation hydrodynamic simulations (Section \ref{sec:thesan}).

The 49 $\OIII$ emitters across five quasar fields represent a significant increase in sample size compared to previous spectroscopic galaxy surveys in quasar fields. For instance, \citet{Meyer2020} identified 21 LAEs in six $z\sim6$ quasar fields and 13 spectroscopically-confirmed Lyman-break galaxies in three fields. Our sample represents a 2 to 4-fold increase in sample size for cross-correlation analysis. This boost results from NIRCam/WFSS's approximately 8-fold larger field-of-view than VLT/MUSE, along with increased efficiency in detecting galaxies using rest-frame optical $\OIII4960,5008$ lines compared to the Ly$\alpha$ emission line. Before moving on to the statistical cross-correlation analysis, we first highlight significant individual associations between galaxies and IGM transmission spikes, illustrating the contribution of $\OIII$ emitters to reionization.    

\subsection{Individual associations between $\OIII$ emitters and Ly$\alpha$ \& Ly$\beta$ transmission spikes}\label{sec:individual}

We define individual 'spike-galaxy associations' if the line-of-sight distance in redshift space between a transmission spike and an $\OIII$ emitter is less than $20\,\mathrm{cMpc}$ ($\Delta v\approx2000\rm\,km\,s^{-1}$). This choice of the line-of-sight separation is somewhat arbitrary and is chosen to reflect visual associations between $\OIII$ emitters and transmission spikes found in Figure \ref{fig:map}. For comparison, Subaru/HSC narrow-band surveys of LAEs in quasar fields span an $\approx40\,\mathrm{cMpc}$ window around the transmissive Ly$\alpha$ forest regions (\citealt{Ishimoto2022,Christenson2023}, see also \citealt{Becker2018,Christenson2021}). This working definition also includes the spike-galaxy associations previously reported in the literature \citep{Kakiichi2018,Kashino2023}, as well as the associations between spikes and metal absorbers, indicative of faint galaxies below the detection limit, within a $\pm2000\,\mathrm{km\,s^{-1}}$ window corresponding to $\approx20\,\mathrm{cMpc}$ \citep{Meyer2019,Christensen2023}.

\subsubsection{$z=6.215$ transmission spike and $\OIII$ overdensity in the J0224-4711 quasar field}\label{sec:J0224}

\begin{figure}
    \centering
	\includegraphics[width=1.05\columnwidth]{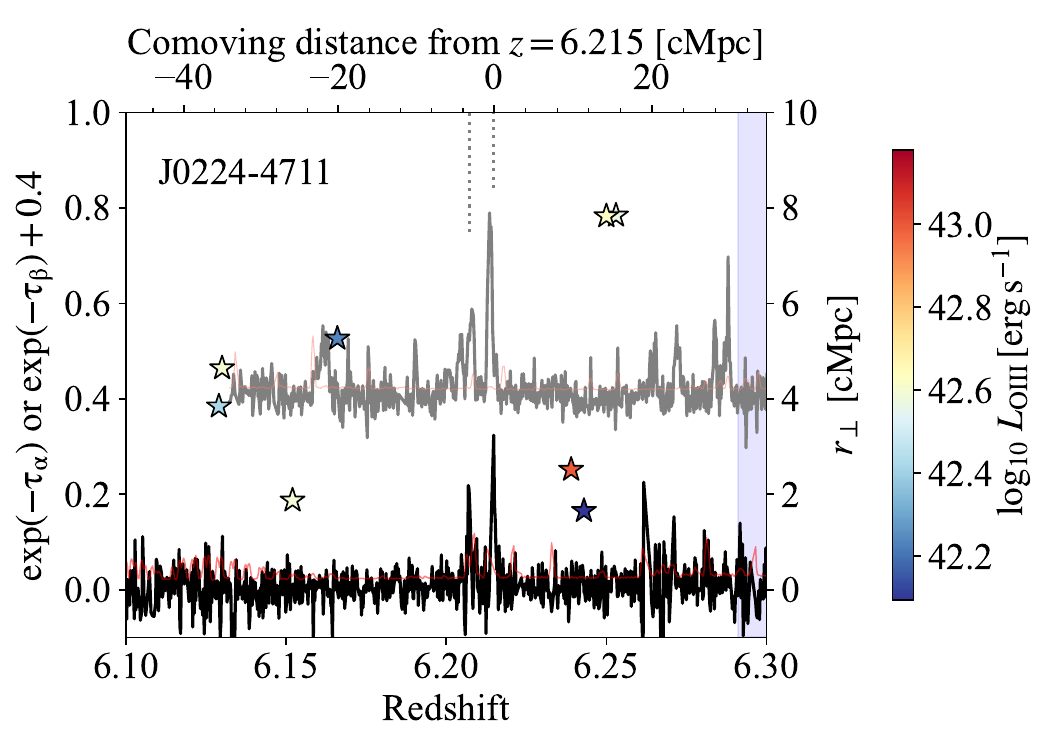}
    \vspace{-0.4cm}
    \caption{Zoom-in around the $z=6.215$ transmission spike in the IGM in the J0224-4711 quasar field. The spatial distribution of $\OIII$ emitters is shown in the same way as in Figure \ref{fig:map}. The locations of Ly$\alpha$ forest transmission spikes are shown by vertical dotted lines. The top x-axis indicates the line-of-sight comoving distance from the transmission spike. The region shows an overdensity of $\OIII$ emitters.}
    \label{fig:0224}
    \centering
    \vspace{0.3cm}
    \hspace*{-0.2cm}
	\includegraphics[width=1.05\columnwidth]{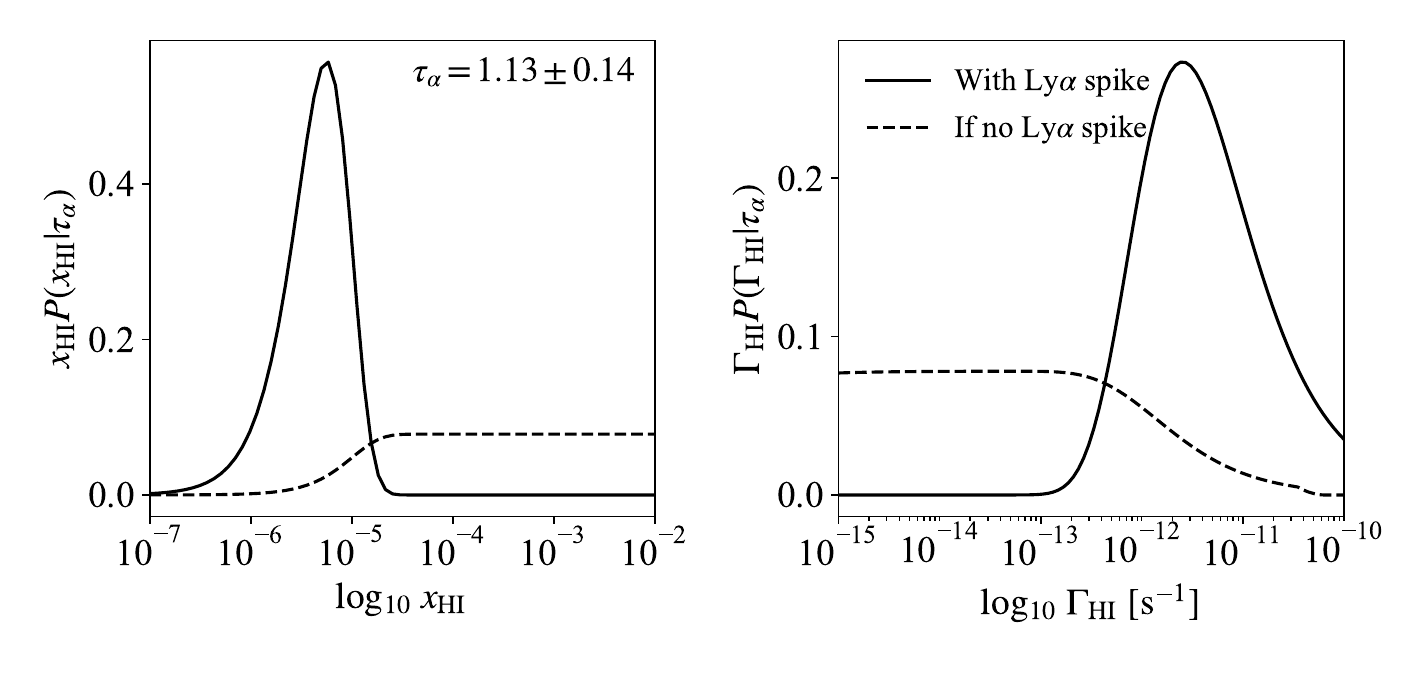}
    \vspace{-0.4cm}
    \caption{({\bf Left}): The probability distribution function of the IGM $\HI$ fraction at the location of the transmission spike at $z=6.215$ given the observed Ly$\alpha$ optical depth at the peak (solid line). The hypothetical case in which no Ly$\alpha$ transmission spike is observed is shown by the dashed line, which corresponds to a $3\sigma$ upper limit on the optical depth. ({\bf Right}): The corresponding probability distribution function of the photoionization rate $\Gamma_{\rm HI}$ at the location of the spike assuming a region in photoionization equilibrium and at a temperature of $T=10^4\rm\, K$. } 
    \label{fig:PDF}
\end{figure}

First, focusing on $z>6$, the most notable association between IGM transmission spikes and an $\OIII$ emitter overdensity is found at $z=6.125$ in the J0224-4711 quasar field. Figure \ref{fig:0224} shows a zoomed-in plot of the region. The region exhibits an overdensity of $\OIII$ emitters within a radius of $r\lesssim30-40\rm\,cMpc$ around the transmission spikes. The transmission spikes are observed both in the Ly$\alpha$ and Ly$\beta$ forests at a coinciding redshift. This strongly suggests the presence of highly ionized IGM at this location.

To estimate the probable value of the IGM $\HI$ fraction at the location of the transmission spike, it is convenient to assume the fluctuating Gunn-Peterson approximation,
\begin{equation}
    \tau_\alpha \simeq \frac{\sigma_\alpha c \bar{n}_{\rm H}(z)}{\nu_\alpha H(z)} \xHI \Delta_b \approx 3.9 \times 10^5 \xHI \Delta_b \left(\frac{1+z}{7}\right)^{3/2}.
\end{equation} 
The observed Ly$\alpha$ optical depth at the peak of the transmission spike corresponds to a combination of the $\HI$ fraction $\xHI$ and the overdensity $\Delta_b$ in the IGM. High Ly$\alpha$ forest transmission may arise either due to a high ionization fraction or low-density fluctuations. The range of probable density fluctuations can be estimated using cosmological hydrodynamic simulations of the IGM, which can be characterized by the volume-weighted density probability distribution function (PDF), $P_V(\Delta_b)$. We use an analytic fit to the NyX simulation at $z=6$ \citep{Lukic2015}. The location of the $z=6.215$ transmission spike is sufficiently far away ($\gtrsim10 \rm\,cMpc$) that we can assume the region is unaffected by the immediate gas overdensities around the observed $\OIII$ emitters and is sufficiently described by the overdensity PDF at the mean IGM. Then, the conditional PDF of the $\HI$ fraction given the observed Ly$\alpha$ optical depth can be expressed as \citep{Kakiichi2018},
\begin{equation}
    P(\xHI | \tau_\alpha) = \int \delta_D \left[\xHI - \frac{\tau_\alpha}{\tau_{\rm GP}(z)}\Delta_b^{-1}\right] P_V(\Delta_b) \, d\Delta_b,
\end{equation}
where $\delta_D(\xHI)$ is the Dirac Delta function and $\tau_{\rm GP}(z) = 3.9 \times 10^5 [(1+z)/7]^{3/2}$ is the Gunn-Peterson optical depth for fully neutral IGM at mean density. The resulting conditional PDF of the $\HI$ fraction is shown in Figure \ref{fig:PDF} (left). The presence of a Ly$\alpha$ transmission spike with $e^{-\tau_\alpha} = 0.324 \pm 0.044$ at the peak, which corresponds to $\tau_\alpha = 1.13 \pm 0.14$, indicates the region is indeed highly ionized to $\xHI \simeq 4.2^{+1.6}_{-1.3} \times 10^{-6}$. If no transmission spikes were detected down to the $3\sigma$ upper limit, it would have favoured a higher neutrality of the IGM.

We can repeat the same argument to estimate the probable value of the photoionization rate assuming the IGM is in photoionization equilibrium. Assuming the likely range of the IGM temperature is given by the prior $P(T)$, the conditional PDF of the photoionization rate $\Gamma_{\rm HI}$ given the observed Ly$\alpha$ optical depth is 
\begin{align}
    &P(\Gamma_{\rm HI}|\tau_\alpha)= \nonumber \\
    &\int\!dT P(T) \int\!\delta_D\!\left[\Gamma_{\rm HI}-\frac{\tau_{\rm GP}(z)}{\tau_\alpha}\frac{\Delta_b^{2}}{\bar{t}_{\rm rec}(T)}\right]P_V(\Delta_b)d\Delta_b,    
\end{align}
where $\bar{t}_{\rm rec}(T)=[\alpha_{\rm A}(T)\bar{n}_{\rm H}(z)]^{-1}$ is the recombination rate at the mean density $\bar{n}_{\rm H}(z)$ and temperature $T$ with $\alpha_A(T)$ being the case A recombination rate coefficient. The estimated photoionization rate at the location of the transmission spike is shown in Figure \ref{fig:PDF} (right). The required photoionization rate at the transmission spike is $\Gamma_{\rm HI}\simeq 3.3^{+3.2}_{-1.5} \times 10^{-12}\,\rm s^{-1}$. 

Compared with the photoionization rate estimated for a typical region of the Universe at $z\approx6$, which yields a mean value of $\bar{\Gamma}_{\rm HI}=0.147^{+0.097}_{-0.044}\times10^{-12}\rm\,s^{-1}$ \citep{Gaikwad2023,Davies2023}, our inferred value near the transmission spike is a factor of 20 larger than the mean value. This indicates that the region marks a part of the Universe with an early completion of the reionization process. The observed association between $\OIII$ emitters and transmission spikes suggests that the large-scale intergalactic environment around the observed $\OIII$ emitters is highly ionized. The completion of reionization has been likely accelerated by the galaxy overdensity.

\subsubsection{Contribution of observed $\OIII$ emitters to reionization}\label{subsec:role_of_galaxies}

The observed $\OIII$ emitters are only the tip of the iceberg of all galaxies that may be present in the environment. How much do the observed $\OIII$ emitters contribute to the ionizing background at the location of the transmission spike? The contribution to the photoionization rate from the observed $\OIII$ emitters can be estimated by
\begin{align}
    &\Gamma_{\rm HI}^{\rm OIII}(\boldsymbol{r})= \nonumber \\
    &~~~\frac{\alpha_g\sigma_{912}}{3+\alpha_g}\sum_{i=1}^{N_{\rm OIII}}\frac{f_{{\rm esc},i}\xi_{{\rm ion},i}L_{{\rm UV},i}}{4\pi|\boldsymbol{r}-\boldsymbol{r}_i|^2(1+z_{{\rm s},i})^{-2}}e^{-|\boldsymbol{r}-\boldsymbol{r}_i|/\lambda_{\rm mfp}^{\rm c}},
\end{align}
where $\alpha_g$ is the EUV ($>13.6\rm\,eV$) spectral slope of galaxies, $\sigma_{912}$ is the photoionization cross section at 912\,\A, $f_{{\rm esc},i}$ is the LyC escape fraction, $\xi_{{\rm ion},i}$ is the ionizing photon production efficiency, $L_{{\rm UV},i}$ is the UV luminosity, $\boldsymbol{r}_i$ is the comoving position of the $i$-th galaxy,  and $\lambda_{\rm mfp}^{\rm c}$ is the comoving mean free path of ionizing photons. 

Using the observed UV magnitudes and positions of the $\OIII$ emitters, we find that their total contribution to the photoionization rate at the location of $z=6.215$ transmission spike in the J0224-4711 field is
\begin{equation}
    \Gamma_{\rm HI}^{\rm OIII}\approx1.8\times10^{-15}\left(\frac{f_{\rm esc}^{\rm OIII}}{0.10}\right)\left(\frac{\xi_{\rm ion}^{\rm OIII}}{10^{25.5}{\rm\,erg^{-1} Hz}}\right)\rm\,s^{-1}.
\end{equation}
assuming the LyC escape fraction and ionizing photon production efficiency of $f_{\rm esc}^{\rm OIII}=0.10$ and $\xi_{\rm ion}^{\rm OIII}=10^{25.5}\rm\,erg^{-1} Hz$ for all the $\OIII$ emitters, the EUV spectral index $\alpha_g=3$, and the proper ionizing mean free path $\lambda_{\rm mfp}=1\rm\,pMpc$ \citep{Becker2021,Zhu2023}. Compared to the required photoionization rate for the transmission spike $\Gamma_{\rm HI}\simeq 3.3^{+3.2}_{-1.5} \times 10^{-12}\,\rm s^{-1}$, this is only $\Gamma_{\rm HI}^{\rm OIII}/\Gamma_{\rm HI}\approx5.5\times10^{-4}$. Even assuming the infinite mean free path, the fractional contribution from the observed $\OIII$ emitters is at most
\begin{equation}
\frac{\Gamma_{\rm HI}^{\rm OIII}}{\Gamma_{\rm HI}} < 2.7\times10^{-3}\left(\frac{f_{\rm esc}^{\rm \scriptscriptstyle OIII}}{0.10}\right)\left(\frac{\xi_{\rm ion}^{\rm \scriptscriptstyle  OIII}}{10^{25.5}{\rm\,erg^{-1} Hz}}\right).    
\end{equation}

The observed $\OIII$ emitters contribute only sub-dominantly ($\lesssim0.3\,\%$) to the total photoionization rate required to maintain the IGM reionized at the location of the transmission spike. Even assuming extreme values of $f_{\rm esc}=100\,\%$ and $\log_{10}\xi_{\rm ion}\simeq26.0$ which corresponds to extremely-metal poor ($Z=10^{-5}$) galaxies with young ages $<10^{6.5}\rm\,yr$ including binary stellar population (\citealt{Eldrige2017}, see also \citealt{Robertson2022}), they would account for only $\sim9\%$ of the total photoionization rate. 



\subsubsection{Role of AGN to reionization}

As our $\OIII$ sample are identified only via $\OIII4960,5008$ emission, we still do not know whether their ionizing radiation is dominated by star formation or by an embedded faint AGN. We estimate the potential AGN contribution to reionization. Assuming the SED of AGN follows a broken power-law \citep{Telfer2002} with FUV and EUV slopes of $\alpha_{\rm FUV}=0.5$ and $\alpha_{\rm EUV}=1.5$, the ionizing photon production efficiency of the AGN is $\log_{10}\xi_{\rm ion}^{\rm AGN}\simeq25.84$. In the extreme case where all $\OIII$ emitters are AGN, assuming the \citet{Telfer2002} SED and 100\,\% escape fractions (even though faint AGN may show lower values, \citealt{Grazian2018}), their maximum contribution to the photoionization rate is still $\sim6\,\%$, i.e. 
\begin{equation}
\frac{\Gamma_{\rm HI}^{\rm AGN}}{\Gamma_{\rm HI}} < 5.9\times10^{-2}\left(\frac{f_{\rm esc}^{\rm \scriptscriptstyle AGN}}{1.0}\right)\left(\frac{\xi_{\rm ion}^{\rm \scriptscriptstyle  AGN}}{10^{25.84}{\rm\,erg^{-1} Hz}}\right),    
\end{equation}
insufficient to raise the photoionization rate to the observed value.

As the $z=6.215$ transmission spikes reside just outside of the the proximity zone of the bright background quasar, one may wonder whether the background quasar may contribute to the photoionization rate. The distance between the transmission spike and the background quasar J0224-4711 at $z_Q=6.5222$ is $R=122\rm\,cMpc$. The photoionization rate from the background quasar is 
\begin{align}
    &\Gamma_{\rm HI}^{\rm QSO}(\boldsymbol{r})= \nonumber\\
    &~~~\frac{\alpha_{\rm EUV}\sigma_{912}}{3+\alpha_{\rm EUV}}\frac{f_{\rm esc}^{\rm QSO}\xi_{\rm ion}^{\rm QSO}L_{\rm UV}^{\rm QSO}}{4\pi|\boldsymbol{r}-\boldsymbol{r}_Q|^2(1+z_{Q})^{-2}}e^{-|\boldsymbol{r}-\boldsymbol{r}_Q|/\lambda_{\rm mfp}^{\rm c}},
\end{align}
which gives the maximum contribution of
\begin{equation}
    \frac{\Gamma_{\rm HI}^{\rm QSO}}{\Gamma_{\rm HI}}
    <3.3\times10^{-2} \left(\frac{f_{\rm esc}^{\rm QSO}}{1.0}\right)\left(\frac{\xi_{\rm ion}^{\rm QSO}}{10^{25.84}{\rm\,erg^{-1} Hz}}\right),
\end{equation}
for $M_{1450}=-26.8$ of the quasar \citep{Dodorico2023} and assuming infinite mean free path, without any Lyman-limit systems to absorb the ionizing photons along the way. Given that mean free path at $z\sim6$ is $\simeq7-14\rm\,cMpc$ \citep{Becker2021, Zhu2023}, i.e.~more than 8 times shorter than the quasar-spike distance, the contribution from the background quasar is likely much smaller. The contribution from the background quasar cannot explain the ionizing background at the transmission spike. 

\smallskip
Thus, we conclude that this association between the $z=6.215$ transmission spike and the $\OIII$ overdensity in J0224-4711 quasar field requires a different population of galaxies to ionize the IGM to the observed level. The main contribution to the required photoionization rate could come from either fainter galaxies within and outside the field-of-view, a luminous population residing outside the field-of-view of the ASPIRE NIRCam single-pointing quasar field, and/or galaxies not selected as $\OIII$ emitters.


\subsection{Other individual associations between Ly$\alpha$ \& Ly$\beta$ transmission spikes and $\OIII$ emitters}

To consolidate the above conclusion, we also investigate other individual associations between transmission spikes and $\OIII$ emitters. Ly$\beta$ transmission spikes are particularly useful as the Ly$\beta$ optical depth is smaller than Ly$\alpha$ ($\tau_\beta=0.16\tau_\alpha$ as it is more sensitive to the ionization state of the IGM.
At $z>6$, we find other prominent Ly$\beta$ transmission spikes:
\begin{itemize}
    \item $z=6.295$ Ly$\beta$ transmission spike - $\OIII$ emitter in the J0226+0302 field,
    \item $z=6.426$ Ly$\beta$ transmission spike - $\OIII$ emitter in the J1526-2050 field.
\end{itemize}
Each transmission spike is associated with one $\OIII$ emitter separated by $r=11$ and $22\,\rm{cMpc}$ from the spike, respectively. While the latter is located slightly outside of our nominal $20\,\rm{cMpc}$ window, we included it as it is the only other prominent transmission spike at $z>6$.

Assuming that they are located inside the implied ionized bubbles, the separation between $\OIII$ emitter and transmission spike can be interpreted as the the lower limits on the size of ionized bubbles $R_b$ as the existence of the transmission spike indicates the highly ionized IGM, that is, 
\begin{equation}
    R_b\gtrsim11-22\rm\,cMpc
\end{equation}
around the $\OIII$ emitters in the J0226+0302 and J1526-2050 fields, respectively. These limits are consistent with the typical size of ionized bubbles ($R_b\sim40-60\rm\,cMpc$) at the final stages of reionziation from simulations \citep[e.g.][]{Wyithe2004,Neyer2023,Lu2024}.

To keep the IGM ionized, following the same argument as above, the observed Ly$\alpha$ optical depth at the locations of the transmission spikes ($\tau_\alpha=2.72\pm0.17$ and $2.17\pm0.17$ for J0226+0302 and J1526-2050 fields) indicate that the required photoionization rates should be $\Gamma_{\rm HI}=1.48^{+1.50}_{-0.68}$ and $1.98^{+1.98}_{-0.91}\times10^{-12}\,\rm{s^{-1}}$, respectively. On the other hand, the contribution from the observed $\OIII$ emitter around the transmission spikes are only $\Gamma_{\rm HI}^{\rm OIII}\approx3.0$ and $0.05\times10^{-15}(f_{\rm esc}^{\rm \scriptscriptstyle OIII}/0.10)(\xi_{\rm ion}^{\rm \scriptscriptstyle  OIII}/10^{25.5}{\rm\,erg^{-1} Hz})$ in the respective fields at most (assuming an infinite mean free path) for the observed UV luminosities and positions of the $\OIII$ emitters. This represents much less than $1\,\%$ 
contribution to the total ionizing background. Thus, the observation of the other individual associations reinforce the conclusion that a hidden unseen population of galaxies around the $\OIII$ emitters is required to maintain the high ionization state of the surrounding IGM.


We also find a plethora of Ly$\alpha$ transmission spikes appearing at $z<6.0$. For example, the spike-galaxy associations include:
\begin{itemize}
    \item $z=5.685$ Ly$\alpha$ transmission spike - $\OIII$ emitters in the J1104+2134 field,
    \item $z=5.832$ Ly$\alpha$ transmission spike - $\OIII$ emitters in the J1526-2050 field, 
    \item $z=5.842$ Ly$\alpha$ transmission spike - $\OIII$ emitters in the J0226+0302 field,
\end{itemize}
suggesting that the general tendency to find $\OIII$ emitters around transmission spikes continue at $z<6.0$. Of course, not all $\OIII$ emitters are located exactly at the redshifts of transmission spikes. For example, 
\begin{itemize}
    \item $z\simeq5.42$ $\OIII$ emitter overdensity in the J0226+0302 field, and
    \item $z\simeq5.70$ $\OIII$ emitter overdensity in the J1526-2050 field
\end{itemize}
are located in the absorbing region between the transmission spikes. This is already seen in previous work \citep{Kakiichi2018,Meyer2020,Kashino2023} and in simulations \citep{Garaldi2022}. This can be easily explained by the absorption due to gas overdensities associated with the $\OIII$ emitters. These $\OIII$ emitter overdensities are bracketed by clusters of Ly$\alpha$ transmission spikes, suggesting that at larger scales, they are also residing in highly transmissive regions of the IGM. 

Clearly there is a large variation in the individual IGM-galaxy associations. This calls for a statistical analysis to quantify the spatial clustering between $\OIII$ emitters and Ly$\alpha$ forest transmission as we will present below.




\section{Galaxy-Ly$\alpha$ forest cross-correlation}\label{sec:xcorr}

\subsection{Mean Ly$\alpha$ forest transmission around $\OIII$ emitters}

In order to quantify the statistical cross-correlation between $\OIII$ emitters and the IGM, we measure the mean Ly$\alpha$ forest transmission $\langle\exp(-\tau_\alpha(r))\rangle$ around $\OIII$ emitters as a function of comoving distance $r$ from each $\OIII$ emitter to Ly$\alpha$ forest pixels,
\begin{equation}
\langle \TIGM(r)\rangle = \frac{\displaystyle \sum_{i\in{{\rm pair}(r)}} w_i e^{-\tau_{\alpha,i}}}{\displaystyle \sum_{i\in{{\rm pair}(r)}} w_i},\label{eq:xcorr}
\end{equation}
where the summation runs over all Ly$\alpha$ forest pixels $i$ within the radial bin $r$ around each $\OIII$ emitter. $e^{-\tau_{\alpha,i}}$ is the Ly$\alpha$ forest transmission in the $i$-th pixel and $w_i$ is the weight. We use a uniform (no) weighting\footnote{
This choice is made because the background is very dark, and the noise at the observed wavelengths of transmission spikes could become the Poisson photon noise limited. In this case, down-weighting by the inverse variance of the noise may underestimate the contribution of the transmission spikes to the final mean Ly$\alpha$ forest transmission around $\OIII$ emitters. We have computed the cross-correlation using both uniform and inverse-variance weighting based on the quasar spectrum noise, $w_i=1/\sigma_i^2$. We found a consistent result independent of the weighting schemes. Since the uniform weighting provides a more conservative estimate, we have chosen the uniform weighting as our fiducial method.
}, i.e. $w_i=1$. The radial separation $r$ between each $\OIII$ emitter and Ly$\alpha$ forest pixel is computed as $r=\sqrt{r_\perp^2+r_\parallel^2}$ where $r_\parallel=D_c(z_{\alpha,i})-D_c(z_{\rm OIII})$ and $r_\perp=\theta D_c(z_{\rm OIII})$ with $D_c(z)$ being the comoving distance to redshift $z$ and $\theta$ being the angular separation between the $\OIII$ emitter and quasar sightline. The noise in the quasar spectrum propagates to the error in the mean Ly$\alpha$ forest transmission around $\OIII$ emitters, 
\begin{equation}
    \sigma^2_{T_{\rm IGM}}(r) = \left.\displaystyle \sum_{i\in{{\rm pair}(r)}} w_i^2 \sigma_i^2\right/\displaystyle\left( \sum_{i\in{{\rm pair}(r)}} w_i\right)^2=\frac{\langle \sigma_N^2\rangle}{N_{\rm pair}(r)},\label{eq:noise_err}
\end{equation}
where $\sigma_i$ is the noise in the continuum normalised quasar spectrum at $i$-th pixel. The second equality assumes the uniform weighting where $\langle\sigma_N^2\rangle$ is the average of the squares of the noise and $N_{\rm pair}(r)$ is the number of $\OIII$ emitter-Ly$\alpha$ forest pixel at each radial bin. 

\begin{figure}
    \centering
    \hspace*{-0.4cm}
	\includegraphics[width=1.05\columnwidth]{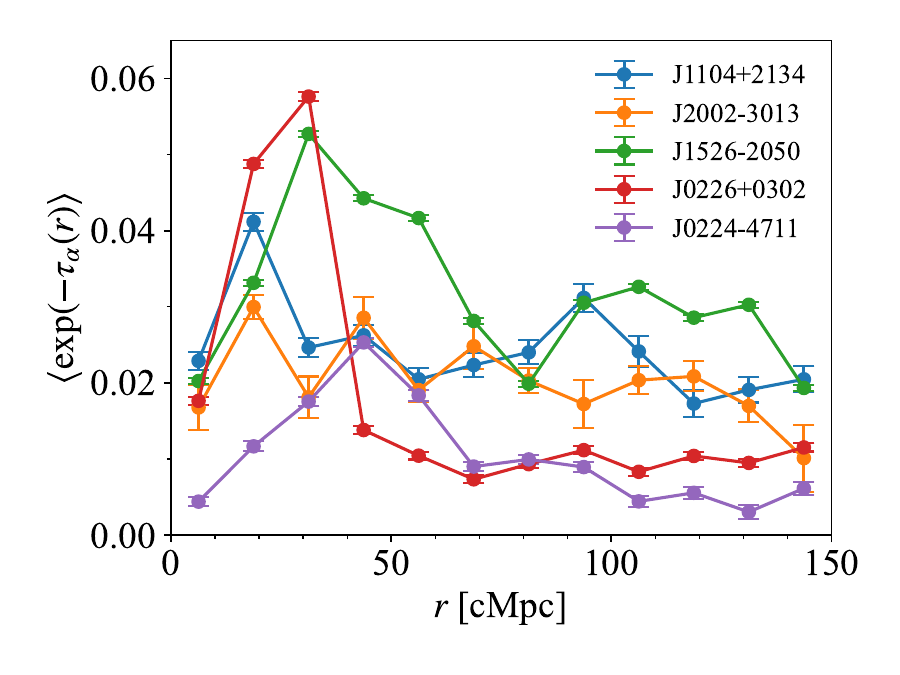}
 \vspace{-0.6cm}
    \caption{Individual measurements of the mean Ly$\alpha$ forest transmission around $\OIII$ emitters in each quasar field. Different colours indicate different quasar fields. The $1\sigma$ error bar takes into account the noise from the quasar spectra. }
    \label{fig:measurement}
\end{figure}

\begin{figure*}
	\includegraphics[width=\textwidth]{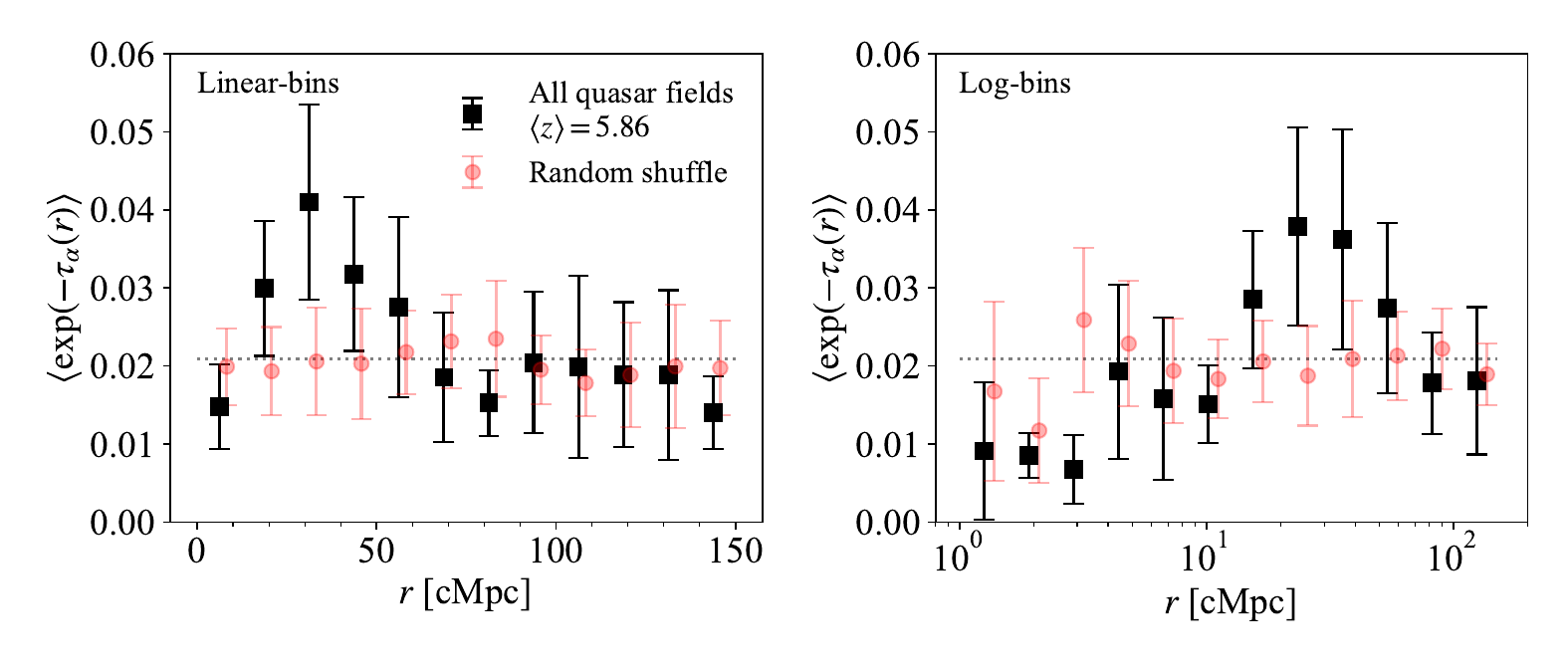}
 \vspace{-0.8cm}
    \caption{({\bf Left}): The full measurement of the mean Ly$\alpha$ forest transmission around $\OIII$ emitters in all 5 quasar fields in our sample (black squares). The median (mean) redshift of $\OIII$ emitters is $\langle z\rangle=5.861 (5.895)$. The $1\sigma$ error is estimated from the Jackknife method. The horizontal dashed line indicates the mean Ly$\alpha$ forest transmission estimated from \citet{Bosman2022}. The measurement includes the $\OIII$ emitters in the redshift range $5.40<z<6.50$ excluding those that lie outside the Ly$\alpha$ forest region of each quasar field. The red circles show the result from the shuffling method (see text). The typical realisation of the random shuffle is shown.  ({\bf Right}): The same measurement, but measured using the logarithmically-space radial bins. In both linearly- and logarithmically-spaced bins, we find $2.2\sigma$ evidence for the excess Ly$\alpha$ forest transmission around $\OIII$ emitters on scales of $10<r<50\rm\,cMpc$.}
    \label{fig:shuffling_analysis}
\end{figure*}

In Figure \ref{fig:measurement}, we show the mean Ly$\alpha$ forest transmission around $\OIII$ emitters in each quasar field. Some fields (e.g. J0226+0302) show clear excess Ly$\alpha$ forest transmission on scales of $r\sim10-40\rm\,cMpc$. We also observe the mean Ly$\alpha$ forest transmission around $\OIII$ emitters varies significantly from field to field.  This field-to-field variation is significantly larger than the error $\sigma_{T_{\rm IGM}}(r)$ from the quasar spectrum noise, confirming the previous claim that the field-to-field variance is the major source of error when measuring the mean Ly$\alpha$ forest transmission around galaxies \citep{Meyer2020}. The scatter in ``baseline'' IGM transmission at large scales between the fields is due to known optical depth fluctuations in the IGM at the end of reionisation, which persist on scales $>70$\,cMpc down to $z\sim5.3$ \citep{Becker2015, Bosman2018, Bosman2022}. 

In order to take into account the field-to-field variation (cosmic variance) in the mean Ly$\alpha$ forest transmission around $\OIII$ emitters across all quasar fields, we use the Jackknife method. We first create Jackknife sample by omitting all $\OIII$ emitters in each quasar field once, providing a total of $N_{\rm JK}=N_{\rm QSO}=5$ Jackknife samples. We then compute $\langle \TIGM(r)\rangle_k$ for each Jackknife sample, $k=1,\dots,N_{\rm JK}$. The Jackknife covariance is then estimated by (e.g.~\citealt{Norberg2009}):
\begin{align}
&{\rm Cov}\left[ \langle\TIGM(r)\rangle, \langle\TIGM(r')\rangle \right]=
\nonumber \\
&~~~~~\frac{N_{\rm JK}-1}{N_{\rm JK}}\sum_{k=1}^{N_{\rm JK}}
\left[\langle\TIGM(r)\rangle_k-\langle\overline{T}_{\rm IGM}(r)\rangle_{\rm JK}\right]\times \nonumber \\
&~~~~~ ~~~~~~~~~~~~~~~~~\left[\langle\TIGM(r)\rangle_k-\langle\overline{T}_{\rm IGM}(r')\rangle_{\rm JK}\right],
\end{align}
where
\begin{equation}
\langle\overline{T}_{\rm IGM}(r)\rangle_{\rm JK}=\frac{1}{N_{\rm JK}}\sum_{k=1}^{N_{\rm JK}}\langle\TIGM(r)\rangle_k,
\end{equation}
is the average over the Jackknife resampled statistics and $\langle\TIGM(r)\rangle_k$ denotes the mean Ly$\alpha$ forest transmission around $\OIII$ emitters in the $k$-th Jackknife sample. The Jackknife error is then the diagonal elements of the Jackknife covariance matrix, $\sigma_{T_{\rm IGM}}(r)=\sqrt{{\rm Cov}\left[ \langle\TIGM(r)\rangle, \langle\TIGM(r)\rangle \right]}$. As we will show in Section \ref{sec:error}, we find a consistent estimate of error between the Jackknife and Bootstrap methods.\footnote{The Bootstrap covariance matrix is estimated by, 
\begin{align}
&{\rm Cov}\left[ \langle\TIGM(r)\rangle, \langle\TIGM(r')\rangle \right]=
\frac{1}{N_{\rm BS}-1}\times\nonumber \\
&~~\sum_{k=1}^{N_{\rm BS}}
\left[\langle\TIGM(r)\rangle_k-\langle\overline{T}_{\rm IGM}(r)\rangle_{\rm BS}\right]
\left[\langle\TIGM(r)\rangle_k-\langle\overline{T}_{\rm IGM}(r')\rangle_{\rm BS}\right],\nonumber
\end{align}
where $\langle\overline{T}_{\rm IGM}(r)\rangle_{\rm BS}$ is the average over the all Bootstrap samples. We create $N_{\rm BS}=1000$ Bootstrap samples by randomly selecting 5 fields with replacement, and compute $\langle\TIGM(r)\rangle_k$ for each Bootstrap sample, $k=1,\dots,N_{\rm BS}$.
}

Figure \ref{fig:shuffling_analysis} shows the full measurement (black squares) of the mean Ly$\alpha$ forest transmission around $\OIII$ emitters across all quasar fields in both linear and logarithmically-spaced radial bins along with the $1\sigma$ error estimated from the Jackknife method. 
The measurement includes all $\text{OIII}$ emitters in the redshift range of $5.40 < z < 6.50$, excluding those that lie outside the Ly$\alpha$ forest region of each quasar field. We quote the median redshift of the $\OIII$ emitters, $\langle z\rangle = 5.861$ as the representative redshift.  We compare our cross-correlation measurement with the mean Ly$\alpha$ transmission (horizontal dashed line) estimated using the mean effective optical depth $\tau_{\rm eff}(z)$ \citep{Bosman2022},
\begin{equation}
    \overline{T}_{\rm IGM}=\frac{1}{N_{\rm OIII}}\sum_{i=1}^{N_{\rm OIII}}e^{-\tau_{\rm eff}(z_{{\rm OIII},i})},\label{eq:mean}
\end{equation}
where the indices run over all $\OIII$ emitters used in the cross-correlation. Note that this mean estimate is more accurate than the effective optical depth evaluated at the mean or median redshift as the mean Ly$\alpha$ forest transmission evolves rapidly from $z=6.6$ to $5.4$ \citep{Bosman2022}.

As shown in Figure \ref{fig:shuffling_analysis}, we find excess Ly$\alpha$ forest transmission around $\OIII$ emitters on scales of $r\approx20-40\rm\,cMpc$. The excess is evident in both linear and logarithmically-spaced radial bins. Using the diagonal elements of the Jackknife covariance matrix, we find $2.2\sigma$ evidence for the excess Ly$\alpha$ forest transmission around $\OIII$ emitters on the scales of $10<r<50{\rm\,cMpc}$ compared to the mean $\overline{T}_{\rm IGM}$. The statistical significance is consistent for the both measurements with linear and logarithmically-spaced radial bins. As shown in the linearly-spaced bins, the mean Ly$\alpha$ forest transmission around $\OIII$ emitters approaches the mean value at large separations, ensuring that the excess is not due to artefacts. 

At smaller scales below $r\lesssim10\rm\,cMpc$, we find the mean Ly$\alpha$ forest transmission around $\OIII$ emitters becomes preferentially absorbed. While the deviation from mean IGM transmission is subtle, we find $1.1-5.6\sigma$ evidence of preferential absorption at $r<10\rm\,cMpc$ in the linearly- and logarithmically-spaced bins, respectively.  The statistical significance is affected by the binning, reflecting the dilution of the absorption signal by binning. To test the impact of binning, we remeasured the mean Ly$\alpha$ forest transmission around $\OIII$ emitters within $r<10\rm\,cMpc$ using finer linear bins of width $1\rm\,cMpc$. We find $5.5\sigma$ evidence for the preferential absorption at $r\lesssim10\rm\,cMpc$ scales, consistent with the result using logarithmically-spaced bins. This preferential absorption is similar to that found around Lyman-break galaxies within several cMpc at intermediate redshifts $z \sim 2-3$ \citep{Turner2014,Bielby2017,Chen2020}.

\subsection{Shuffling test}

\begin{table}
    \centering
    \caption{The observed mean Ly$\alpha$ forest transmission around $\OIII$ emitters, $\langle T_{\rm IGM}(r)\rangle$, measured in both linearly- logarithmically-spaced bins. The $1\sigma$ error according to the diagonal element of Jackknife covariance matrix is tabulated. The proppaged spectral noise is shown in bracket. The mean $\overline{T}_{\rm IGM}$ is estimated by replacing the observed Ly$\alpha$ forest with $e^{-\tau_{\rm eff}(z)}$.}
    \label{tab:measurement}
    \begin{tabular}{llll}
    \hline 
    $r$          & $\langle T_{\rm IGM}(r)\rangle$  & $\sigma_{T_{\rm IGM}}$  & $\overline{T}_{\rm IGM}^\dagger$  \\
    $[\rm cMpc]$ &                                  & Jackknife (noise)       &  \\
    \hline 
    \multicolumn{4}{c}{\underline{Linearly-spaced bins}}  \\
    6.25  &  1.48$\times10^{-2}$  &  5.45$\times10^{-3}$  $(2.92\times10^{-4})$    &  2.14$\times10^{-2}$ \\
    18.8  &  2.99$\times10^{-2}$  &  8.64$\times10^{-3}$  $(2.83\times10^{-4})$    &  2.11$\times10^{-2}$ \\
    31.2  &  4.10$\times10^{-2}$  &  1.25$\times10^{-2}$  $(2.93\times10^{-4})$    &  2.07$\times10^{-2}$ \\ 
    43.8  &  3.18$\times10^{-2}$  &  9.87$\times10^{-3}$  $(3.04\times10^{-4})$    &  2.09$\times10^{-2}$ \\
    56.2  &  2.75$\times10^{-2}$  &  1.15$\times10^{-2}$  $(2.96\times10^{-4})$    &  2.09$\times10^{-2}$ \\  
    68.8  &  1.86$\times10^{-2}$  &  8.33$\times10^{-3}$  $(3.05\times10^{-4})$    &  2.11$\times10^{-2}$ \\ 
    81.2  &  1.53$\times10^{-2}$  &  4.19$\times10^{-3}$  $(2.84\times10^{-4})$    &  2.13$\times10^{-2}$ \\
    93.8  &  2.04$\times10^{-2}$  &  9.04$\times10^{-3}$  $(3.18\times10^{-4})$    &  2.15$\times10^{-2}$ \\
    106   &  1.99$\times10^{-2}$  &  1.17$\times10^{-2}$  $(3.12\times10^{-4})$    &  2.12$\times10^{-2}$ \\
    119   &  1.89$\times10^{-2}$  &  9.27$\times10^{-3}$  $(3.05\times10^{-4})$    &  2.00$\times10^{-2}$ \\
    131   &  1.88$\times10^{-2}$  &  1.09$\times10^{-2}$  $(3.16\times10^{-4})$    &  2.01$\times10^{-2}$ \\
    144   &  1.40$\times10^{-2}$  &  4.64$\times10^{-3}$  $(3.51\times10^{-4})$    &  1.63$\times10^{-2}$ \\
    \multicolumn{4}{c}{\underline{Logarithmically-spaced bins}}  \\
    1.26  & 9.10$\times10^{-3}$   & 8.77$\times10^{-3}$  $(1.82\times10^{-3})$     &  3.00$\times10^{-2}$         \\
    1.91  & 8.54$\times10^{-3}$   & 2.84$\times10^{-3}$  $(1.35\times10^{-3})$     &  2.72$\times10^{-2}$         \\
    2.90  & 6.75$\times10^{-3}$   & 4.36$\times10^{-3}$  $(7.96\times10^{-4})$     &  2.41$\times10^{-2}$         \\
    4.41  & 1.93$\times10^{-2}$   & 1.12$\times10^{-2}$  $(6.59\times10^{-4})$     &  2.27$\times10^{-2}$         \\
    6.69  & 1.58$\times10^{-2}$   & 1.04$\times10^{-2}$  $(6.41\times10^{-4})$     &  2.01$\times10^{-2}$         \\
    10.2  & 1.51$\times10^{-2}$   & 5.03$\times10^{-3}$  $(4.67\times10^{-4})$     &  2.01$\times10^{-2}$         \\
    15.4  & 2.85$\times10^{-2}$   & 8.82$\times10^{-3}$  $(4.04\times10^{-4})$     &  2.09$\times10^{-2}$         \\
    23.4  & 3.79$\times10^{-2}$   & 1.27$\times10^{-2}$  $(3.13\times10^{-4})$     &  2.13$\times10^{-2}$         \\
    35.5  & 3.62$\times10^{-2}$   & 1.41$\times10^{-2}$  $(2.74\times10^{-4})$     &  2.05$\times10^{-2}$         \\
    54.0  & 2.74$\times10^{-2}$   & 1.09$\times10^{-2}$  $(2.34\times10^{-4})$     &  2.09$\times10^{-2}$         \\
    81.9  & 1.78$\times10^{-2}$   & 6.48$\times10^{-3}$  $(1.80\times10^{-4})$     &  2.13$\times10^{-2}$         \\
    124   & 1.81$\times10^{-2}$   & 9.44$\times10^{-3}$  $(1.58\times10^{-4})$     &  1.95$\times10^{-2}$         \\
    \hline
    \multicolumn{4}{l}{{$^\dagger$ The mean estimated by equation ($\ref{eq:mean}$) is $\overline{T}_{\rm IGM}=2.10\times10^{-2}$.}}  
    \end{tabular}
    \end{table}

To test whether excess Ly$\alpha$ forest transmission around $\OIII$ emitters is indeed real, we need to estimate the mean Ly$\alpha$ forest transmission around $\OIII$ emitters in the case of no spatial correlation. We do this by the ``shuffling method''.  We randomly shuffle the Ly$\alpha$ forest spectrum of the $n$-th quasar field with a Ly$\alpha$ spectrum from the other $m=1,\dots,N_{\rm QSO}$ ($m\neq n$) quasar fields. Since all quasar fields are widely separated on the sky, shuffling the Ly$\alpha$ forest along different lines-of-sight artificially de-correlates the spatial distribution of galaxies and the IGM. We then compute the mean Ly$\alpha$ forest transmission around $\OIII$ emitters using the shuffled Ly$\alpha$ forest spectra, 
\begin{equation}
    \langle \TIGM^{\rm shuffle}(r)\rangle = \frac{\displaystyle \sum_{i\in{{\rm pair}(r)}} w^{\rm \scriptscriptstyle shuffle}_i e^{-\tau^{\rm shuffle}_{\alpha,i}}}{\displaystyle \sum_{i\in{{\rm pair}(r)}} w^{\rm \scriptscriptstyle 
 shuffle}_i}\approx \overline{T}_{\rm IGM},
\end{equation}
where $e^{-\tau^{\rm shuffle}_{\alpha,i}}$ is the Ly$\alpha$ forest transmission from the shuffled quasar spectrum and $w_i^{\rm \scriptscriptstyle 
 shuffle}$ is the corresponding weight of the shuffled spectrum. In the limit of infinitely many galaxy-Ly$\alpha$ signtline pairs and no systematics, we expect that this shuffled cross-correlation should approach the mean IGM transmission $\overline{T}_{\rm IGM}$.

This shuffling method circumvents the issue of modelling the selection function of $\OIII$ emitters, as required for generating a simulated catalogue of $\OIII$ emitters. Furthermore, shuffling observed Ly$\alpha$ forest spectra among different quasar sightlines correctly captures the redshift evolution of Ly$\alpha$ forest optical depth over the redshift interval where we perform the cross-correlation analysis.

In Figure \ref{fig:shuffling_analysis}, we compare the mean Ly$\alpha$ forest transmission around $\OIII$ emitters with the shuffled measurement. The shuffled measurement is consistent with the mean IGM transmission $\overline{T}_{\rm IGM}$, confirming that the mean $\overline{T}_{\rm IGM}$ correctly captures the limit of no spatial correlation. Compared with the shuffle measurement at $0<r<150\rm\,cMpc$, the observed mean Ly$\alpha$ forest transmission around $\OIII$ emitters shows clear departure from the random shuffles at $5.2\sigma$ significance, indicating evidence for the statistical spatial correlation between $\OIII$ emitters and IGM at $\langle z \rangle\simeq5.86$. 

To conclude our cross-correlation analysis, we have also compared the shuffled measurement with the more careful estimate of the mean IGM transmission. We have computed the mean by artificially replacing the observed Ly$\alpha$ forest transmission with the mean value, i.e. $e^{-\tau_{\alpha,i}}\rightarrow e^{-\tau_{\rm eff}(z_{\alpha,i})}$, when calculating the mean Ly$\alpha$ forest transmission around $\OIII$ emitters using equation (\ref{eq:xcorr}). This method accounts for the impact of the gradual change of the mean IGM transmission over the redshift interval ($5.40<z<6.50$) for each radial bin more accurately than equation (\ref{eq:mean}). We tabulate the estimate of the global mean, $\overline{T}_{\rm IGM}$, along with our observed mean Ly$\alpha$ forest transmission around $\OIII$ emitters, $\langle T_{\rm IGM}(r)\rangle$ in Table \ref{tab:measurement}. Both the shuffled measurement and the global mean agree very well at all scales.\footnote{The slightly higher values of the mean IGM transmission in the three inner logarithmically-spaced radial bins are simply because the $\OIII$ emitters contributing to the inner radial bins are located at slightly lower redshifts than the mean redshift of the sample. They are consistent with the random shuffles within the statistical uncertainty.} This ensures that our estimate of $\overline{T}_{\rm IGM}$ represents the correct mean of the sample, and $\langle T_{\rm IGM}(r)\rangle/\overline{T}_{\rm IGM}-1$ can be interpreted as the spatial cross-correlation between $\OIII$ emitters and Ly$\alpha$ forest transmission.

In summary, we conclude that the observed excess Ly$\alpha$ forest transmission around $\OIII$ emitters is genuine and is not due to either systematics or misplacement of the global mean.
    
\subsection{Comparison with previous work}

\begin{figure}
    \centering
	\includegraphics[width=\columnwidth]{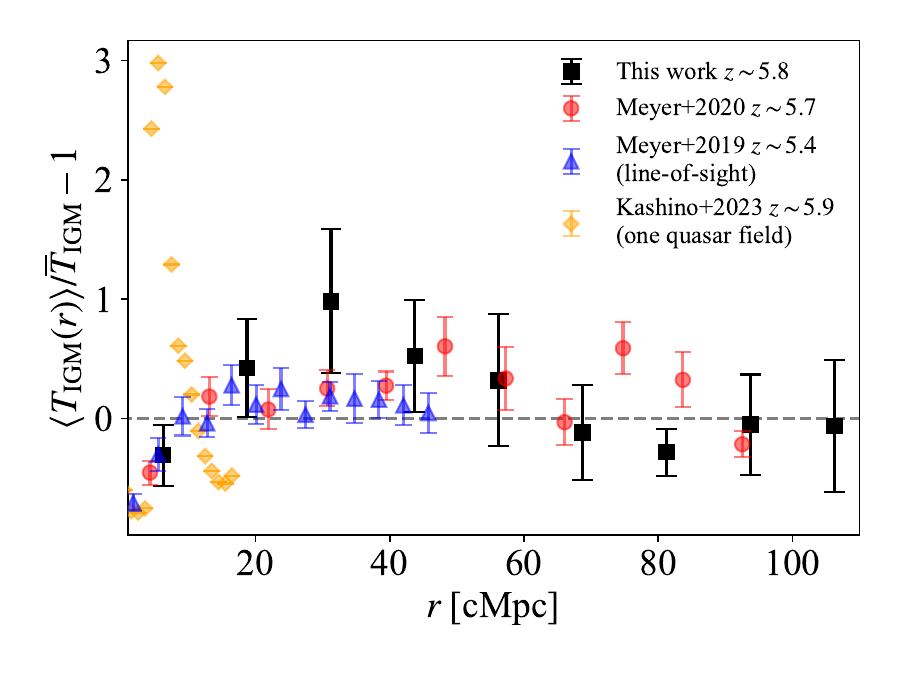}
 \vspace{-0.7cm}
    \caption{Comparison of our $\OIII$ emitter-Ly$\alpha$ forest cross-correlation from 5 ASPIRE QSO fields with previous work. The red circles indicate the LAE-Ly$\alpha$ forest cross-correlation from \citet{Meyer2020}. The blue triangles show the line-of-sight $\CIV$ absorber-Ly$\alpha$ forest cross-correlation from \citet{Meyer2019}. The yellow diamonds show the $\OIII$ emitter-Ly$\alpha$ forest cross-correlation measured from a single QSO field J0100+2802 from \citet{Kashino2023}.}
    \label{fig:comparison}
\end{figure}

We compare our measurement with previous work measuring the galaxy-Ly$\alpha$ forest cross-correlations from \citet{Meyer2019,Meyer2020} and \citet{Kashino2023} in Figure \ref{fig:comparison}. For our ASPIRE result, we adopt the mean IGM transmission $\overline{T}_{\rm IGM}$ tabulated in Table \ref{tab:measurement} and compute the fluctuations $\langle \TIGM(r)\rangle/\overline{T}_{\rm IGM}-1$ around the mean.

\citet{Meyer2020} measured the mean Ly$\alpha$ forest transmission around LAEs based on the MUSE spectroscopic survey of multiple quasar fields. While the survey did not find a statistically significant correlation in the $\langle \TIGM(r)\rangle/\overline{T}_{\rm IGM}-1$ measurement, they reported $\sim3\sigma$ evidence for an excess of Ly$\alpha$ transmission spikes at $\sim10-60\rm\,cMpc$ using LAEs at $z\sim5.7$ by cross-correlating the spatial distribution of LAEs with the identified location of Ly$\alpha$ transmission spikes. The scale of our observed excess in Ly$\alpha$ forest transmission around $\OIII$ emitters reassuringly coincides with the reported physical scales of the excess by \citet{Meyer2020}. The apparently small uncertainties in the \citet{Meyer2020} measurement of $\langle T_{\rm IGM}(r)\rangle/\overline{T}_{\rm IGM}-1$ likely reflect the fact that their error is estimated by bootstrapping the sample of individual galaxies instead of quasar fields, as well as the small sample size, which makes it challenging to robustly estimate the size of uncertainties internally within the data. 

In comparison with the \citet{Meyer2019} measurement of the $\CIV$ absorber-Ly$\alpha$ forest cross-correlation at $z\sim5.4$ along lines-of-sight, the spatial scale of excess transmission is also broadly in agreement although the amount of excess transmission is smaller in \citet{Meyer2019} than in this work. Although our current error on the excess transmission is still large, as we will show in Section \ref{sec:thesan} this can be explained by the difference in the redshifts where these measurements are made. Cosmological simulations indicate that the excess Ly$\alpha$ forest transmission around galaxies evolves as a function of redshift \citep{Garaldi2022}. As we go towards higher redshifts, excess transmission becomes higher due to the larger fluctuations in the Ly$\alpha$ forest transmission around galaxies. 

\citet{Kashino2023} recently measured the mean Ly$\alpha$ forest transmission around $\OIII$ emitters in a single field towards the $z=6.32$ quasar J0100+2806. They reported significant excess transmission at $r\sim5-10\rm\,cMpc$ around $\OIII$ emitters. This is smaller than the scale at which we found excess transmission ($r\sim20-40\rm\,cMpc$) in this work. This is not surprising given that the field-to-field variation is very large. The same is also true when compared with \citet{Kakiichi2018} where the measurement is made in a single quasar field. In our sample, we similarly find that the J1104+2134 quasar field shows excess Ly$\alpha$ forest transmission at smaller scales than the statistical average (Figure \ref{fig:measurement}). This reinforces the fact that a large number of quasar fields need to be surveyed in order to robustly measure the galaxy-Ly$\alpha$ forest cross-correlation. 


\subsection{Redshift evolution}\label{sec:observed_redshift_evolution}

\begin{figure*}
    \centering
    \includegraphics[width=0.95\textwidth]{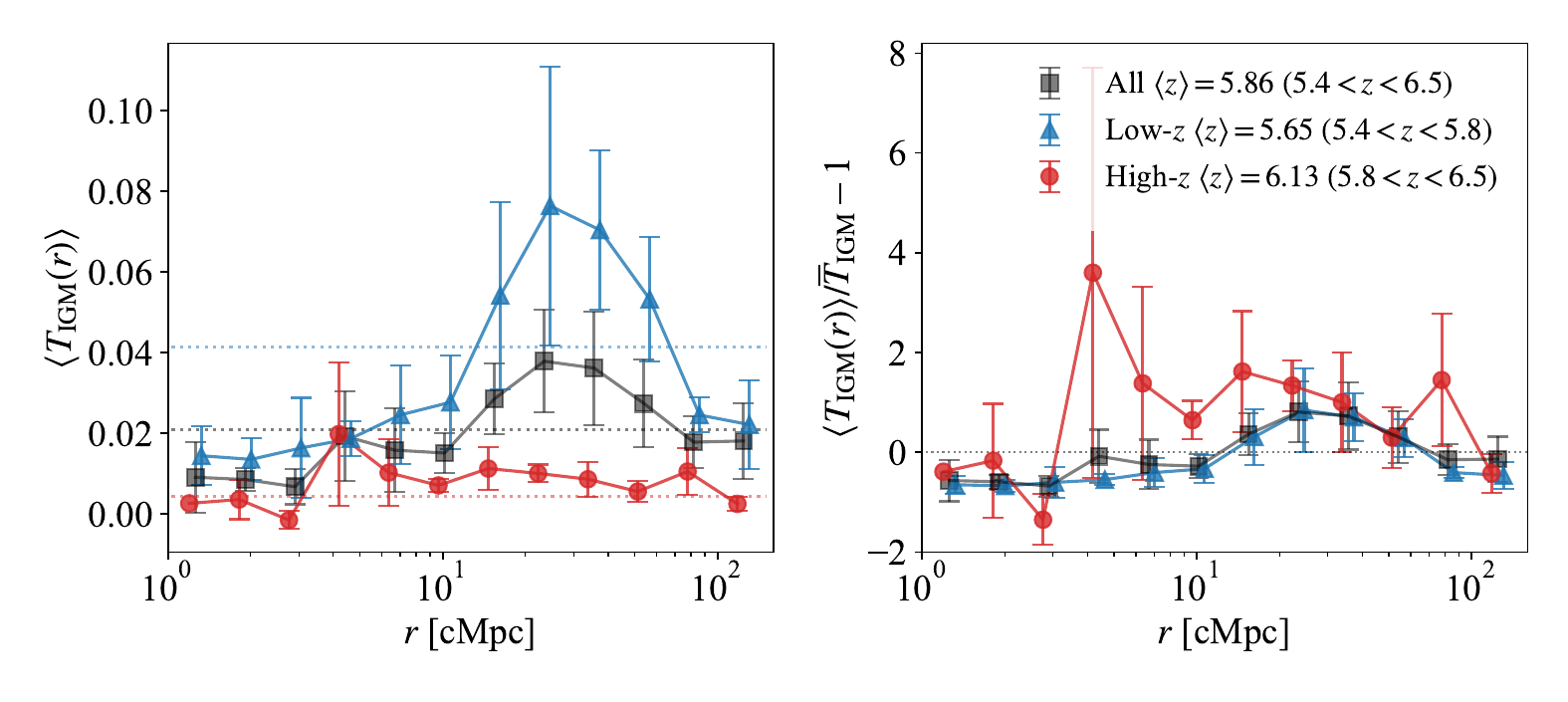}
 \vspace{-0.4cm}
    \caption{Observed redshift evolution of $\OIII$ emitter-Ly$\alpha$ forest cross-correlation. ({\bf Left}): The observed mean Ly$\alpha$ forest transmission around $\OIII$ emitters for three different redshift bins (All: $\langle z\rangle=5.86$ $(5.4<z<6.5)$ (black square), low-$z$:  $\langle z\rangle=5.65$ $(5.4<z<5.8)$ (blue triangle), high-$z$: $\langle z\rangle=6.13$ $(5.8<z<6.5)$ (red circle)). The horizontal lines show the estimated mean IGM transmission $\overline{T}_{\rm IGM}$ according to equation \ref{eq:mean}. ({\bf Right}): The observed $\OIII$ emitter-Ly$\alpha$ forest cross-correlation, $\langle T_{\rm IGM}(r)\rangle/\overline{T}_{\rm IGM}-1$, for the three different redshift bins. The errorbars show the $1\sigma$ error estimated from the Jackknife method.}\label{fig:observed_redshift_evolution}
\end{figure*}

In order to examine the redshift evolution of the galaxy-Ly$\alpha$ forest cross-correlation, we divide our sample into two different redshift bins: the low-$z$ sample ($5.4<z<5.8$) and the high-$z$ sample ($5.8<z<6.5$). This divides our entire $\OIII$ emitter sample used in the full analysis ($5.4<z<6.5$) into approximately half. Figure \ref{fig:observed_redshift_evolution} shows the observed cross-correlation signals in the different redshift bins. As shown in the left panel, the overall normalisation of the mean Ly$\alpha$ forest transmission around $\OIII$ emitters increases with decreasing redshift. This is expected, as the mean Ly$\alpha$ forest transmission is higher at lower redshift. We also observe the excess Ly$\alpha$ forest transmission around $\OIII$ emitters in the lower-$z$ sample at a mean redshift of $\langle z\rangle=5.65$. For the high-$z$ sample at $\langle z\rangle=6.13$, the excess Ly$\alpha$ forest transmission is more difficult to see, although there is consistent excess IGM transmission from $R\sim4$ to $80\rm\,cMpc$ around $\OIII$ emitters. We repeated the shuffling test to check if the excess is still significant. We observe a similar excess compared to the randomly shuffled measurement, although the statistical significance remains low. A more careful quantitative conclusion requires the full analysis of all JWST quasar fields. Here, we note that large-scale excess IGM transmission could persist over a wide range of distance around $\OIII$ emitters at higher redshift.

Figure \ref{fig:observed_redshift_evolution} (right) divides out the redshift evolution of the overall normalisation and shows the redshift evolution of the observed $\OIII$ emitter-Ly$\alpha$ forest cross-correlation signals, $\langle T_{\rm IGM}(r)\rangle/\overline{T}_{\rm IGM}-1$. The possible extended excess IGM transmission around $\OIII$ emitters for the high-$z$ sample is more visible in the cross-correlation. For the low-$z$ sample, although there is a significant redshift evolution of the overall normalisation between the low-$z$ and all samples, their cross-correlation signals appear almost identical. Both show a peak of excess IGM transmission at $R\approx20-40\rm\,cMpc$ around $\OIII$ emitters. The statistical significance for the low-$z$ sample is smaller due to the reduced sample size. Observationally, this is not surprising, as the measurement of the mean Ly$\alpha$ forest transmission around $\OIII$ emitters averages the residual transmitted fluxes in the Ly$\alpha$ forest transmission spikes; the measurement is thus naturally weighted towards a lower redshift where the transmission spikes are higher. Physically, if the slow (or lack of) redshift evolution of the galaxy-Ly$\alpha$ forest cross-correlation signal is real, it could provide valuable insight into how the state of the IGM evolves around galaxies at the final stages of reionization. Further studies with a larger quasar field sample are required to examine the redshift evolution in greater detail.

\section{Understanding the error budget}\label{sec:error}

Since the present analysis uses only a subset of the ASPIRE sample (5 out of 25 quasar fields), it is crucial to understand the error budget of the galaxy-Ly$\alpha$ forest cross-correlation measurement to avoid potential obstacles or unidentified systematics in future analyses.

\subsection{Theoretical covariance matrix}

We do this by comparing the observationally estimated error with the theoretical expectation. We follow the well-established formalism from galaxy and Ly$\alpha$ forest surveys to derive the theoretical covariance matrix for the galaxy-Ly$\alpha$ forest cross-correlation function, $\langle T_{\rm IGM}(r)\rangle/\overline{T}_{\rm IGM}-1$.

We measure the mean Ly$\alpha$ forest transmission around $\OIII$ emitters at the mean redshift $\langle z\rangle=5.86$ from a collection of pencil beam surveys centred on $N_{\rm QSO}=5$ background quasars. Each field covers a comoving area of $A_{\rm QSO}=\Omega_{\rm FoV}D^2_c(\langle z \rangle)$ where $\Omega_{\rm FoV}=1.59\rm\,arcmin^2$ is the single-pointing field of view of the NIRCam WFSS. 
The total comoving survey volume is then given by $V_{\rm s}=A_{\rm QSO}\sum_{n=1}^{N_{\rm QSO}}\int_{z_{n,\rm min}}^{z_{n,\rm max}}cdz/H(z)$ where $z_{{\rm min},n}$ and $z_{{\rm max},n}$ are the minimum and maximum redshifts of the Ly$\alpha$ forest region of $n$-th quasar field.

Writing the covariance matrix in terms of the mean Ly$\alpha$ forest transmission around galaxies $\langle T_{\rm IGM}(r_i)\rangle$ measured with radial bin $r_i$ of width $\Delta r$, we find that the covariance matrix is given by \citep[e.g.][]{Sanchez2008, White2010, White2015,Grieb2016}
\begin{align}
    &{\rm Cov}[\langle\TIGM(r_i)\rangle,\langle\TIGM(r_j)\rangle]=\nonumber \\
    &~~~~~        
    \frac{\overline{T}_{\rm IGM}^2}{V_{\rm s}}\int_0^\infty\frac{k^2dk}{2\pi^2}\bar{j_0}(k|r_i)\bar{j_0}(k|r_j){\rm Var}[P_{g\alpha}(k)],      
\end{align}
where $\bar{j_0}(k|r_i)$ is the radial-bin averaged spherical Bessel function of the first kind\footnote{The radial-bin averaged spherical Bessel function of the first kind is $\bar{j_0}(k|r_i)=\frac{1}{V_i}\int_{V_i} j_0(kr)d^3r$, which simplifies to
\begin{equation}
    \bar{j_0}(k|r_i)=3\frac{\sin(x_+)-\sin(x_-)-x_+\cos(x_+)+x_-\cos(x_-)}{x_+^3-x_-^3}
\end{equation}
with $x_\pm=k(r_i\pm\Delta r/2)$.     
} 
and ${\rm Var}[P_{\rm g\alpha}(k)]$ is the monopole of the variance of the 3D galaxy-Ly$\alpha$ forest cross-power spectrum $P_{g\alpha}(\bm{k})$,
\begin{equation}
    {\rm Var}[P_{g\alpha}(k)]=\frac{1}{2}\int_{-1}^{1}{\rm Var}[P_{g\alpha}(\bm{k})]d\mu,
\end{equation}
where $\mu=k_{\parallel}/k$. Under the assumption of Gaussian random fields, \citet{McQuinn2011} show that the variance of galaxy-Ly$\alpha$ forest cross-power spectrum is given by (see also \citealt{FontRibera2014})
\begin{align}
&  {\rm Var}[P_{g\alpha}(\bm{k})]= \nonumber \\
&P_{g\alpha}({\bm{k}})^2+\left(P_{g}(\bm{k})+n_{\rm g,3D}^{-1}\right)\left(P_{\alpha}(\bm{k})+P^{\rm 1D}_{\alpha}(k_\parallel) n_{\rm eff,2D}^{-1}\right),\label{eq:var_Pk}
\end{align}
where $P_g(\bm{k})$ and $P_\alpha(\bm{k})$ is the 3D auto-power spectrum of galaxies and Ly$\alpha$ forest and $P^{\rm 1D}_{\alpha}(k_\parallel)$ is the line-of-sight Ly$\alpha$ forest power spectrum. We estimate the power spectra using the linear perturbation theory including the effect of the UV background fluctuations \citep{Pontzen2014, Gontcho2014}, which is sufficient for the order-of-magnitude esitmate of the theoretical covariance matrix. The explict forms of the power spectra are shown in Appendix \ref{app:linear_theory}. The $n_{\rm g,3D}^{-1}$ term is the Poisson shot noise of the foreground galaxy sample where $n_{\rm g,3D}=N_{\rm OIII}/V_s$ is the number density of $\OIII$ emitters. The $P^{\rm 1D}_{\alpha}(k_\parallel) n_{\rm eff,2D}^{-1}$ term, the so-called `aliasing term', arises because the Ly$\alpha$ forest is sampling the underlying IGM fluctuations along discrete lines of sight \citep{McDonald2007}. The expression of equation (\ref{eq:var_Pk}) follows \citet{McQuinn2011} where the contribution from both the aliasing term and spectral noise term of the instrument are combined. They define the noise-weighted surface number density of background quasars,
\begin{equation}
    n_{\rm eff,2D}=\frac{1}{A_{\rm s}}\sum_{n=1}^{N_{\rm QSO}}\frac{P^{\rm 1D}_{\alpha}(k_\parallel)}{P^{\rm 1D}_{\alpha}(k_\parallel)+P_{N,n}},\label{eq:neff2D}
\end{equation}
where $A_{\rm s}=N_{\rm QSO}A_{\rm QSO}$ is the total comoving survey area and $P_{N,n}$ is the noise power spectrum of the $n$-th quasar spectrum. In terms of the rms noise per pixel, $\sigma_N$, of the continuum-normalised spectrum, the noise power spectrum can be written as $P_{N,n}=\Delta x_{\rm pixel}\sigma_N^2/\overline{T}_{\rm IGM}^2$, where $\Delta x_{\rm pixel}=\frac{c(1+z)}{H(z)R}$ is the pixel size in unit of comoving length and $R$ is the spectral resolution. In reality, quasar spectra comes from various instruments with different spectral resolution and noise properties. Here we adopt the mean rms noise $\sigma_N=0.026$ measured from our observed spectra (corresponding to $\rm SNR\approx38$ per pixel at continuum) and the nominal spectral resolution $R=8900$ of X-Shooter. As we will see below, the spectral noise is sub-dominant contribution to the error budget and the results are not sensitive to the choice of the spectral parameters.

The above theoretical estimate of the covariance matrix illuminates two important limits. First, in the limit of cosmic variance dominated regime, ${\rm Var}[P_{g\alpha}(\bm{k})]= P_{g\alpha}({\bm{k}})^2+P_{g}(\bm{k})P_{\alpha}(\bm{k})$, the error in the mean Ly$\alpha$ forest transmission around galaxies can only be reduced by increasing the survey volume $V_{\rm s}\propto N_{\rm QSO}\Omega_{\rm FoV}$, equivalent to targeting more quasar fields. In the limit of observational noise dominated regime where the shot-noise and spectral noise dominate the error budget, ${\rm Var}[P_{g\alpha}(\bm{k})]=P_N/(n_{\rm g,3D} n_{\rm QSO,2D})$, the covariance matrix becomes ${\rm Var}[\langle\TIGM(r)\rangle]=\sigma_N^2 N_{\rm pair}^{-1}$. This means that the increased WFSS depth and higher signal-to-noise ratio of quasar spectra will reduce the error as the number of galaxy-Ly$\alpha$ forest pairs increases and the spectral noise decreases. 

\subsection{The observed error is dominated by cosmic variance}

\begin{figure}
	\includegraphics[width=\columnwidth]{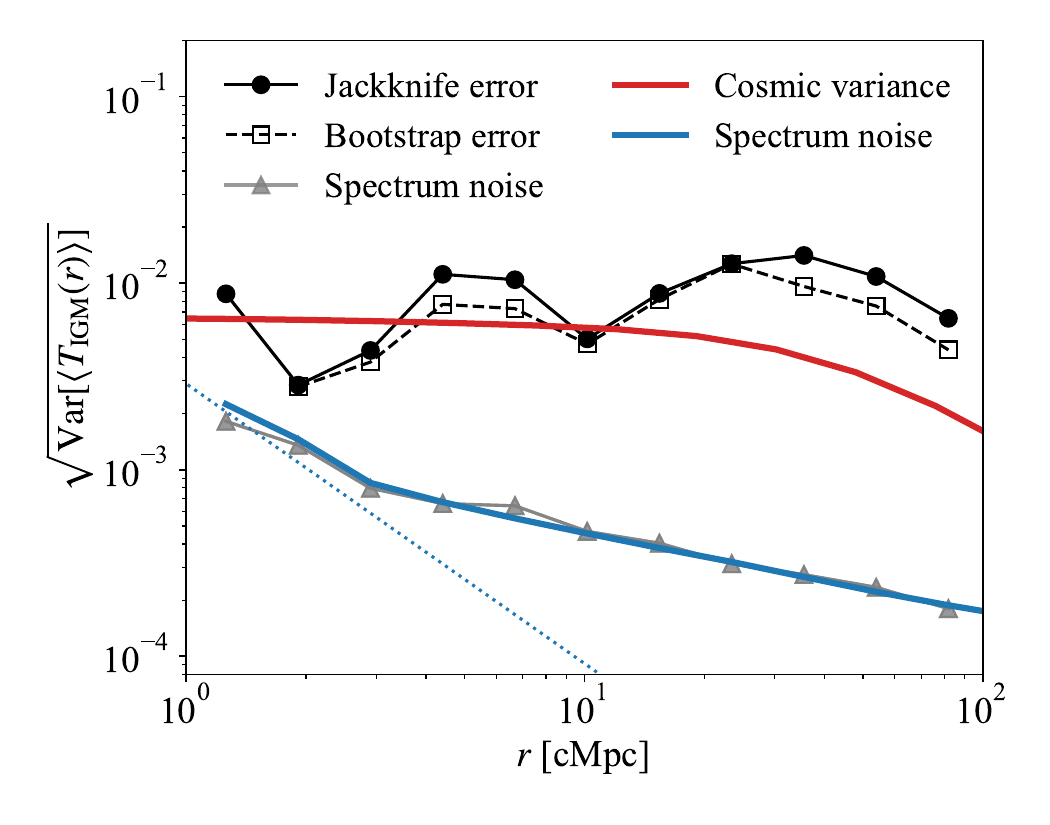}
 \vspace{-0.6cm}
    \caption{Comparison between the observationally estimated variance of mean Ly$\alpha$ forest transmission around $\OIII$ emitters at $\langle z\rangle=5.86$ (black circles: Jackknife error, open squares: Bootstrap error) and the theoretical covariance matrix (red: cosmic variance dominated regime, blue: spectrum noise dominated regime). The diagonal elements of the covariance matrix are shown. The gray triangles show the observationally propagated error from quasar spectra noise. The blue dotted line corresponds to the scaling of spectrum noise assuming $N_{\rm pair}\propto 4\pi r^2\Delta r$. Due to the pencil beam survey geometry centred on multiple discrete quasar fields, the number of galaxy-Ly$\alpha$ forest pixel pairs does not scale as $\propto 4\pi r^2\Delta r$. The solid blue line uses the observed number of pairs which takes into account the observed survey geometry.}
    \label{fig:error_budget}
\end{figure}

In Figure \ref{fig:error_budget}, we compare the variance from the observationally estimated covariance matrix using Jackknife and Bootstrapping methods with the theoretical expectation. The figure shows that both the Jackknife and Bootstrapping errors are consistent with the theoretical expectation of cosmic variance. This suggests that the observed error budget from our initial 5 ASPIRE quasar fields is primarily dominated by cosmic variance. \citet{Meyer2020,Garaldi2024b} also argued that the observed galaxy-Ly$\alpha$ forest cross-correlation function is dominated by cosmic variance. The agreement between our observationally estimated covariance matrix and the theoretical cosmic variance further supports that the error budget in our galaxy-Ly$\alpha$ forest cross-correlation measurement is indeed dominated by cosmic variance.

As we target quasar sightlines where high signal-to-noise quasar spectra are available, the error from spectral noise is sub-dominant compared to cosmic variance. Figure \ref{fig:error_budget} shows that the observational spectral noise is quickly reduced as we average many galaxy-Ly$\alpha$ forest pixel pairs, which scales as $\propto \sigma_N / \sqrt{N_{\rm pair}}$. The propagated error due to the spectral noise (equation \ref{eq:noise_err}) agrees with the theoretical limit of the observational noise-dominated regime, indicating that this noise is a sub-dominant contribution to the overall error budget. Therefore, considering only the observational error from the spectra would underestimate the total error budget of the cross-correlation measurement.

In fact, since only the sum of the spectral noise power spectrum $P_{N,n}$ and the intrinsic 1D line-of-sight Ly$\alpha$ forest power spectrum $P^{\rm 1D}_{\alpha}(k_\parallel)$ contributes to the total covariance matrix (see equation \ref{eq:neff2D}), the quasar spectra only need to be deep enough to ensure that the spectral noise is sub-dominant compared to the intrinsic line-of-sight IGM fluctuations. This is achieved with a signal-to-noise ratio of ${\rm SNR} > \overline{T}_{\rm IGM}^{-1} \sqrt{\Delta x_{\rm pixel} / P^{\rm 1D}_\alpha(k_\parallel)}$ for the quasar spectra. Assuming $R = 8900$ and $P_\alpha^{\rm 1D}(k_{\parallel}) \approx \mathcal{O}(1) \, h^{-1} \, \rm cMpc$, which we find to be a good estimate for $k_\parallel \sim 0.01-1 \, h \, \rm cMpc^{-1}$, we find that ${\rm SNR} \gtrsim 36 / \sqrt{\mathcal{O}(1)}$ per pixel ensures that the line-of-sight error, i.e., the aliasing term, is dominated by the intrinsic IGM fluctuations. Note that for our cross-correlation measurement, both the errors from spectral noise and intrinsic 1D IGM fluctuations are comparable. However, since both errors decrease as $\propto 1 / \sqrt{N_{\rm pair}}$, the final error budget is still dominated by the cosmic variance error.

Of course, this requirement is esimated for the statistical measurement of the galaxy-Ly$\alpha$ forest cross-correlation signal. In other words, this does not mean that a higher signal-to-noise ratio of the quasar spectra is unimportant. There is immense scientific and practical value in studying the individual direct associations between galaxies and Ly$\alpha$ forest transmission spikes (e.g. Section \ref{sec:individual}) and in ensuring that the observed cross-correlation signal is robust against potential systematics and noise from the Ly$\alpha$ forest spectra. Nonetheless, having established the detection of the cross-correlation signal, the modest impact of the spectral noise on the final error budget implies that we would benefit from a wide-area survey with more quasar fields, even if the signal-to-noise ratio of the individual quasar Ly$\alpha$ forest is not high. This approach has already been recognised in cosmological Ly$\alpha$ forest surveys.

In summary, the current observed error budget of the galaxy-Ly$\alpha$ forest cross-correlation measurement can be understood in terms of cosmic variance. There is no excess error from unknown systematics, which is good news for future analyses. We expect that the error budget can be reduced by adding more quasar fields, which should be reduced by a factor of $\sqrt{25/5}\approx2$ with the total 25 ASPIRE quasar fields, and further in the future by combining ASPIRE with other JWST surveys such as EIGER \citep{Kashino2023}.

\section{Interpreting the galaxy-Ly$\alpha$ forest cross-correlation: MODELS}\label{sec:interpretation}

The galaxy-Ly$\alpha$ forest cross-correlation probes the physical state of the IGM around $\OIII$ emitters. In order to interpret the observed signal, we compare the observations with cosmological radiative transfer models based on the conditional luminosity function (CLF) framework \citep{Kakiichi2018,Meyer2020}. In this framework, the mean Ly$\alpha$ forest transmission around $\OIII$ emitters is modelled as,
\begin{align}
    &\langle T_{\mathrm{IGM}}(r) \rangle = \nonumber \\
    &~\int d\Delta_b P_V(\Delta_b | r) \exp\left[-\tau_{0} \Delta_b^\beta \left(\frac{\langle \Gamma_{\mathrm{HI}}(r) \rangle}{10^{-12}\,\mathrm{s}^{-1}}\right)^{-1}\right], \label{eq:TIGM}
\end{align}
where $\beta = 2 - 0.72(\gamma - 1)$ with $\gamma$ being the power-law slope of the temperature-density relation, $\langle \Gamma_{\mathrm{HI}}(r) \rangle$ is the average photoionization rate at radius $r$ from $\OIII$ emitters, and $P_V(\Delta_b | r)$ is the volume-weighted PDF of gas overdensities $\Delta_b$ as a function of radial distance $r$ from host dark matter haloes of mass $> M_{\rm min}/\rm\,M_\odot$. We use the density PDF measured from NyX cosmological hydrodynamic simulations (see Appendix \ref{app:PDF}). For convenience, we denote the Ly$\alpha$ optical depth at mean density $\Delta_b = 1$ and $\Gamma_{\mathrm{HI}} = 10^{-12}\,\mathrm{s}^{-1}$ by,
\begin{equation}
    \tau_{0} \simeq 10(1 + f_{\mathrm{He}}) \left(\frac{T_0}{10^4\,\mathrm{K}}\right)^{-0.72} \left(\frac{1+z}{7}\right)^{9/2},
\end{equation}
where $T_0$ is the IGM temperature at mean density and $f_{\mathrm{He}}$ is the fraction of electrons released per helium atom (for singly ionized helium atom, $f_{\mathrm{He}} \simeq 0.0789$).  The IGM temperature at mean density is assumed to be $T_0=1.2\times10^4\rm\,K$ and the temperature-density relation of $\gamma=1.04$ \citep{Gaikwad2020,Villasenor2022}. 

We assume the HOD model for $\OIII$ emitters with a step function with a smooth transition, $\langle N(M_{\rm h})\rangle=\frac{1}{2}\left[1+{\rm erf}\left(\frac{\log_{10}\Mh-\log_{10}M_{\rm min}}{\sigma_{\rm log M}}\right)\right]$, with a fixed scatter $\sigma_{\rm log M}=0.2$ \citep[e.g.][]{Zheng2005}. We set a minimum host halo mass for $\OIII$ emitters to be $\log_{10} M_{\rm min}/{\rm M_\odot}=11.0$. Recently, \citet{Eilers2024,Pizzati2024} measured the host halo mass of $\OIII$ emitters through the clustering analysis with the EIGER survey and report the host halo mass of $\OIII$ emitters to be $\log_{10} M_{\rm min}/{\rm M_\odot}=10.56_{-0.03}^{+0.05}$. 
Since the average $\OIII$ luminosity of our ASPIRE $\OIII$ emitters are slightly brighter than the EIGER sample due to the shallower depth, our value of $M_{\rm min}=10^{11}\rm\,M_\odot$ should be reasonable. We also set the average UV magnitude of $\OIII$ emitters in the model to be $\langle M_{\rm UV}\rangle=-20$ consistent with the average value of our sample. 

\begin{figure}
    \centering
    \includegraphics[width=0.9\columnwidth]{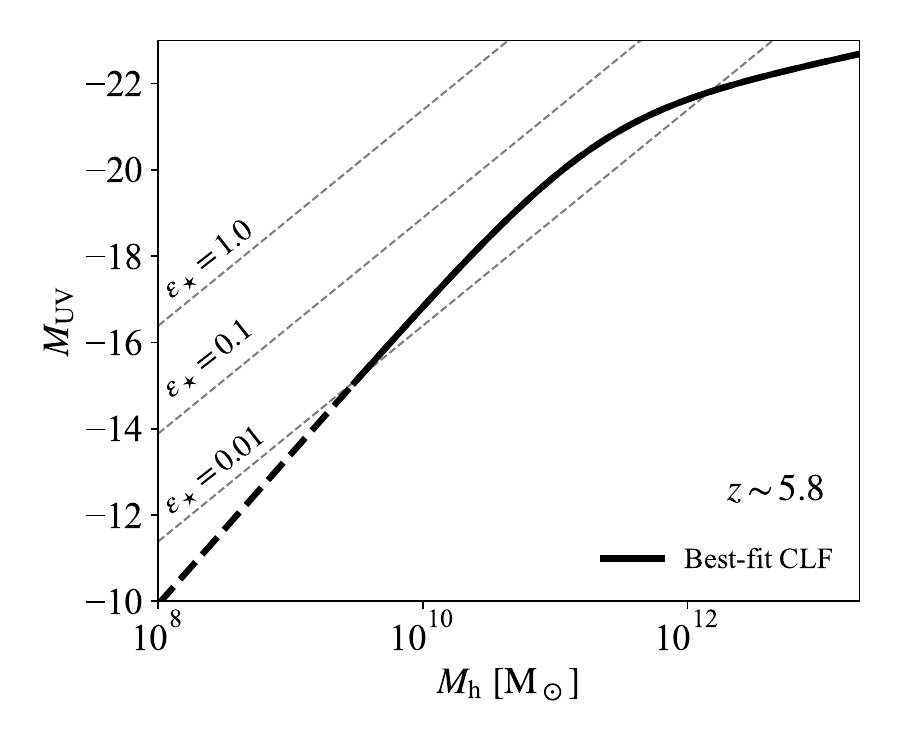}
 \vspace{-0.4cm}
    \caption{The best-fit $M_{\rm UV}-M_{\rm h}$ relation (red) from the CLF framework used in our model. The dashed line indicate the extrapolation below the observed limit of the UV magnitude. For comparison, we overlaid the simple SFR model ${\rm SFR}=\epsilon_\star f_{\rm b} M_{\rm h}/t_{\rm ff}$ from \citet{Ferrara2023} where $\epsilon_\star$ is th star formation efficiency, $f_{\rm b}$ is the cosmic baryon fraction, and $t_{\rm ff}$ is the free-fall timescale. We assumed $L_{\rm UV}=({\rm SFR}/1.15\times10^{-28}{\rm\,M_\odot\,yr^{-1}})\,{\rm erg\,s^{-1}\,Hz^{-1}}$.}\label{fig:Muv-Mh}
\end{figure}

To predict star-forming galaxies clustered around the $\OIII$ emitters, we populate dark matter haloes with galaxies with UV magnitudes according to the $M_{\rm UV}-M_{\rm h}$ relation based on the CLF framework. We use the best-fit CLF parameters matched to the observed UV luminosity function \citep{Bouwens2021} and the angular auto-correlation functions of Lyman-break galaxies at $z\sim6$ \citep{Harikane2022}. Figure \ref{fig:Muv-Mh} shows the best-fit $M_{\rm UV}-M_{\rm h}$ relation for our fiducial model. We extrapolate the $M_{\rm UV}-M_{\rm h}$ relation down to $M_{\rm UV}^{\rm lim}=-10$ to account for the faint population. This limiting UV magnitude corresponds to the atomic cooling limit of halo mass $\sim10^8\rm\,M_\odot$ \citep[e.g.][]{Greif2008}. At this redshift, the best-fit CLF is consistent with a low star formation efficiency of \(\epsilon_\star \sim 0.01\), indicating that galaxies completing reionization are already in a self-regulated regime, unlike the extreme starburst galaxies found at $z \gtrsim 10$.

Each galaxy then emits the ionizing photons to the surrounding IGM according to the LyC leakage $f_{\rm esc}\xi_{\rm ion}$ with the ionizing luminosity, 
\begin{equation}
    \dot{N}_{\rm ion}=f_{\rm esc}\xi_{\rm ion}L_{\rm UV},
\end{equation}
where $f_{\rm esc}$ is the LyC escape fraction and $\xi_{\rm ion}$ is the ionizing photon production efficiency. The total ionizing photon luminosity density above the limiting UV magnitude is,
\begin{equation}
    \dot{n}_{\rm ion}=\langle f_{\rm esc}\xi_{\rm ion}\rangle\,\rho_{\rm UV}(<M_{\rm UV}^{\rm lim}),
\end{equation} 
where the resulting total UV luminosity density from our CLF framework is $\rho_{\rm UV}(<M_{\rm UV}^{\rm lim})\simeq3.3\times10^{26}\rm\,erg\,s^{-1}\,cMpc^{-3}\,Hz^{-1}$ for $M_{\rm UV}^{\rm lim}=-10$, which is consistent with the $1\sigma$ upper limit on the extragalactic background light at $z\sim6$ from the gamma-ray attenuation to high-redshift sources \citep{FermiLAT2018}. For our fiducial model, We assume that the LyC leakage of all galaxies is the same and constant, with a population-averaged LyC leakage of $\log_{10}\langle f_{\rm esc}\xi_{\rm ion}\rangle/\mathrm{[erg^{-1}Hz]} = 24.5$. The mean photoionization rate is $\bar{\Gamma}_{\rm HI}=[\alpha_g/(\alpha_g+3)]\sigma_{912}\lambda_{\rm mfp}\dot{n}_{\rm ion}(1+z)^3\simeq2.2\times10^{-13}\rm\,s^{-1}$ where we assume the EUV slope of $\alpha_g=3$ for all galaxies. 
The mean free path of ionizing photons $\lambda_{\rm mfp}$ is fixed to be $2\rm\,pMpc$ consistent with the rapid evolution of the mean free path at $5.4<z<6.0$ \citep{Becker2021,Zhu2023}.

The average photoionization rate around $\OIII$ emitters is given by the collective population of galaxies including both the observed $\OIII$ emitters and the surrounding unseen galaxies,
\begin{equation}
    \langle \Gamma_{\rm HI}(r)\rangle =\langle \Gamma_{\rm HI}^{\rm OIII}(r)\rangle + \langle \Gamma^{\rm CL}_{\rm HI}(r)\rangle
\end{equation}
where $\langle \Gamma_{\rm HI}^{\rm OIII}(r)\rangle$ is the contribution from the observed $\OIII$ emitters and the contribution from the surrounding galaxies, $\langle \Gamma^{\rm CL}_{\rm HI}(r)\rangle$, is characterized by the ionizing luminosity-weighted correlation function $\langle\xi_g(r')\rangle_L$ between the $\OIII$ emitters and galaxies with UV magnitudes brighter than $M_{\rm UV}^{\rm lim}$ (equivalently, $L_{\rm UV}^{\rm lim})$, that is,
\begin{align}
\langle\Gamma_{\rm HI}^{\rm CL}(r)\rangle&=\frac{\bar{\Gamma}_{\rm HI}}{\lambda_{\rm mfp}}\int\frac{e^{-|\br-\br'|/\lambda_{\rm mfp}}}{4\pi|\br-\br'|^2}\left[1+\langle\xi_g(r')\rangle_L\right] d^3\br', \nonumber \\
&=\bar{\Gamma}_{\rm HI}\left[1+\int_0^\infty\frac{k^2dk}{2\pi}R(k\lambda_{\rm mfp})\langle P_g(k)\rangle_L\frac{\sin kr}{kr}\right],
\end{align}
where $R(k\lambda_{\rm mfp})=\arctan(k\lambda_{\rm mfp})/(k\lambda_{\rm mfp})$ is the Fourier transform of the radiative transfer kernel $e^{-r/\lambda_{\rm mfp}}/(4\pi r^2\lambda_{\rm mfp})$. The ionizing luminosity-weighted galaxy power spectrum is given by,
\begin{align}
    &\langle P_g(k)\rangle_L= \nonumber \\
    &~~\frac{1}{\dot{n}_{\rm ion}}
    \int^\infty_{L_{\rm UV}^{\rm lim}}\dot{N}_{\rm ion}\Phi(L_{\rm UV})P_g(k,L_{\rm UV}|\!>\!M_{\rm min})dL_{\rm UV},
\end{align}
Here, $\Phi(L_{\rm UV})$ is the UV luminosity function of galaxies and $P_g(k,L_{\rm UV}|\!>\!M_{\rm min})$ is the galaxy cross-power spectrum between $\OIII$ emitters with host-halo mass $>M_{\rm min}$ and galaxies with luminosity $L_{\rm UV}$, which is fully characterized by the HOD and CLF model parameters.

In what follows, we first discuss the observed $\OIII$ emitter-Ly$\alpha$ forest cross-correlation in the context of this simple (flexible) model to build our physical understanding. The comparison with full cosmological radiation hydrodynamic simulations is discussed in Section \ref{sec:thesan}.

\begin{figure*}
	\includegraphics[width=\textwidth]{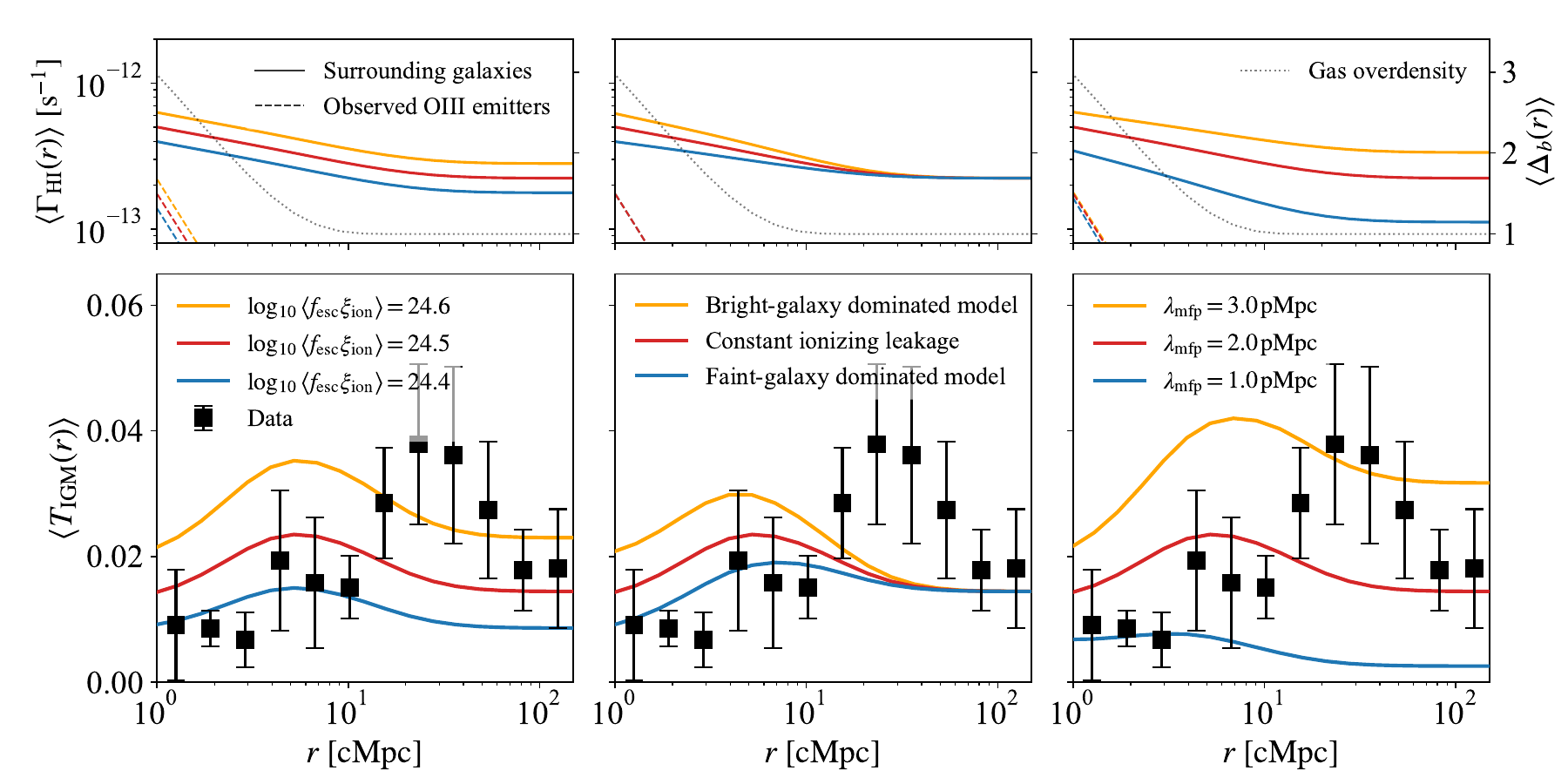}
    \caption{Comparison of the observed mean Ly$\alpha$ forest transmission around $\OIII$ emitters at $\langle z\rangle=5.86$ with the theoretical model based on analytic radiative transfer + CLF framework \citep{Kakiichi2018}. The top panels show the average photoionization rate around $\OIII$ emitters, $\langle \Gamma_{\rm HI}(r) \rangle$ (solid line: contribution from surrounding galaxies; dashed line: central $\OIII$ emitters) on the left y-axis. The average gas overdensity profile around $\OIII$ emitters is indicated on the right y-axis. The bottom panels indicate the mean Ly$\alpha$ forest transmission around $\OIII$ emitters, $\langle T_{\rm IGM}(r)\rangle$ as a function of radial distance from $\OIII$ emitters. (Left): The model prediction with varying average LyC leakage $\langle f_{\rm esc}\xi_{\rm ion}\rangle=24.4,24.5,24.6$. The other parameters are fixed to the fiducial values as indicated in the text. (Middle): The model prediction with varying relative contribution from bright vs faint galaxies (see text). (Right): The model prediction with varying mean free path $\lambda_{\rm mfp}=1.0,2.0,3.0\rm\,pMpc$. The black squares show the observed mean Ly$\alpha$ forest transmission around $\OIII$ emitters with $1\sigma$ error estimated from the Jackknife method.}\label{fig:model}
\end{figure*}

\subsection{LyC leakage and qualitative explanation of the observed mean Ly$\alpha$ forest transmission around $\OIII$ emitters}\label{sec:dens}

Figure \ref{fig:model} (left) shows a comparison of the observed mean Ly$\alpha$ forest transmission around $\OIII$ emitters with the fiducial model where the IGM is kept ionized by galaxies with $M_{\rm UV}<-10$ and constant LyC leakage $\langle f_{\rm esc}\xi_{\rm ion}\rangle=10^{24.5}\rm\,erg^{-1}Hz$, along with varying values of the average LyC leakage. The top panel in Figure \ref{fig:model} (left) shows that the average photoionization rate $\langle\Gamma_{\rm HI}(r)\rangle$ and the gas overdensity $\langle\Delta_b(r)\rangle$ around $\OIII$ emitters explain the origin of this $\langle T_{\rm IGM}(r)\rangle$ profile. 

The model generally predicts excess Ly$\alpha$ forest transmission at large scales and preferential absorption close to $\OIII$ emitters. While the peak location of the excess IGM transmission is not reproduced by the model, the model qualitatively captures the shape of the observed mean Ly$\alpha$ forest transmission around $\OIII$ emitters. The observed normalisation of the average IGM transmission is broadly consistent with the standard value of the average LyC leakage,
\begin{equation}
    \log_{10}\langle f_{\rm esc}\xi_{\rm ion}\rangle/[\rm erg^{-1}Hz]\approx24.5,
\end{equation}
corresponding to, for example, $f_{\rm esc}=0.10$ and $\log_{10}\langle \xi_{\rm ion}\rangle\approx25.5$. However, varying the LyC leakage does not shift the scale of the peak excess IGM transmission. 

The large-scale excess Ly$\alpha$ transmission is caused by the highly ionized environment of the IGM around $\OIII$ emitters. As the enhanced UV background is driven by the collective population of galaxies clustered around $\OIII$ emitters, it extends to larger scales. Note that the contribution from the observed $\OIII$ emitters themselves is sub-dominant to the large-scale excess transmission as we can see from the dashed lines in the top panel. This is consistent with the fact that the observed $\OIII$ emitters contribute only sub-dominantly ($\lesssim0.3\,\%$) to the inferred photoionization rate required to maintain the IGM reionized at the location of the $z=6.215$ transmission spike (Section \ref{sec:individual}). The preferential absorption at small scales is caused by the gas overdensities around $\OIII$ emitters. This occurs because optical depth is highly sensitive to the gas density, $\tau_\alpha\propto \Delta_b^{1.97}/\Gamma_{\rm HI}$ where we assumed $\gamma=1.04$, meaning that at small scales, the enhanced UV background is compensated by the gas overdensities around $\OIII$ emitters.

We emphasize that the excess transmission is caused by the increased occurrence probability of Ly$\alpha$ transmission spikes around $\OIII$ emitters. As the Ly$\alpha$ optical depth (e.g. \citealt{Fan2006,Becker2015}), 
\begin{align}
    &\tau_\alpha\approx \nonumber \\
    &~~33\Delta_b^{\beta}\left(\frac{\Gamma_{\rm HI}}{3\times10^{-13}\rm\,s^{-1}}\right)^{-1}\left(\frac{T_0}{10^4\rm\,K}\right)^{-0.72}\left(\frac{1+z}{6.86}\right)^{4.5},
\end{align}
is already high at mean density, the observable transmission spikes (with height $>e^{-\tau_\alpha^{\rm th}}$, $\tau_\alpha^{\rm th}=3$ for $e^{-\tau_\alpha^{\rm th}}=0.05$) require coincidental {\it underdense fluctuations} in the enhanced patch of the UV background, satisfying
\begin{equation}
    \Delta_b<0.29\left(\frac{\tau_\alpha^{\rm th}}{3}\right)^{1/2}\left(\frac{\Gamma_{\rm HI}}{3\times10^{-13}\rm\,s^{-1}}\right)^{1/2}\left(\frac{T}{10^4\rm\,K}\right)^{0.36},\label{eq:spike_condition}
\end{equation}  
at $z=5.86$. The probability of occurence of such underdense fluctuations is low, but finite, and increases gradually toward galaxies as the UV background increases. On the other hand, the probability plummets when it comes too close to galaxies where the increased average gas density will diminish the probablity of underdense fluctuations to occur, explaining why we observe large field-to-field variance in the individual associations between $\OIII$ emitters and Ly$\alpha$ transmission spikes and why galaxies are not located exactly at the peak of the transmission spikes. 

It is worth noting that highly ionized regions of the IGM are exactly where the mean free path is expected to be the longest. As we will discuss below, this helps explain why the model (with a constant mean free path) underestimates the peak location of the excess IGM transmission. The mean free path could be longer than average in the highly ionized IGM around $\OIII$ emitters, allowing ionizing photons to penetrate further, potentially leading to the excess IGM transmission at larger scales.

\subsection{Contribution of bright and faint galaxies to reionization}\label{sec:source_bias}

\begin{figure}
	\includegraphics[width=\columnwidth]{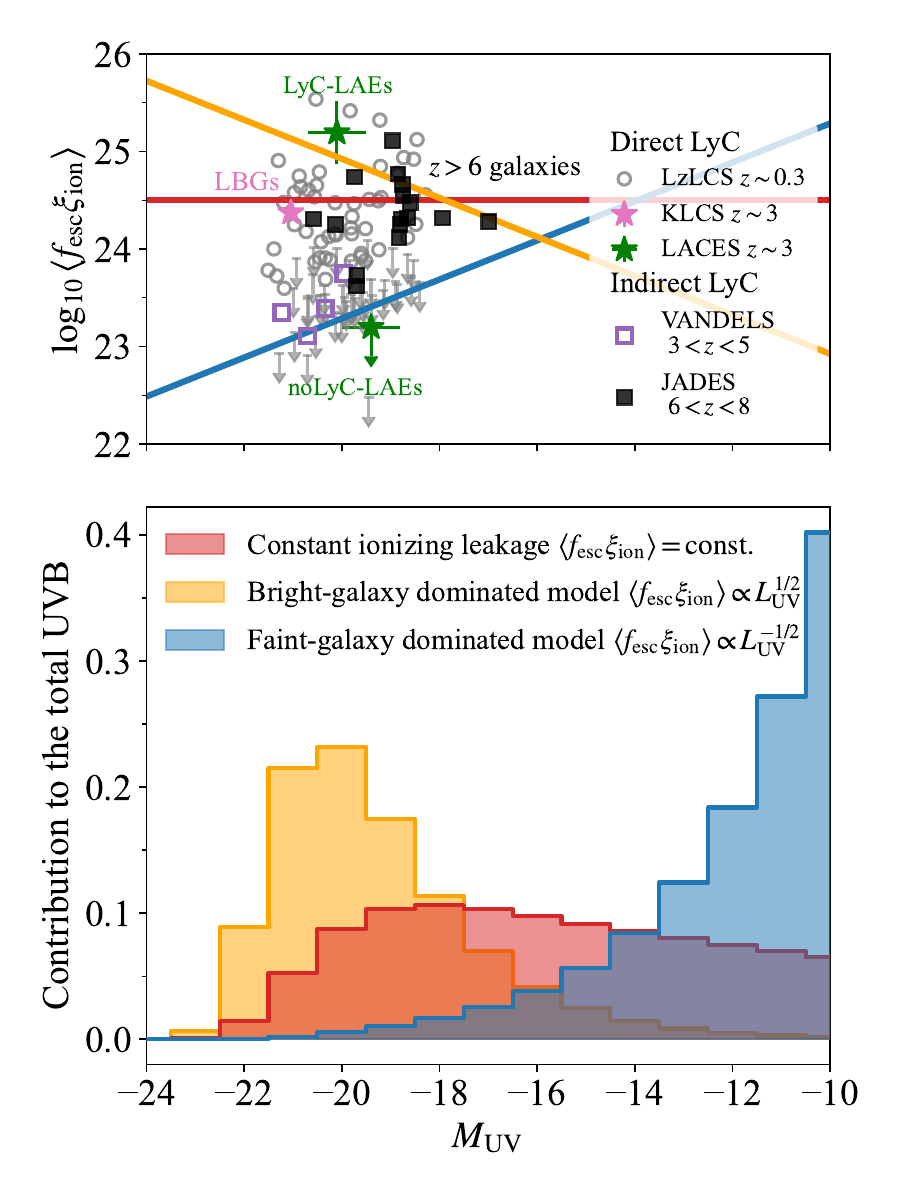}
 \vspace{-0.6cm}
    \caption{Relative contribution of galaxies to the total ionizing budget at $z\sim5.8$. ({\bf Top}): Different average LyC leakage as a function of UV magnitudes of galaxies (red: constant LyC leakage, yellow: bright galaxy-dominated model, blue: faint galaxy-dominated model). Symbols indicates the observational estimate of LyC leakage from the direct LyC detection (gray circles: LzLCS \citep{Flury2022}, pink stars: KLCS \citep{Steidel2018}, green stars: LACES \citep{Nakajima2020}) and the indirect method using UV/optical spectral features (purple squares: VANDELS \citep{Saldana-Lopez2023}, black squares: JADES \citep{Saxena2023b}). ({\bf Bottom}): The fractional contribution of galaxies to the total UV background, $\bar{\Gamma}_{\rm HI}$, at $z\sim5.8$ for the three different models.}\label{fig:UVB_contribution}
\end{figure}

Does the different contribution of bright and faint galaxies to the UV background affect the observed mean Ly$\alpha$ forest transmission around $\OIII$ emitters? In Figure \ref{fig:model} (middle), we show the model prediction where the average LyC leakage of galaxies varies as a function of UV luminosities, where the ionizing budget can be dominated by bright ($M_{\rm UV}<-20$) or faint ($M_{\rm UV}>-14$) galaxies. We model this by assuming a simple power-law dependence of the average LyC leakage on the UV luminosity, i.e. $\langle f_{\rm esc}\xi_{\rm ion}\rangle\propto L_{\rm UV}^{\zeta}$, where $\zeta=1/2$ for the bright galaxy-dominated model and $\zeta=-1/2$ for the faint galaxy-dominated model. All the models are normalised to give the same average photoionization rate. Figure \ref{fig:UVB_contribution} illustrates the relative contribution of galaxies to the total ionizing budget at $z\sim5.8$ for the three different models. 


Figure \ref{fig:model} (middle) shows the impact of varying contributions of galaxies to the total ionizing budget on the mean Ly$\alpha$ forest transmission around $\OIII$ emitters. The figure indicates that the observed peak location of the excess IGM transmission around $\OIII$ emitters cannot be explained by the different contributions of bright and faint galaxies to the total ionizing budget. Although the bright galaxy-dominated model predicts a more pronounced excess in Ly$\alpha$ forest transmission compared to the faint galaxy-dominated model, it still fails to explain the observed large-scale excess transmission.

It is informative to see the dependence of the radial photoionization rate profile on the ionizing sources in the linear limit $\langle P_g(k) \rangle_L\approx b_{\rm OIII}\langle b_g\rangle_L P_m(k)$ where $P_m(k)$ is the linear matter power spectrum, which yields 
\begin{align}
&    \langle\Gamma^{\rm CL}_{\rm HI}(r)\rangle\approx \nonumber \\
&~~\bar{\Gamma}_{\rm HI}\left[
        1+b_{\rm OIII}\langle b_g\rangle_L\int_0^\infty\frac{k^2dk}{2\pi^2}
        R(k\lambda_{\rm mfp})P_m(k)\frac{\sin kr}{kr},
    \right]
\end{align}
where $b_{\rm OIII}$ is the bias of $\OIII$ emitters and $\langle b_g\rangle_L$ is the luminosity-weighted bias of ionizing galaxies, which corresponds to the source bias in the linear theory of \citet{Pontzen2014,Gontcho2014},
\begin{equation}
  \langle b_g\rangle_L=\frac{\int^\infty_{L^{\rm min}_{\rm UV}} \langle f_{\rm esc}\xi_{\rm ion}\rangle L_{\rm UV}b_g(L_{\rm UV})\Phi(L_{\rm UV})dL_{\rm UV}}{\int^\infty_{L^{\rm min}_{\rm UV}} \langle f_{\rm esc}\xi_{\rm ion}\rangle L_{\rm UV}\Phi(L_{\rm UV})dL_{\rm UV}}.
\end{equation}
For the constant LyC leakage model, our source bias is $\langle b_g\rangle_L=2.87$ for $M_{\rm UV}^{\rm lim}=-10$. Note that this is larger than the average galaxy bias $b_g=1.58$ brighter than $M_{\rm UV}^{\rm lim}=-10$ because more luminous galaxies in massive haloes contribute more to the total ionizing budget than the faint galaxies even in the case of constant LyC leakage (see Figure \ref{fig:UVB_contribution}). Galaxies with $M_{\rm UV}\approx-18$ in haloes of $M_{\rm h}\sim 10^{10-11}\rm\,M_\odot$ in fact contribute most to the total ionizing budget. The biases of bright- and faint-galaxy dominated models are $\langle b_g\rangle_L=4.11$ and $\langle b_g\rangle_L=1.87$, respectively. While our models bracket the reasonable variation of LyC leakage from different galaxies expected from previous observations \citep{Steidel2018,Nakajima2020,Flury2022,Saldana-Lopez2023,Saxena2023b}, the model still cannot explain the observed large-scale excess transmission around $\OIII$ emitters. This hints that additional physical processes beyond the relative contribution of galaxies to reionization are required to explain the observed mean Ly$\alpha$ forest transmission around $\OIII$ emitters.




\subsection{Mean free path of ionizing photons and the spatial fluctuations of absorbers}\label{sec:absorbers}

The rapid evolution of the mean free path of ionizing photons and its spatial fluctuations are an obvious possibility for explaining the large-scale excess transmission around $\OIII$ emitters. In Figure \ref{fig:model} (right), we show the model prediction where the values of the constant mean free path of ionizing photons are varied from $1\rm\,pMpc$ to $3\rm\,pMpc$. This illustrates the possible range at $z\sim5.8$ measured by \citet{Becker2021,Zhu2023} from stacked quasar spectra (see also \citealt{Bosman21, Satyavolu2023, Roth2023}). While a larger mean free path increases the excess IGM transmission at larger scales, this model variation alone still cannot explain the observed large-scale excess transmission. As all three of these model variations have the same total ionizing photon budget $\dot{n}_{\rm ion}$, an increasing mean free path gives rise to an increased UV background $\bar{\Gamma}_{\rm HI} \propto \dot{n}_{\rm ion} \lambda_{\rm mfp}$, resulting in an overestimate of the IGM transmission at larger scales. We have explored models with lower LyC leakage to compensate for the increase in the UV background. However, even under the extreme assumption of a large mean free path $\gg3\rm\,pMpc$, we found it difficult to reconcile the observed large-scale excess transmission.

Thus, in order to explain the observed large-scale excess, we need to consider the spatial fluctuations of the mean free path of ionizing photons. This arises from the spatial fluctuations of $\HI$ absorbers, i.e. the sink of ionizing photons \citep{Davies2016,DAloisio2020}. The linear perturbation of the cosmological radiative transfer equation provides an illustrative modification to the UV background fluctuations. By Fourier transforming the cross-power spectrum between $\OIII$ emitters and the UV background, $P_{\rm O{\scriptscriptstyle \,III},\Gamma}(k)=b_{\rm OIII}b_{\Gamma}(k)P_m(k)$ where $b_{\Gamma}(k)=\frac{(\langle b_g\rangle_L-b_\kappa)R(k\lambda_{\rm mfp})}{1+b_{\kappa,\Gamma}R(k\lambda_{\rm mfp})}$ \citep{Pontzen2014,Gontcho2014}, we find,
\begin{align}
    & \langle\Gamma^{\rm CL}_{\rm HI}(r)\rangle/\bar{\Gamma}_{\rm HI}-1\approx b_{\rm OIII}\langle b_g\rangle_L\times\nonumber \\
    & \int_0^\infty\frac{k^2dk}{2\pi^2}
        \left[\frac{1-b_\kappa/\langle b_g\rangle_L}{1+b_{\kappa,\Gamma}R(k\lambda_{\rm mfp})}\right]R(k\lambda_{\rm mfp})P_m(k)\frac{\sin kr}{kr},\label{eq:linear_theory}
\end{align}
where $b_\kappa$ is the bias of absorbers and $b_{\kappa,\Gamma}$ is the linear response of the bias of absorbers with respect to the perturbation of photoionization rate. The latter takes the value between $-1<b_{\kappa,\Gamma}<0$. This linear limit  is mathematically consistent and self-consistently takes into account the impact of both absorbers and sources of ionizing photons on the UV background fluctuations (with a cost of introducing two additional bias parameters, $b_\kappa$ and $b_{\kappa,\Gamma}$). 

We have experimented with the spatial fluctuations of absorbers and how they could increase the excess transmission on large scales. We find that, using the linear theory, while the spatial variation of absorbers can indeed increase the excess transmission on large scales, the peak location of the excess IGM transmission around $\OIII$ emitters still cannot be fully reproduced in the context of the linear theory. A full non-linear treatment of the spatial fluctuations of absorbers in radiative transfer simulations is necessary to quantitatively predict the observed mean Ly$\alpha$ forest transmission around $\OIII$ emitters. In fact, \citet{Meyer2020} required the introduction of that effect to explain the large-scale excess transmission in the galaxy-transmission spike cross-correlation function. 
Recall from Section \ref{sec:dens} that the regions of the IGM giving rise to transmission spikes are likely highly ionized, such that they should represent regions with particularly long mean free paths. It is therefore logical to expect that fluctuations in the mean free path should impact the shape of the transmission excess. Nonetheless, a fully self-consistent quantitative prediction of the effect demands resolving the self-shielding absorbers in radiative transfer or radiation hydrodynamic simulations (e.g.~\citealt{Cain2023}).
We leave this for future work. Here, we simply conclude by noting that the spatial fluctuations of absorbers would be important to explain the observed mean Ly$\alpha$ forest transmission around $\OIII$ emitters. 

\begin{figure*}
    \centering
	\includegraphics[width=1.8\columnwidth]{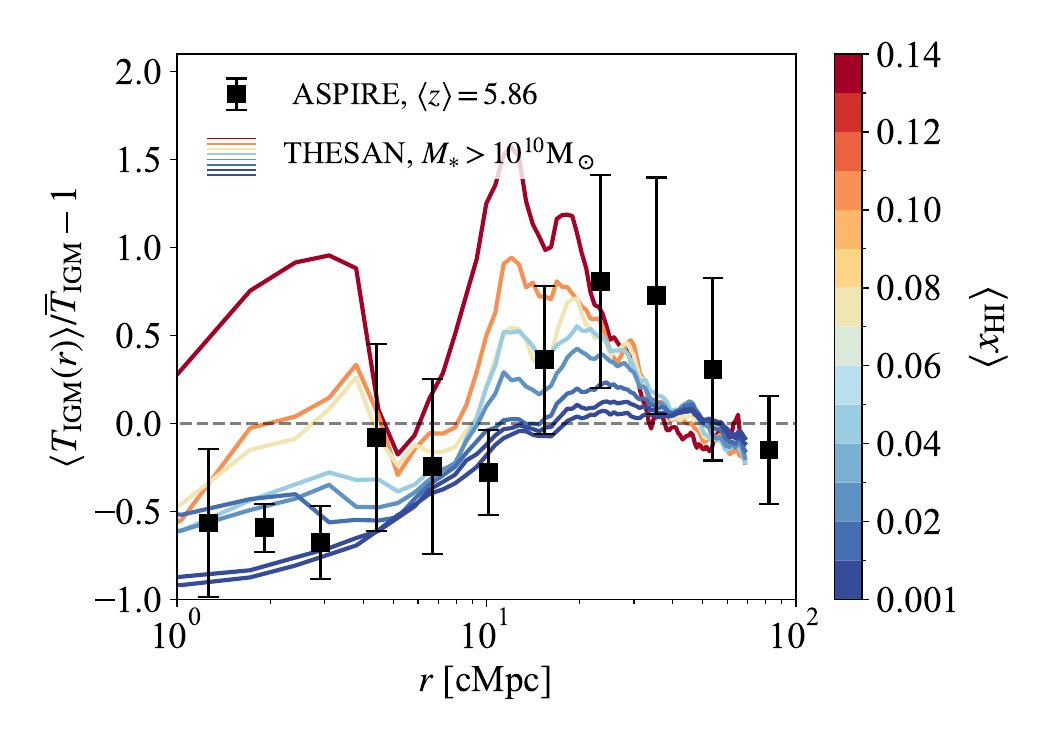}
    \vspace{-0.6cm}
    \caption{Comparison of the observed $\OIII$ emitter-Ly$\alpha$ forest cross-correlation, $\langle T_{\rm IGM}(r)\rangle/\overline{T}_{\rm IGM}-1$, at $\langle z\rangle=5.86$ (black squares) with the results from the THESAN cosmological radiation hydrodynamic simulation. The coloured curves show the results form THESAN-1 snapshots from $z=6.2$ to $5.5$ corresponding to global $\HI$ fractions of $\langle\xHI\rangle=0.14$ to $3.4\times10^{-3}$. We chose the central galaxies with stellar mass of $M_\ast>10^{10}\,\rm M_\odot$ from the THESAN simulation. The black errorbars show the $1\sigma$ error estimated from the Jackknife method.}\label{fig:thesan}
\end{figure*}

\section{Galaxy-Ly$\alpha$ forest cross-correlation in the context of cosmological radiation hydrodynamic simulations}\label{sec:thesan}

The above analysis highlights various physical processes shaping the observed $\OIII$ emitter-Ly$\alpha$ forest cross-correlation. Could the observed cross-correlation be explained by the cosmological radiation hydrodynamic simulations which self-consistently take into account the all these effects? 

In Figure \ref{fig:thesan} we compare the predictions of the galaxy-Ly$\alpha$ forest cross-correlation, $\langle T_{\rm IGM}(r)\rangle/\overline{T}_{\rm IGM}-1$, from the THESAN cosmological radiation hydrodynamic simulations (\citealt{Kannan2022, Garaldi2022, Smith2022}, see also \citealt{Garaldi2024} for the public data release) with our ASPIRE result. $\OIII$ emitters are represented by the central galaxies with stellar mass of $M_\ast>10^{10}\,\rm M_\odot$ in the THESAN simulation. The selection based on star formation rate or $\OIII$ flux gives the similar result \citep{Garaldi2024b}. We then measure the mean Ly$\alpha$ forest transmission around them in the same way as we did for the ASPIRE data. The coloured curves shows the resulting cross-correlation from the THESAN-1 simulation at redshifts from $z=6.2$ to $5.5$ corresponding to global $\HI$ fractions of $\langle\xHI\rangle=0.14$ to $3.4\times10^{-3}$.

We find that the observed $\OIII$ emitter-Ly$\alpha$ forest cross-correlation agrees generally well with the THESAN's prediction around central galaxies with stellar masses above $M_\ast>10^{10}\,\rm M_\odot$. This is surprising, given that the simulation was run {\it before} the observation was made. The only adjustable parameter is the selection of $\OIII$ emitters in the THESAN simulation. We discuss in Section \ref{sec:halomass} a more in-depth analysis of the different models of $\OIII$ emitters based on halo mass, using a more cosmologically sound measurement of their spatial correlation function.

THESAN shows the peak of the excess IGM transmission around the central galaxies is at $r\sim10-30\,\rm cMpc$. While the observed peak location in ASPIRE is slightly larger than in THESAN, they are in reasonable agreement within the error bars. The excess IGM transmission increases with the global $\HI$ fraction, with the peak locations shifting gradually from small to large scales as reionization progresses \citep{Garaldi2022}. THESAN also shows excess absorption due to gas overdensities around the central galaxies at $r\lesssim10\,\rm cMpc$.\footnote{The apparent excess transmission for the three highest redshift snapshots (corresponding to $\langle\xHI\rangle>0.07$) is likely an artifact due to the small number of galaxies with $M_\ast>10^{10}\,\rm M_\odot$. These galaxies are not tracing a representative IGM structure but rather particular structures within the simulation box. The same is true for the innermost radial bin of Figure \ref{fig:tension} (right).} Overall, THESAN captures the observed large-scale excess transmission around $\OIII$ emitters, which is not reproduced by the analytic model in Section \ref{sec:interpretation}.

\subsection{Late end of reionization and neutral islands at $z<6$}\label{sec:late_reion}

\begin{figure*}
	\includegraphics[width=\textwidth]{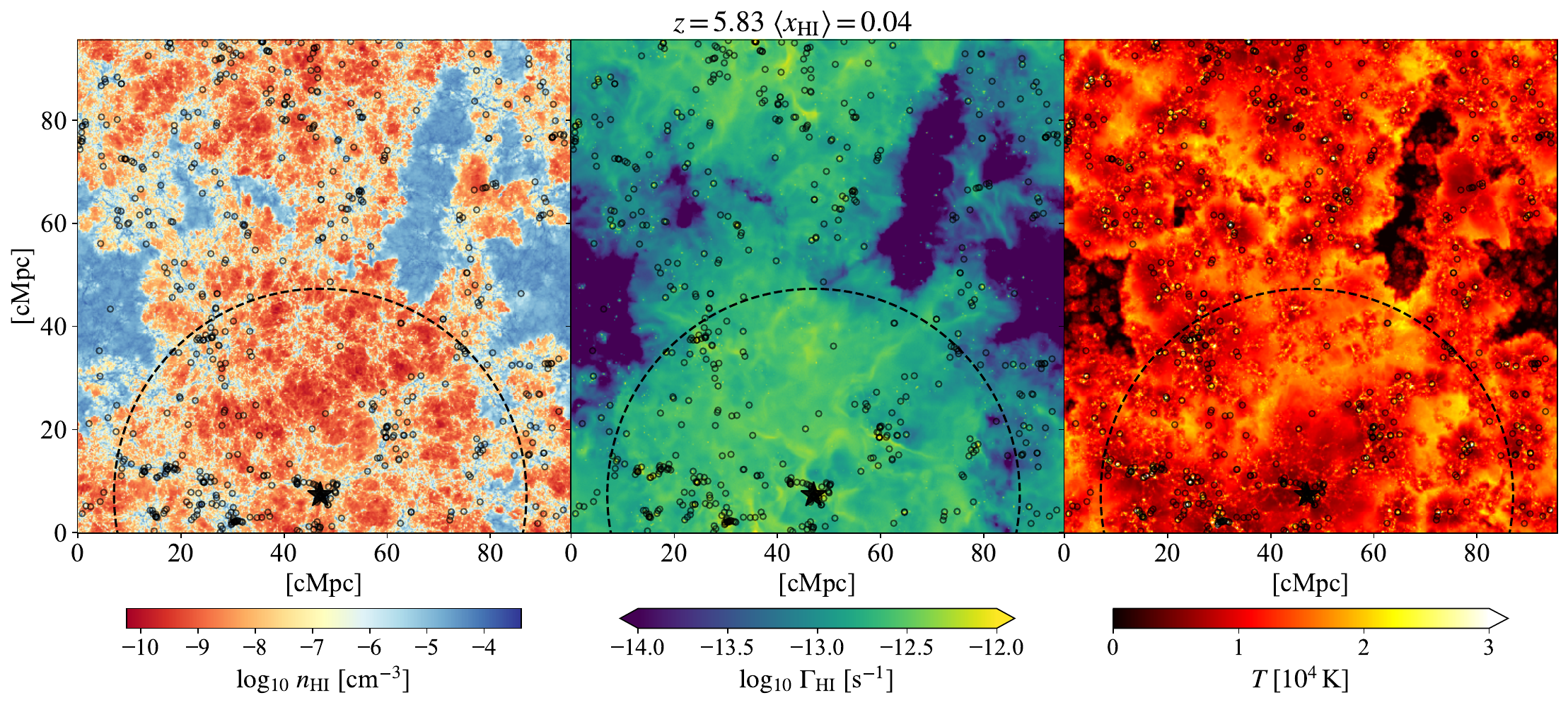}
 \vspace{-0.45cm}
    \caption{Sliced maps of the $\HI$ number density $n_{\HI}$ (left), the photoionization rate $\Gamma_{\rm HI}$ (middle), and the temperature $T$ (right) around a central galaxy (black star symbol) with a stellar mass of $M_\ast>10^{10}\,\rm M_\odot$ in the THESAN-1 snapshot at $z=5.83$, corresponding to the fourth bluest galaxy-Ly$\alpha$ forest cross-correlation in Figure \ref{fig:thesan}. The open circles show the distribution of surrounding galaxies with stellar masses of $M_\ast>5\times10^7\,\rm M_\odot$. The large dashed circle indicates a radius of $40\rm\,cMpc$ around the central galaxy. All sliced maps have a width of $3.7\,\rm cMpc$.}\label{fig:thesan_map}
\end{figure*}

In order to understand the origin of the agreement between ASPIRE and THESAN in the large-scale excess IGM transmission around galaxies, in Figure \ref{fig:thesan_map} (left), we show the sliced map of the $\HI$ number density $n_{\HI}$ around a central galaxy with a stellar mass of $M_\ast>10^{10}\,\rm M_\odot$ at $z=5.83$. The map corresponds to the THESAN galaxy-Ly$\alpha$ forest cross-correlation (the fourth bluest curve from the bottom) that best reproduces the ASPIRE result in Figure \ref{fig:thesan}. 

The most notable physical feature in THESAN missed by the analytic model is the presence of neutral islands in the IGM at $z<6$. The neutral islands are regions where the ionization fronts (I-fronts) have not yet reached, leaving the IGM fully neutral. The neutral islands are clearly visible in the $\HI$ number density map in Figure \ref{fig:thesan_map} (left). Since these regions (with a Gunn-Peterson optical depth of $\tau_{\rm GP}\sim10^5$) completely absorb the Ly$\alpha$ forest transmission, the observable transmission occurs only within the ionized bubbles. To examine the impact of neutral islands on large-scale excess IGM transmission, we compare the prediction of the galaxy-Ly$\alpha$ forest cross-correlation at $z=5.83$ with the masked cross-correlation that excludes Ly$\alpha$ forest pixels where the underlying IGM is predominantly neutral ($\xHI > 0.1$), as detailed in Appendix \ref{app:CCF_test}. We find that excluding neutral islands has little impact on the shape of the galaxy-Ly$\alpha$ forest cross-correlation, suggesting that the presence of neutral islands does not directly contribute to the excess IGM transmission around galaxies. Therefore, the IGM fluctuations {\it inside} bubbles must also be present to produce the excess IGM transmission.

This means that while ionized bubbles are necessary, they are not a sufficient condition for excess IGM transmission. As illustrated in Figure \ref{fig:thesan_map} by the dashed circle of radius $R=40\rm\,cMpc$ around a central galaxy, the outermost extent of the excess Ly$\alpha$ forest transmission around galaxies coincides with the typical size of the ionized bubbles. ($R_b \sim 40-60\,\rm cMpc$ at the end of reionization,~\citealt{Wyithe2004, Neyer2023, Lu2024}). As the excess IGM transmission should occur within the typical radius of ionized bubbles, the outermost radius at which $\langle T_{\rm IGM}(r)\rangle/\overline{T}_{\rm IGM}-1 \approx 0$ is the lower limit for the typical bubble size around galaxies.\footnote{It is important to remember that the peak location of the excess IGM transmission results from the two competing effects: the enhanced UV background from surrounding galaxies and the gas overdensities around the central galaxy. Therefore, the peak location should not be mistaken for the typical scale of ionized bubbles around galaxies.}

In this interpretation, the observed $\OIII$ emitter-Ly$\alpha$ forest cross-correlation in ASPIRE suggests that $\OIII$ emitters at $\langle z \rangle = 5.86$ must be surrounded by large ionized bubbles exceeding $\sim 50 \rm\,cMpc$ in radius. Additionally, Figure \ref{fig:thesan} indicates that the simulated excess IGM transmission diminishes rapidly towards the end of reionization. This implies that the substantial excess in IGM transmission observed in ASPIRE supports the notion of a late end to reionization at $z < 6$, which is consistent with quasar absorption studies \citep{Bosman2022,Zhu2021,Zhu2022, Becker2024}.

\subsection{Spatial fluctuations of UV background, IGM opacities,  and temperature}

\begin{figure}
\centering
	\includegraphics[width=0.9\columnwidth]{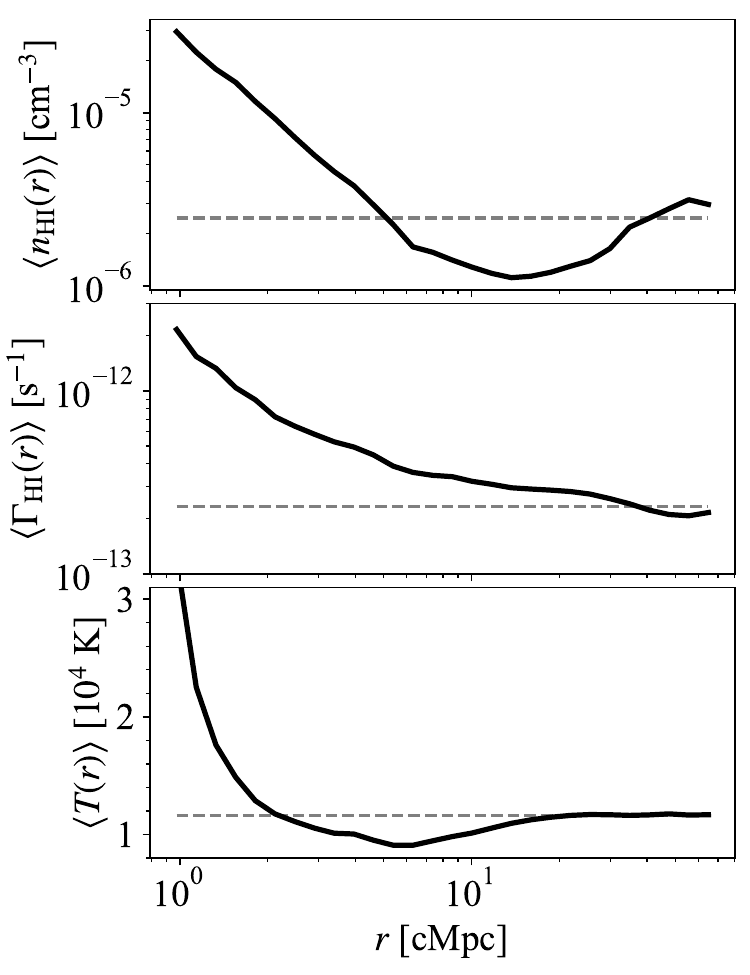}
    \vspace{-0.1cm}
    \caption{Spherically-averaged profile of $\HI$ number density (top), photoionization rate (middle), and temperature (bottom) around galaxies with stellar mass of $M_\star>10^{10}\rm\,M_\odot$ at $z=5.83$ in the THESAN-1 snapshot. The average values are indicated by horizontal dashed lines. }
    \label{fig:spherical_profile}
\end{figure}

The spatial fluctuations of the physical state of the IGM inside the ionized bubbles must be significant to produce the large excess Ly$\alpha$ forest transmission around $\OIII$ emitters as observed by ASPIRE. As shown in Figure \ref{fig:thesan_map}, the THESAN simulation supports the idea that fluctuations in all IGM properties -- gas density, temperature, and UV background modulated by the distribution of ionizing sources and self-shielded absorbers -- underlie the shape the galaxy-Ly$\alpha$ forest cross-correlation.

To further examine the origin of the excess IGM transmission, we show the spherically averaged profiles of $\HI$ number density, photoionization rate, and temperature around galaxies with stellar masses of $M_\ast > 10^{10}\,\rm M_\odot$ at $z = 5.83$ in Figure \ref{fig:spherical_profile}.

The enhancement of the spherically averaged photoionization rate $\langle\Gamma_{\rm HI}(r)\rangle$ extends out to $\sim30\,\rm cMpc$. The distribution of surrounding galaxies with stellar masses of $M_\ast > 5 \times 10^7\,\rm M_\odot$ (open circles) shows numerous galaxies around a central galaxy, whose leaked LyC radiation collectively enhances the UV background on large scales, contributing to the excess IGM transmission around $\OIII$ emitters \citep{Garaldi2024b}. 

This large-scale enhancement of the UV background fluctuations exceeds expectations from the analytic RT framework with a fixed mean free path in Section \ref{sec:interpretation}, suggesting additional impacts from the spatial fluctuations of absorbers within the bubbles. Our experimentation with the linear theory \citep[equation \ref{eq:linear_theory},][]{Pontzen2014,Gontcho2014} also supports the hypothesis that spatially varying absorbers play an important role in the large-scale enhancement of the UV background.

This results from the self-consistent treatment of absorbers and ionizing sources in cosmological radiation hydrodynamic simulations. In highly ionized regions, the mean free path is longer than average, which allows galaxies in these regions to have longer mean free paths. Because of this spatially varying mean free path, resulting from the response of the absorbers' distribution to the local photoionization, the surrounding galaxies collectively create a higher and more extended UV background around the central galaxies. As a result, the effect contributes to the excess Ly$\alpha$ forest transmission around $\OIII$ emitters, bringing the simulation closer to the observed signal.

Interestingly, the temperature fluctuations may have a non-negligible impact on the shape of excess IGM transmission around galaxies. Figure \ref{fig:spherical_profile} shows that the spherically averaged temperature increases from the inner region of bubbles at $\sim 6 \, \rm cMpc$ to larger radii at $\gtrsim 20\, \rm cMpc$. This is because the outer region has been reionized more recently than the inner region. We find the average temperature of $T \approx 1.2 \times 10^4 \, \rm K$ at the outskirts of the bubbles. This is broadly consistent with the estimate of photoheating due to photoionization across the I-fronts, resulting in a temperature increase of
\begin{equation}
T \simeq \frac{2}{3k_{\rm B}} \frac{G_{\rm HI}/\Gamma_{\rm HI}}{2 + Y/2X} \approx 1.15 \times 10^4 \, \rm K \left( \frac{2 + \alpha_{\rm eff}}{5} \right)^{-1},
\end{equation}
to leading order, where the numerical factor includes both photoheating of $\HI$ and $\HeI$. Here, $\alpha_{\rm eff}$ is the effective EUV slope of ionizing sources at the position of the I-front, and $G_{\rm HI}$ is the thermal energy injected by $\HI$ photoionization with $G_{\rm HI}/\Gamma_{\rm HI} \approx h_{\rm p}\nu_{\rm HI}/(2 + \alpha_{\rm eff})$ \citep[e.g.][]{Abel1999,Kakiichi2017}. The increase in temperature below $r \approx 3 \, \rm cMpc$ is due to the increasing contribution from heated gas resulting from shocks, feedback, and adiabatic compression in the gas around galaxies.

As the Ly$\alpha$ optical depth scales as $\tau_\alpha \propto \Delta_b^2 \Gamma_{\rm HI}^{-1} T^{-0.72}$, an increase in temperature at the outskirts of bubbles may contribute to the enhancement of IGM transmission on large scales. While the temperature contrast between the inside and the outskirts of bubbles results in only a small decrease in the Ly$\alpha$ optical depth by a factor of $\sim (12000\,\rm K/8000\,\rm K)^{-0.72} \approx 0.75$, this is sufficient to elevate low Ly$\alpha$ transmission from, e.g., $e^{-\tau_\alpha} = 0.018$ ($\tau_\alpha = 4$) to a sizeable transmission spike of $e^{-\tau_\alpha} = 0.05$ ($\tau_\alpha = 3$). Although small, this effect is comparable to the change in the Ly$\alpha$ optical depth due to the photoionization rate in the same region, $\sim (3 \times 10^{-13} \, \rm s^{-1}/2 \times 10^{-13} \, \rm s^{-1})^{-1} \approx 0.67$. Thus, the impact of the IGM temperature fluctuations on the exact shape of the excess Ly$\alpha$ forest transmission may not be ignored.

\begin{figure}
\centering
\hspace{-0.4cm}
	\includegraphics[width=1.05\columnwidth]{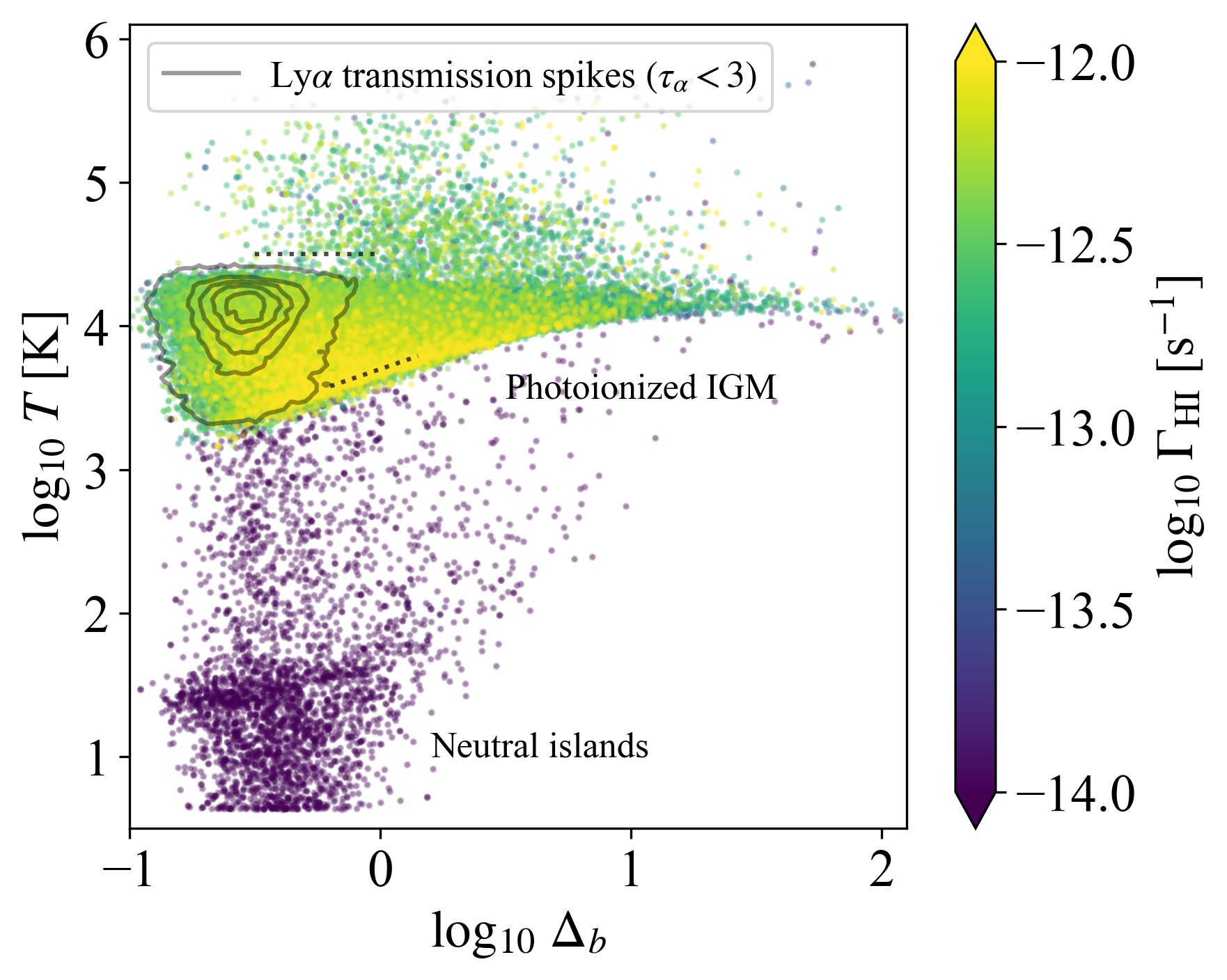}
    \caption{The distribution of Ly$\alpha$ forest pixels in temperature-density-photoionization space in the THESAN-1 simulation at $z=5.83$. Each point represents the temperature $T$, gas overdensity $\Delta_b$, and photoionization rate $\Gamma_{\rm HI}$ at the spatial position of a Ly$\alpha$ forest pixel. Random Ly$\alpha$ forest pixels are drawn from random 300 skewers through the simulation box.
    The contours show the region of high Ly$\alpha$ transmission pixels with $\tau_\alpha<3$ ($e^{-\tau_\alpha}>0.05$). The outermost contour encloses 99\% of the total high Ly$\alpha$ forest transmission, and each subsequent contour encloses top 80\%, 60\%, 35\%, and 15\% of the total high Ly$\alpha$ forest transmission pixels. The horizontal and diagonal dotted lines show the slopes of $T \propto \mathrm{const.}$ and $\Delta_b^{-0.6}$, indicating the expected relation of the recently photoheated IGM and the asymptotic limit after cooling, respectively.}
    \label{fig:phase_diagram}
\end{figure}

Figure \ref{fig:phase_diagram} further clarifies the interplay between density, UV background, and thermal fluctuations of the IGM, and the origin of high Ly$\alpha$ transmission spikes ($\tau_\alpha < 3$) in terms of the `phase diagram' -- the temperature-density-photoionization relation of Ly$\alpha$ forest pixels. We find that the majority of high Ly$\alpha$ forest transmission spikes arise from underdense ($\Delta_b < 1$), photoionized IGM with a high UV background. The Ly$\alpha$ forest pixels at transmission spikes satisfy the condition:
\begin{equation}
\Gamma_{\rm HI} \gtrsim 10^{12.5} \, {\rm s^{-1}} \left(\frac{\tau_\alpha^{\rm th}}{3}\right)^{-1} \left(\frac{\Delta_b}{0.3}\right)^{2} \left(\frac{T}{10^4 \, \rm K}\right)^{-0.72},
\end{equation}
confirming our analysis in Section \ref{sec:interpretation} (equation \ref{eq:spike_condition}) \citep[also][]{Kakiichi2018,Meyer2020}. Note that in Figure \ref{fig:phase_diagram}, the high occurrence probability of transmission spikes is shifted towards recently photo-heated gas with a temperature just above $T \approx 10^{4} \, \rm K$, whereas the IGM after cooling satisfying the asymptotic temperature-density relation ($T \propto \Delta_b^{0.6}$, e.g. \citealt{McQuinn2016}) contributes little to the transmission spikes. This means that the gas photo-heated by recent reionization to $T > 10^4 \, \rm K$ at the outskirts of bubbles creates a more favourable condition for Ly$\alpha$ transmission spikes, hence increasing the contribution to the large-scale excess Ly$\alpha$ forest transmission.

In summary, the better agreement between ASPIRE and THESAN should arise from the significant fluctuations of the UV background generated by both the distribution of ionizing sources and absorbers, as well as the thermal fluctuations of the IGM inside ionized bubbles. Although, due to the apparent insensitivity of the galaxy-Ly$\alpha$ forest cross-correlation to the presence of neutral islands, we can only place a lower limit on the typical bubble size around $\OIII$ emitters, the large excess Ly$\alpha$ forest transmission requires significant fluctuations in the IGM properties. Such large fluctuations in the UV background and temperature are more naturally produced through reionization. Thus, the observed large excess Ly$\alpha$ forest transmission around $\OIII$ emitters strongly supports the idea that reionization is still ongoing, characterized by ionized bubbles of $\gtrsim 50 \, \rm cMpc$ around $\OIII$ emitters, and is on the verge of completion at $\langle z \rangle = 5.86$.

\subsection{Redshift evolution}

\begin{figure}
\hspace*{-0.25cm}
	\includegraphics[width=1.051\columnwidth]{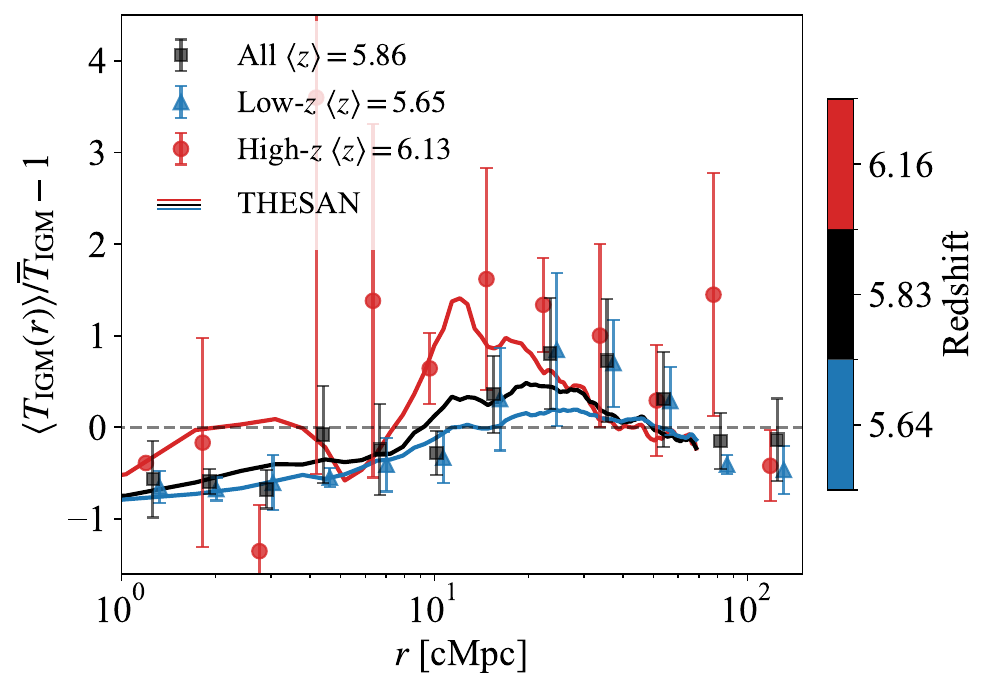}
    \caption{Redshift evolution of galaxy-Ly$\alpha$ forest cross-correlation in ASPIRE and THESAN. The symbols with error bars show the observed signals around $\OIII$ emitters at three different redshift bins (All: $\langle z\rangle=5.86$ $(5.4<z<6.5)$ (black square), low-$z$:  $\langle z\rangle=5.65$ $(5.4<z<5.8)$ (blue triangle), high-$z$: $\langle z\rangle=6.13$ $(5.8<z<6.5)$ (red circle)). The curves show the simulated signals around galaxies with stellar masses of $M_\star>10^{10}\,\rm M_\odot$ at the closest redshifts ($z=6.16$ (red), $5.83$ (black), $5.64$ (blue)) to the observed values.}
    \label{fig:redshift_evolution_full}
\end{figure}

In Figure \ref{fig:redshift_evolution_full}, we compare the redshift evolution of the galaxy-Ly$\alpha$ forest cross-correlation between observation and simulation. Our tentative indication of the observed redshift evolution of the signal in ASPIRE from $z=6.13$ to $5.65$ is broadly consistent with the theoretical trend in THESAN. Although the current large error bars preclude making any definitive statements, the higher excess IGM transmission around galaxies suggests much larger IGM fluctuations and a higher global $\HI$ fraction at higher redshift (e.g. $\langle\xHI\rangle=0.12$ at $z=6.16$ for THESAN, corresponding to ASPIRE's $\langle z\rangle=6.13$ data). This potential redshift evolution is very rapid. We only have $\sim50\,(100)\,\rm Myr$ between $z=6.13$ and $5.83\,(5.65)$. The observed galaxy-Ly$\alpha$ forest cross-correlation sensitively depends on the evolution of the IGM around galaxies at the tail end of reionization.

\begin{figure}
\centering
	\includegraphics[width=\columnwidth]{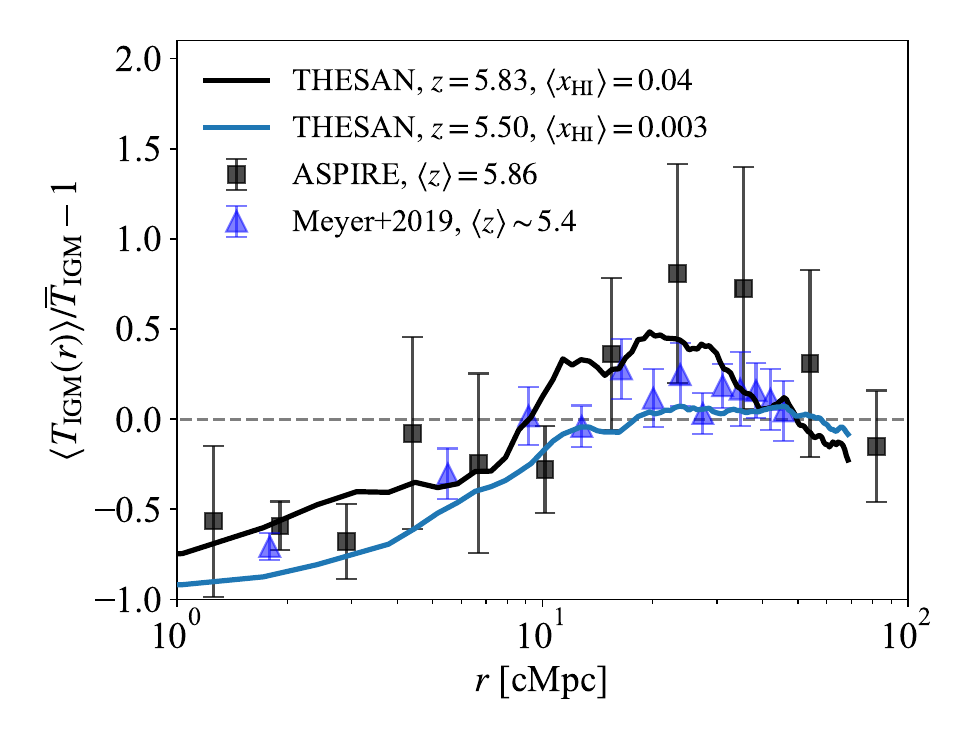}
 \vspace{-0.5cm}
    \caption{Comparison of the redshift evolution of galaxy-Ly$\alpha$ forest cross-correlation. The observed cross-correlation signals at $\langle z\rangle=5.86$ and $5.4$ from ASPIRE (black squares: $\OIII$ emitter-Ly$\alpha$ forest cross-correlation) and \citet{Meyer2019} (blue triangles: line-of-sight $\CIV$ absorber-Ly$\alpha$ forest cross-correlation) are compared with simulated cross-correlation at the close redshift snapshot (red: $z=5.83$, blue: $z=5.50$).}
    \label{fig:redshift_evolution}
\end{figure}

This potential redshift evolution of the galaxy-Ly$\alpha$ forest cross-correlation to $z\sim5$ can also be seen in comparison with previous observations. In Figure \ref{fig:redshift_evolution}, we show the comparison of the ASPIRE result with the 1D line-of-sight $\CIV$ absorber-Ly$\alpha$ forest cross-correlation \citep{Meyer2019} as a proxy for the full galaxy-Ly$\alpha$ forest cross-correlation. The observed redshift evolution from $z\simeq5.8$ to $5.4$ is in agreement with the THESAN simulation. The figure indicates that the large-scale excess Ly$\alpha$ forest transmission around galaxies disappears rapidly at $z<6$. This observed redshift evolution is well explained by THESAN. The rapid disappearance of the excess IGM transmission is due to the completion of the reionization process. The IGM fluctuations, such as the UV background, are smoothed out due to the increasing mean free path and lack of neutral islands in the post-reionization epoch. The decreasing excess IGM transmission is consistent with the smaller observed excess transmission reported by \citet{Meyer2019} at $z\sim5.4$. The disappearance of the excess cross-correlation signal is rapid during the final stages of reionization, with only approximately $100\,\rm Myr$ between $z=5.86$ and $z=5.4$. This rapid disappearance is also consistent with the null detection -- although the error bars remain large -- of the LAE-Ly$\alpha$ forest cross-correlation at $z\simeq4.9$ from photometric IGM tomography \citep{Kakiichi2023}.

After the disappearance of the excess IGM transmission due to the reionization process, only the excess IGM absorption from gas overdensities around galaxies remains, which is clearly visible from $z = 5.86$ to $5.4$. We may be witnessing a transition in the IGM structure around galaxies from the reionization epoch to cosmic noon, where the observed cross-correlation shifts from large-scale excess transmission to small-scale excess absorption around galaxies \citep{Turner2014,Bielby2017,Chen2020}.

\subsection{Towards a better understanding of reionization}\label{sec:halomass}

\begin{figure*}
    \centering
	\includegraphics[trim=10 0 10 0, clip, width=1.05\columnwidth]{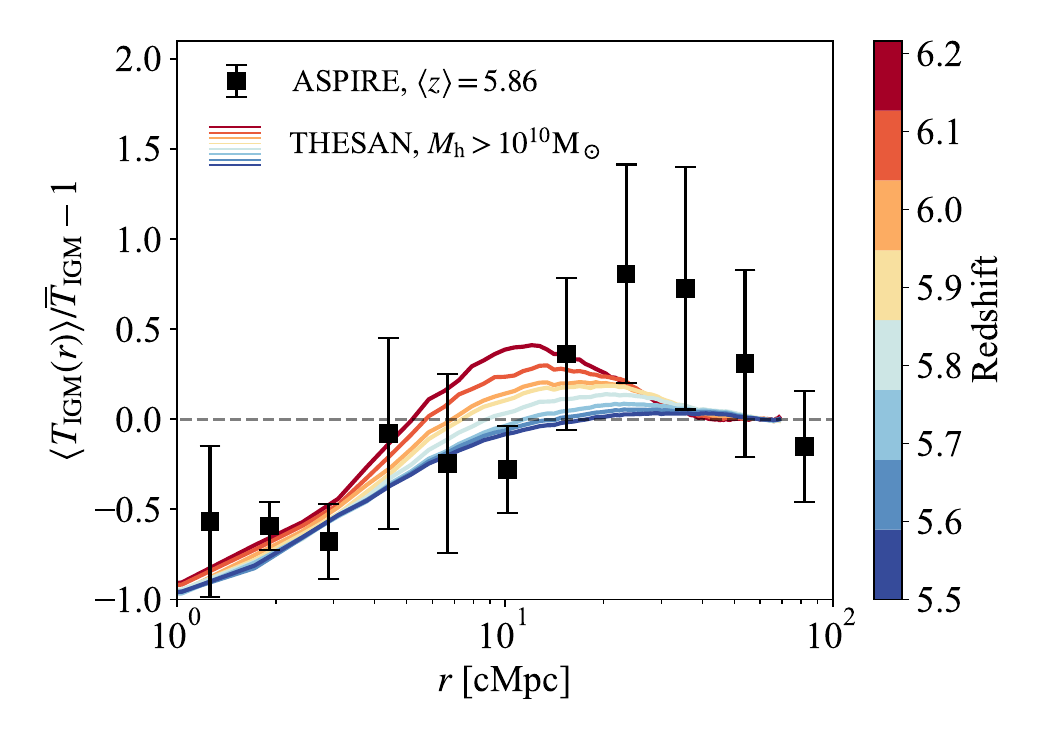}
    \hspace{-0.3cm}
	\includegraphics[trim=10 0 10 0, clip, width=1.05\columnwidth]{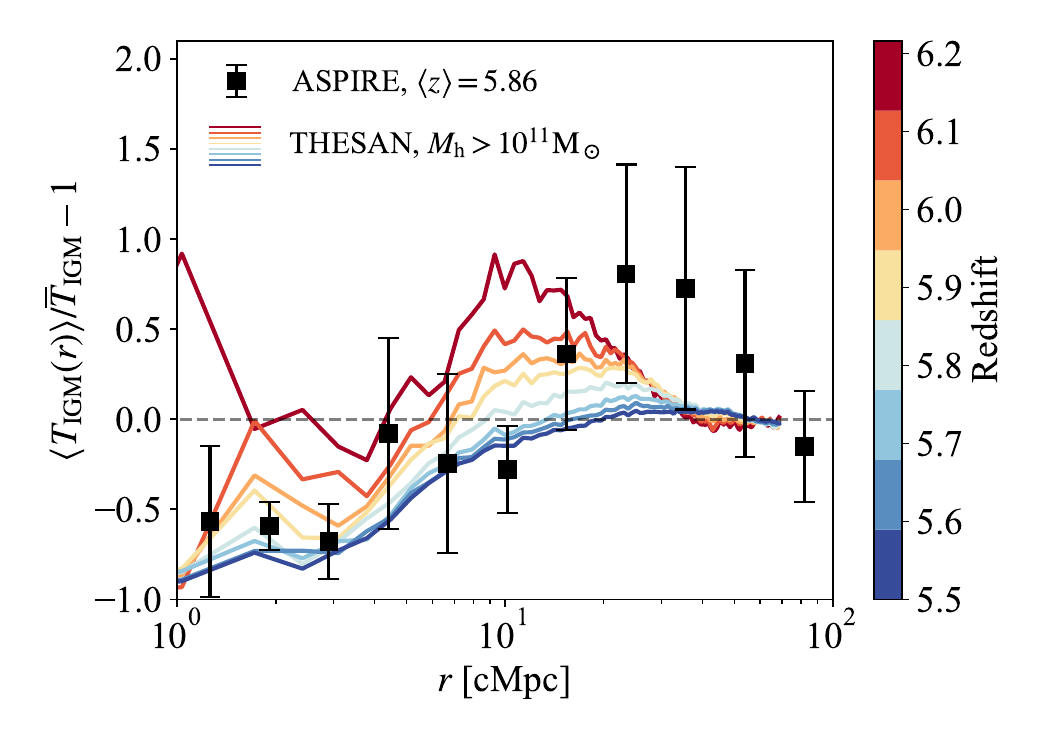}
    \vspace{-0.6cm}
    \caption{Same as Figure \ref{fig:model}, but for different host halo masses and colour-coded by redshift. The left panel shows the comparison with the THESAN simulation where $\OIII$ emitters are represented by galaxies with host halo mass of $M_h>10^{10}\,\rm M_\odot$. The right panel shows the comparison with the THESAN simulation where $\OIII$ emitters are represented by galaxies with host halo mass of $M_h>10^{11}\,\rm M_\odot$. The figure highlights the potential tension between the observed $\OIII$ emitter-Ly$\alpha$ forest cross-correlation and the THESAN simulation.}
    \label{fig:tension}
\end{figure*}

While the present analysis shows generally good agreement between ASPIRE and THESAN within the current observational error bars, this is true only {\it if} we choose $\OIII$ emitters based on stellar masses ($M_\ast>10^{10}\,\rm M_\odot$, see also \cite{Garaldi2024b} for other choice). However, stellar mass is a difficult quantity to estimate observationally, as it is affected by uncertain SED modelling and is also challenging to simulate due to its dependence on the complex stellar mass assembly history. A much cleaner quantity is the host halo mass of $\OIII$ emitters, which can be measured independently from the auto-correlation function \citep{Eilers2024,Pizzati2024}. Here we show that if we model $\OIII$ emitters based on the host halo mass in the simulation, the apparent agreement between observations and simulations worsens, indicating a potential tension between them. 


\cite{Eilers2024,Pizzati2024} found the minimum host halo mass of $\OIII$ emitters to be $\log_{10} M_{\rm min}/{\rm M_\odot} = 10.56^{+0.05}_{-0.03}$. The minimum (average) host halo mass of simulated galaxies with $M_{\star} > 10^{10}\,\rm M_\odot$ at $z = 5.83$ in THESAN used in Figure \ref{fig:thesan} is $2.7\,(5.1) \times 10^{11}\,\rm M_\odot$, which is a factor of 7 higher than the observationally inferred value. As this represents a more biased region of the simulation, the excess IGM transmission around them is shifted to larger scales than those around the host halos of the observationally estimated masses of $\OIII$ emitters.    

Figure \ref{fig:tension} clearly illustrates this. It shows a comparison between the observed $\OIII$ emitter-Ly$\alpha$ forest cross-correlation and the simulated signals from THESAN, where $\OIII$ emitters are represented by galaxies with host halo masses of $M_h > 10^{10}\,\rm M_\odot$ (left) and $M_h > 10^{11}\,\rm M_\odot$ (right), bracketing the range of likely host halo masses for the $\OIII$ emitters. The figure highlights the potential tension between observation and simulation. The comparison in Figure \ref{fig:tension} suggests that the peak location and amplitude of the simulated $\OIII$ emitter-Ly$\alpha$ forest cross-correlation tend to be shifted to smaller scales and lower amplitudes than the ASPIRE result. While the current error bars are still large, they suggest a tension between observations and simulations, indicating the need for further investigation. If this tension is confirmed by future observations, it would imply that the current understanding of the reionization process is incomplete. The observations indicate a larger excess IGM transmission at larger scales around $\OIII$ emitters than the simulations suggest. We discuss possible scenarios to reconcile this.

\medskip
\noindent$\bullet$ {\it $\OIII$ emitters live in more biased regions?}

\smallskip
One possibility is that star-forming activity in galaxies with intense $\OIII$ emission is preferentially enhanced in more biased regions, such as protoclusters or overdense environments, which makes them more likely to be selected as $\OIII$ emitters. In this scenario, $\OIII$ emitters reside preferentially in more biased regions, leading to a larger excess IGM transmission around them. However, this interpretation is at odds with the auto-correlation function measurement of $\OIII$ emitters, as such preferential segregation of $\OIII$ emitters in overdense regions should equally affect the auto-correlation function. Thus, it is unlikely that this is a viable solution. 

\medskip
\noindent$\bullet$ {\it Larger ionized bubbles around $\OIII$ emitters?}
\smallskip

As discussed in Section \ref{sec:late_reion}, the size of ionized bubbles determines the maximum outermost radius within which the excess IGM transmission around $\OIII$ emitters can occur. THESAN represents only one possible reionization morphology within a moderately large simulation box (95.5\,cMpc). Since the bubble size defines the outermost radius at which excess IGM transmission can be observed, larger bubble sizes around galaxies with host halo masses of $M_h\sim10^{10-11}\rm\,M_\odot$ than those in THESAN could lead to excess IGM transmission around $\OIII$ emitters at larger scales. Testing this scenario would require simulations with at least a $\sim200\rm\,cMpc$ box or larger to capture the large bubbles that may be present in the final stages of reionization. \citet{Conaboy2025} recently report the modelling of the cross-correlation in a larger simulation box. The impact of reionization morphology on the galaxy-Ly$\alpha$ forest cross-correlation function needs to be examined to test the viability of this scenario.

\medskip
\noindent$\bullet$ {\it Enhanced bias of the reionizing galaxies?}
\smallskip

A larger (luminosity-weighted) bias of ionizing sources gives rise to a more enhanced UV background around $\OIII$ emitters, potentially leading to a higher excess IGM transmission required to better explain the observation. An observationally reasonable variation of LyC leakage $f_{\rm esc}\xi_{\rm ion}$ from individual galaxies, based on both direct and indirect estimates of $f_{\rm esc}$ and $\xi_{\rm ion}$ \citep{Steidel2018,Nakajima2020,Flury2022,Saldana-Lopez2023,Saxena2023b}, suggests the predicted ionizing source biases range from $\langle b_s\rangle_L\approx2$ for the faint galaxy-dominated scenario to $\langle b_s\rangle_L\approx4$ for the bright galaxy-dominated reionization scenario (Section \ref{sec:source_bias}). Furthermore, JWST observations indicate only a mild increase of $\xi_{\rm ion,0}$ to fainter UV magnitudes in the range of $M_{\rm UV}\sim-22$ to $-15$ \citep{Simmonds2024}. It is unclear how one could significantly increase the bias of ionizing galaxies in the standard picture where ionizing photons are produced from star-forming regions and escape into the IGM.

Some exotic ionizing source models argue that including the non-stellar contribution to ionizing photon production from the conversion of kinetic energy to radiation via shocks may even produce larger ionizing source biases of $\langle b_{\rm shock}\rangle_L\sim8$ \citep{Wyithe2011}. \citet{Johnson2011} considered supernova shocks on galactic scales as a potential source of ionizing photons. While these non-stellar sources do not contribute to the total ionizing budget, their peculiar dependence on the halo mass of the host galaxies may lead to an increased bias of ionizing sources.

As discussed in Section \ref{sec:source_bias}, the change in the source bias alone is unlikely to explain the excess IGM transmission. While it could increase the UV background on large scales, this enhancement would also result in less small-scale excess absorption of the IGM around $\OIII$ emitters, overshooting the small-scale cross-correlation. Although the change in the UV background fluctuations is one of the key factors determining the galaxy-Ly$\alpha$ forest cross-correlation, the effect of source models needs to be examined carefully before a conclusion is made.

\cite{Garaldi2024b} took the first step in examining the impact of source models on the simulated galaxy-Ly$\alpha$ forest cross-correlation. They found an apparent insensitivity of source models on the cross-correlation signal. \cite{Gangolli2024} arrived at a similar conclusion, although in the context of the Ly$\alpha$ forest opacity-galaxy density relation. More studies are encouraged to fully disentangle the impacts of source models and implications for the reionization scenario.

\medskip
\noindent$\bullet$ {\it Large-scale thermal fluctuations of the IGM?}
\smallskip

The spatial fluctuations in the IGM temperature have also been suggested to modulate Ly$\alpha$ forest transmission. \citet{DAloisio2015} show that IGM thermal fluctuations produce an anti-correlation between galaxy densities and Ly$\alpha$ forest transmission, as regions far from galaxies are reionized last and therefore have higher temperatures due to having had less time to cool. This is true after the completion of reionization, i.e., when ionized bubbles completely percolate the entire IGM. On the other hand, during reionization, the temperature is highest at the edges of ionized bubbles, as these regions are just being heated by I-fronts (see Figures \ref{fig:thesan_map} \& \ref{fig:spherical_profile}). This means that Ly$\alpha$ forest transmission ($\tau_\alpha\propto \Gamma_{\rm HI}^{-1}T^{-0.72}$) just inside the ionized bubbles is enhanced, contributing to the large-scale excess IGM transmission around galaxies.

While THESAN self-consistently includes the impact of thermal fluctuations in the IGM, accurately simulating these fluctuations remains numerically challenging. It depends on the spectral hardness of ionizing sources, as well as the spectral hardening of I-fronts, which critically depends on both spatial resolution and the frequency sampling of the radiation field. \citet{DAloisio2019} show that post-I-front temperatures as high as $T\approx25,000-30,000\rm\,K$ may be achieved. A coherent increase in IGM temperature just inside the ionized bubbles could potentially enhance the excess IGM transmission by a factor of two or so. In this scenario, further amplification of the excess IGM transmission around $\OIII$ emitters would bring the simulation closer to the observed signal. 

This picture, involving large fluctuations of the IGM temperature in addition to the UV background fluctuations, is also in line with the suggestion raised by \citet{Christenson2023,Gangolli2024}. \citet{Christenson2023} observed the large scatter in the Ly$\alpha$ forest opacity-galaxy density relation along transmissive IGM sightlines. \citet{Gangolli2024} showed that the elevated gas temperatures from recent reionization at the outskirts of bubbles also lead to transmissive IGM sightlines. The same physical effect may also result in a shift in the peak of the excess IGM transmission around $\OIII$ emitters as observed by ASPIRE. Here, we only note the potential impact. A full investigation of this effect is left for future work.

\medskip
\noindent$\bullet$ {\it  Early onset of reionization?}
\smallskip

The detection of Ly$\alpha$ emission lines in $z\sim10-13$ galaxies \citep{Bunker2023,Witstok2024} indicates the onset of reionization as early as 330 Myr after the Big Bang. Combined with the late end of reionization at $z<6$, a more extended reionization history would result in a larger contrast in the IGM temperature inside the ionized bubbles. The early onset allows the inner radii more time to cool via adiabatic and Compton cooling, while the outer region has just been heated by late reionization. If the regions traced by $\OIII$ emitters are affected by the patchy early onset of reionization, a large temperature variation inside bubbles may naturally explain the high excess IGM transmission at large scales just inside the bubbles while allowing for preferential excess absorption of the IGM in the inner regions with cooler IGM. Further quantitative analysis of this scenario is necessary to assess the implications of the early onset of reionization.

\section{Conclusions}\label{sec:conclusion}

In this paper, we present an analysis of the spatial correlation between $\OIII$ emitters and Ly$\alpha$ forest transmission for a subset of JWST ASPIRE quasar fields. We measure the mean Ly$\alpha$ forest transmission around $5.4 < z < 6.5$ for $\OIII$ emitters identified using the NIRCam/WFSS F356W observation in the foreground of five quasars with $z \gtrsim 6.5$. By cross-correlating the $\OIII$ emitters' positions with the Ly$\alpha$ forest transmitted flux measured high signal-to-noise quasar spectra, we find a large-scale excess in IGM transmission around $\OIII$ emitters at $\langle z \rangle = 5.86$ on scales of $\sim 20-40\, \rm cMpc$. On smaller scales, we also find that the Ly$\alpha$ forest is preferentially absorbed at $r \lesssim 10\, \rm cMpc$ around $\OIII$ emitters, indicating the surrounding gas overdensities. We carefully examine the statistical significance and error budget of the observed $\OIII$ emitter-Ly$\alpha$ forest cross-correlation using both the observed data and theoretical covariance matrix. We find that the large-scale excess IGM transmission is detected at $2.2\sigma$, and the observed cross-correlation over $0 < r < 150\, \rm cMpc$ shows a clear departure from the null hypothesis at $\simeq 5\sigma$ significance, indicating evidence for a statistical spatial correlation between $\OIII$ emitters and the IGM at $\langle z \rangle = 5.86$.

We interpret the observed $\OIII$ emitter-Ly$\alpha$ forest cross-correlation in the context of an analytic RT framework and the THESAN cosmological radiation hydrodynamic simulation, which self-consistently models galaxy assembly and the late end of reionization at $z < 6$. We find that the THESAN simulation can reproduce the observed large-scale excess IGM transmission around $\OIII$ emitters detected by ASPIRE within the observational error bars, if we model $\OIII$ emitters as galaxies with stellar masses of $M_\star > 10^{10}\rm\,M_\odot$. The small-scale excess Ly$\alpha$ absorption can also be naturally explained by the increasing gas overdensities probing the outskirts of the circumgalactic medium around galaxies. The analytic model, which only includes density and UV background fluctuations with a fixed mean free path in the post-reionized IGM, cannot reproduce the observed signal. This indicates that large-scale IGM fluctuations beyond these simple assumptions must exist at $z \simeq 5.8$ to explain the observation.

The improved agreement between ASPIRE and THESAN suggests that the large-scale fluctuations of the IGM -- caused the UV background fluctuations driven by both the distribution of ionizing sources and absorbers and/or thermal fluctuations from reionization -- are necessary to produce the large-scale excess Ly$\alpha$ forest transmission on scales of tens of cMpc around galaxies. Such large-scale fluctuations are most naturally produced by the reionization process, hinting at the existence of ionized bubbles at the observed redshift. In this picture, the outermost extent of the large-scale galaxy-Ly$\alpha$ forest cross-correlation can be interpreted as a lower limit to the typical size of ionized bubbles around galaxies, indicating that $\OIII$ emitters at $\langle z \rangle = 5.86$ must be surrounded by large ionized bubbles exceeding $\sim 50\, \rm cMpc$. Overall, the observed large-scale excess Ly$\alpha$ forest transmission around $\OIII$ emitters supports the notion that reionization is still ongoing at $z < 6$, creating the large-scale fluctuations of the IGM (UV background and thermal fluctuations) inside ionized bubbles. Reionization is on the verge of completion at $z \simeq 5.8$.

This completion of late reionization requires faint galaxies below our detection limit. The enhanced large-scale UV background for the excess Ly$\alpha$ forest transmission demands a collective population of fainter galaxies surrounding the observed $\OIII$ emitters. The observed $\OIII$ emitters fall short of providing the necessary ionizing budget, assuming reasonable values for the LyC escape fraction and ionizing photon production efficiency. Our analysis of the individual associations between $\OIII$ emitters and Ly$\alpha$ forest transmission spikes further indicates that the LyC leakage from these emitters does not generate enough ionizing radiation to maintain the high ionization levels of the surrounding IGM, as evidenced by the presence of Ly$\alpha$ forest transmission spikes. This conclusion holds true even if we assume that all $\OIII$ emitters host AGN activities and exhibit $100\,\%$ LyC escape fractions, suggesting that AGN alone are insufficient to drive reionization. Generally, an average LyC leakage of $\log_{10}\langle f_{\rm esc}\xi_{\rm ion}\rangle/[{\rm erg^{-1}Hz}]\approx24.5$ down to galaxies with $M_{\rm UV}\approx-10$ is required to establish a sufficient UV background. Thus, we conclude that an unseen population of fainter galaxies, or systems not selected as $\OIII$ emitters (or luminous populations residing outside the single NIRCam/WFSS field of view), is responsible for completing reionization.

Despite the broad agreement, a more careful comparison between ASPIRE and THESAN presents challenges to our understanding of reionization and the origin of the observed galaxy-Ly$\alpha$ forest cross-correlation. If we model the $\OIII$ emitters based on halo masses of $\gtrsim 10^{10-11}\rm\,M_\odot$ as suggested from the observed auto-correlation function of $\OIII$ emitters, THESAN underpredicts both the observed peak position and amplitude of the excess Ly$\alpha$ forest transmission around $\OIII$ emitters. This suggests potential shortcomings in state-of-the-art cosmological reionization simulations. If this tension persists, it would require even larger IGM fluctuations at $z \simeq 5.8$ than predicted. The potential scenarios include the existence of larger ionized bubbles around $\OIII$ emitters at $z<6$, further enhancement of the large-scale UV background or thermal fluctuations of the IGM due to different source models and/or improved numerical resolution, and possibly a patchy early onset of reionization at $z\gtrsim 10-13$. The impacts of these scenarios on the observed galaxy-Ly$\alpha$ forest cross-correlation need to be quantitatively examined to understand their physical implications and a way forward with improved measurements of the galaxy-Ly$\alpha$ forest cross-correlation in the future.

On the observational front, we find that the observational error in the $\OIII$ emitter-Ly$\alpha$ forest cross-correlation is dominated by cosmic variance. There is significant field-to-field variation in the spatial correlation between $\OIII$ emitters and Ly$\alpha$ forest transmission, likely resulting from a patchy reionization process where the completion of reionization is inhomogeneous across different parts of the Universe. The observed error can be explained in terms of the theoretical covariance matrix, suggesting that the origin of the error is well understood. The noise in the quasar spectra is a sub-dominant contribution to the error budget. This implies that an increased number of quasar fields observed with JWST should lower the overall error budget. This is promising, as our present analysis only uses a subset of the ASPIRE quasar fields (5 out of 25 fields). Future analyses with all ($>34$) quasar fields observed with JWST, including the six EIGER quasar fields and other quasar fields (GO 4092: \citet{Becker2023GO}, GO 5911: \citet{Simcoe2024GO}), should provide a more robust measurement of the galaxy-Ly$\alpha$ forest cross-correlation signal, providing unique insights into how galaxies complete reionization and the role of galaxies during the final stages of reionization.


\section*{Acknowledgments}

We thank Andreu Font-Ribera for useful discussions and Zaria Luki\'{c} for making the NyX simulation available to us.

KK is supported by the DAWN Fellowship. The Cosmic Dawn Center (DAWN) is funded by the Danish National Research Foundation under grant No. 140. 
FW acknowledges support from NSF award AST-2513040.
SEIB is funded by the Deutsche Forschungsgemeinschaft (DFG) under Emmy Noether grant number BO 5771/1-1.
RAM acknowledges support from the Swiss National Science Foundation (SNSF) through project grant 200020\_207349.
SZ acknowledges support from the National Science Foundation of China (no. 12303011). 
RK acknowledges support of the Natural Sciences and Engineering Research Council of Canada (NSERC) through a Discovery Grant and a Discovery Launch Supplement, funding reference numbers RGPIN-2024-06222 and DGECR-2024-00144.
VD acknowledges financial support from the Bando Ricerca Fondamentale INAF 2022 Large Grant “XQR-30”.

This work is based on observations made with the NASA/ESA/CSA James Webb Space Telescope. The data were obtained from the Mikulski Archive for Space Telescopes (MAST) at the Space Telescope Science Institute, which is operated by the Association of Universities for Research in Astronomy, Inc., under NASA contract NAS 5-03127 for JWST. The specific observations analysed can be accessed via \url{https://doi.org/10.17909/vt74-kd84}. These observations are associated with program \#2078. Support for program \#2078 was provided by NASA through a grant from the Space Telescope Science Institute, which is operated by the Association of Universities for Research in Astronomy, Inc., under NASA contract NAS 5-03127. 

This work is based on observations collected at the European Organisation for Astronomical Research in the Southern Hemisphere under program
IDs 087.A-0890(A), 088.A-0897(A), 097.B-1070(A), 098.A0444(A), 098.B-0537(A), 0100.A-0625(A), 0102.A-0154(A),
1103.A-0817(A), 1103.A-0817(B), and 2102.A-5042(A). The
paper also used data Based on observations obtained at the international Gemini Observatory, a program of NSF NOIRLab, which is managed by the Association of Universities for Research in Astronomy (AURA) under a cooperative agreement with the U.S. National Science Foundation on behalf of the Gemini Observatory partnership: the U.S. National Science Foundation (United States), National Research Council (Canada), Agencia Nacional de Investigaci\'{o}n y Desarrollo (Chile), Ministerio de Ciencia, Tecnolog\'{i}a e Innovaci\'{o}n (Argentina), Minist\'{e}rio da Ci\^{e}ncia, Tecnologia, Inova\c{c}\~{o}es e Comunica\c{c}\~{o}es (Brazil), and Korea Astronomy and Space Science Institute (Republic of Korea). Some of the data presented herein were obtained at Keck Observatory, which is a private 501(c)3 non-profit organization operated as a scientific partnership among the California Institute of Technology, the University of California, and the National Aeronautics and Space Administration. The Observatory was made possible by the generous financial support of the W. M. Keck Foundation. The authors wish to recognize and acknowledge the very significant cultural role and reverence that the summit of Maunakea has always had within the Native Hawaiian community. We are most fortunate to have the opportunity to conduct observations from this mountain.

\section*{Data Availability}

The raw data used in this paper are available from the MAST archive and ESO archive. The $\OIII$ emitter catalogue will be publicly released along with a ASPIRE survey overview paper (Wang et al. in prep) at \url{https://aspire-quasar.github.io}. Reduced X-Shooter spectra of QSOs are available through public github repository of the XQR-30 at \url{https://github.com/XQR-30/Spectra}.

\bibliographystyle{mn2e}
\bibliography{reference} 

\appendix
\section{Linear theory revisited}\label{app:linear_theory}

For completeness, we describe the linear perturbation theory of power spectra between galaxies and the Ly$\alpha$ forest used in the covariance matrix calculation. We follow the formulation of \citet{Pontzen2014} and \citet{Gontcho2014}.

To linear order, the fluctuations in galaxy distribution $\delta_{\rm g}(k,\mu)$ and Ly$\alpha$ forest transmission $\delta_{\alpha}(k,\mu)$ can be expressed as:
\begin{equation}
    \delta_g(k,\mu) = b_g(1 + \beta_g\mu^2)\delta_m(k),
\end{equation}
and
\begin{equation}
    \delta_{\alpha}(k,\mu) = b_{\alpha}(1 + \beta_\alpha\mu^2)\delta_m(k) + b_{\alpha,\Gamma}\delta_{\Gamma}(k),
\end{equation}
where $\delta_m(k)$ represents the matter fluctuations, $b_g$ and $b_\alpha$ are the linear density biases of galaxies and the Ly$\alpha$ forest, respectively. The parameter $\beta_g = f/b_g$ (approximately $1/b_g$ at $z > 4$, where $f$ is the growth rate of structure) is the linear redshift-space distortion (RSD) parameter for galaxies, while $\beta_\alpha$ is the RSD parameter for the Ly$\alpha$ forest. The term $\delta_{g,\rm shot}$ refers to the Poisson shot noise of galaxies. The effect of UV background fluctuations is introduced through the additional term $b_{\alpha,\Gamma}\delta_{\Gamma}(k)$, where $b_{\alpha,\Gamma}$ is the linear bias of the Ly$\alpha$ forest with respect to UV background perturbations, and $\delta_\Gamma(k)$ represents the fluctuations in the photoionization rate. The fluctuations in the photoionization rate are given by 
\begin{equation}
    \delta_{\Gamma}(k) = b_{\Gamma}(k)\delta_m(k) + \delta_{\Gamma,\rm shot},
\end{equation}
where the former term is driven by coherent fluctuations in the matter density $\delta_m(k)$, with $b_{\Gamma}(k)$ being the linear bias of the photoionization rate relative to matter density perturbations, and the latter term represents the shot-noise contribution from ionizing sources.

\citet{Pontzen2014} (see also \citet{Gontcho2014}) find that the linearised radiative transfer equation gives the linear bias of the photoionization rate with respect to the matter density perturbations as
\begin{equation}
    b_\Gamma(k)=\frac{b_s-b_\kappa}{b_{\kappa,\Gamma}+R^{-1}(k\lambda_{\rm mfp})}\approx b_s R(k\lambda_{\rm mfp}),
\end{equation}
where $b_s$ is the bias of ionizing sources, $b_\kappa$ is the bias of absorbers, $b_{\kappa,\Gamma}$ is the linear response of the bias of absorbers with respect to the perturbation of photoionization rate, and $R(k\lambda_{\rm mfp})=\arctan(k\lambda_{\rm mfp})/(k\lambda_{\rm mfp})$ with $\lambda_{\rm mfp}$ being the mean free path of ionizing photons. We assume the source bias $b_s=2.87$ derived from our RT+CLF framework (Section \ref{sec:interpretation}) and the mean free path of $\lambda_{\rm mfp}=2\rm\,pMpc$. For simplicity, we have assumed $b_\kappa=b_{\kappa,\Gamma}=0$, equivalent to ignoring the spatial variation of the absorbers due to the UV background fluctuations for our covariance matrix calculation. While in main text we argue that the spatial variation of absorbers is a important factor to explain the observed large-scale excess IGM transmission, since we only use the linear theory to give an order-of-magnitude estimate of the covariance matrix, this assumption does not affect our main conclusion. 

The resulting 3D auto-power spectra of galaxies and Ly$\alpha$ forest are given by
\begin{equation}
    P_{g}(k,\mu)=b_g^2(1+\beta_g\mu^2)^2P_m(k),\label{eq:Pg}
\end{equation}
and 
\begin{equation}
    P_{\alpha}(k,\mu)=[b_{\alpha}(1+\beta_\alpha\mu^2)+b_{\alpha,\Gamma}b_{\rm \Gamma}(k)]^2P_m(k)+b_{\alpha,\Gamma}^2P_{\Gamma,\rm shot}(k).\label{eq:Palpha}
\end{equation}
The 3D cross-power spectrum between galaxies and Ly$\alpha$ forest is given by
\begin{equation}
    P_{\rm g\alpha}(k,\mu)=b_g(1+\beta_g\mu^2)[b_{\alpha}(1+\beta_\alpha\mu^2)+b_{\alpha,\Gamma}b_{\rm \Gamma}(k)]P_m(k),\label{eq:Pg-alpha}
\end{equation}
where $P_m(k)$ is the 3D linear matter power spectrum. The discrete random nature of the ionizing sources gives the shot-noise contribution $P_{\Gamma,\rm shot}(k)$ to the UV background flctuations. which is given by, assuming constant mean free path and constant LyC leakage for all galaxies,
\begin{equation}
  P_\Gamma^{\rm shot}(k)=\bar{n}_{\rm eff}^{-1}R^2(k\lambda_{\rm mfp}),
  ~~~\bar{n}_{\rm eff}=\frac{\left[\int^\infty_{L_{\rm UV}^{\rm min}} L_{\rm UV}\Phi(L_{\rm UV})dL_{\rm UV}\right]^2}{\int^\infty_{L_{\rm UV}^{\rm min}} L_{\rm UV}^2\Phi(L_{\rm UV})dL_{\rm UV}}.\label{eq:Pshot}
\end{equation}
We assume the minium UV luminosity corresponding to $M_{\rm UV}=-10$. In the case of the UV background fluctuations driven by galaxies, this shot-noise contribution is sub-dominant and can be ignored. 

The 3D Ly$\alpha$ forest power spectrum relates to the line-of-sight 1D counterpart by \citep[e.g.][]{McDonald2000,Palanque-Delabrouille2013}
\begin{equation}
P_\alpha^{\rm 1D}(k_v)=\frac{H(z)e^{-k_v^2 v_{\rm th}^2}}{(1+z)}\int_{k_\parallel}^\infty\frac{dk k}{2\pi}P_\alpha(k),\label{eq:P1D}
\end{equation}
where $v_{\rm th}=\sqrt{k_B T/\mu m_{\rm p}}\approx13.0(T/1.2\times10^4\,\rm K)^{1/2}\rm\,km\,s^{-1}$ is the thermal velocity of the gas that affects the line-of-sight smoothing of the observed Ly$\alpha$ forest power spectrum and $k_v=(1+z)k_\parallel/H(z)$ is the wavenumber in Fourier velocity space.

The bias and RSD parameters of galaxies ($\OIII$ emitters) are chosen based on the best-fit parameters of the HOD modelling. For the HOD parameters used in Section \ref{sec:interpretation}, we find $b_g=5.71$ and $\beta_g=0.17$ for $\OIII$ emitters. The bias and RSD parameters of Ly$\alpha$ forest are uncertain and we simply assume $b_\alpha=-1.2$ and $\beta_\alpha=1.5$. We have tested with other values of $b_\alpha$ and $\beta_\alpha$ and found that our conclusion is not affected by the exact choice of these parameters.

The response the Ly$\alpha$ forest transmission with respect to the change in the UV background is captured by the bias factor $b_{\alpha,\Gamma}$. We can analytically estimate the value of $b_{\alpha,\Gamma}$ by realising that $b_{\alpha,\Gamma}=\left.\frac{d\ln\overline{T}_{\rm IGM}}{d\ln \Gamma}\right|_{\Gamma=\bar{\Gamma}}$, from which we find
\begin{equation}
  b_{\alpha,\Gamma}=
  \frac{1}{\overline{T}_{\rm IGM}}\int d\Delta_b P(\Delta_b)\tau_0\Delta_b^\beta\bar{\Gamma}_{\rm HI,-12}^{-1} e^{-\tau_0\Delta_b^\beta\bar{\Gamma}_{\rm HI,-12}^{-1}},
\end{equation}
where $P(\Delta_b)$ is the density PDF at mean IGM, $\bar{\Gamma}_{\rm HI,-12}=\bar{\Gamma}_{\rm HI}/10^{-12}\rm\,s^{-1}$ is mean photoionization rate in units of $10^{-12}\rm\,s^{-1}$. Our choice of bias parameters ensures that the linear theory approach is consistent with our RT+CLF framework on the large scale.

These auto- and cross-power spectra (equations \ref{eq:Pg}, \ref{eq:Palpha}, and \ref{eq:Pg-alpha}) are used to estimate the covariance matrix of the mean Ly$\alpha$ forest transmission around galaxies, as shown in Figure \ref{fig:error_budget}. Note that the linear prediction underestimates the Jackknife (Bootstrap) error at $r\sim20-40\,\rm cMpc$. This is understandable since the linear theory, i.e. the large-scale limit of the RT+CLF framework, cannot fully reproduce the peak location of the excess IGM transmission at $r\sim20-40\,\rm cMpc$ around $\OIII$ emitters. However, it remains clear that cosmic variance is the dominant error source in the galaxy-Ly$\alpha$ forest cross-correlation measurement.

\begin{figure}
    \centering
        \includegraphics[width=0.46\columnwidth]{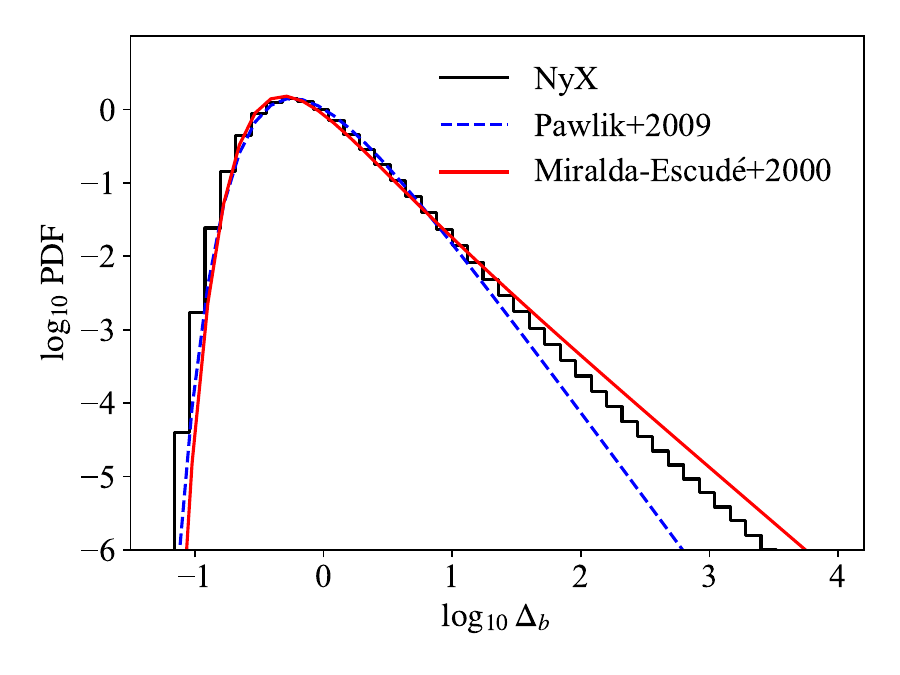}
        \raisebox{0.1cm}{\includegraphics[width=0.5\columnwidth]{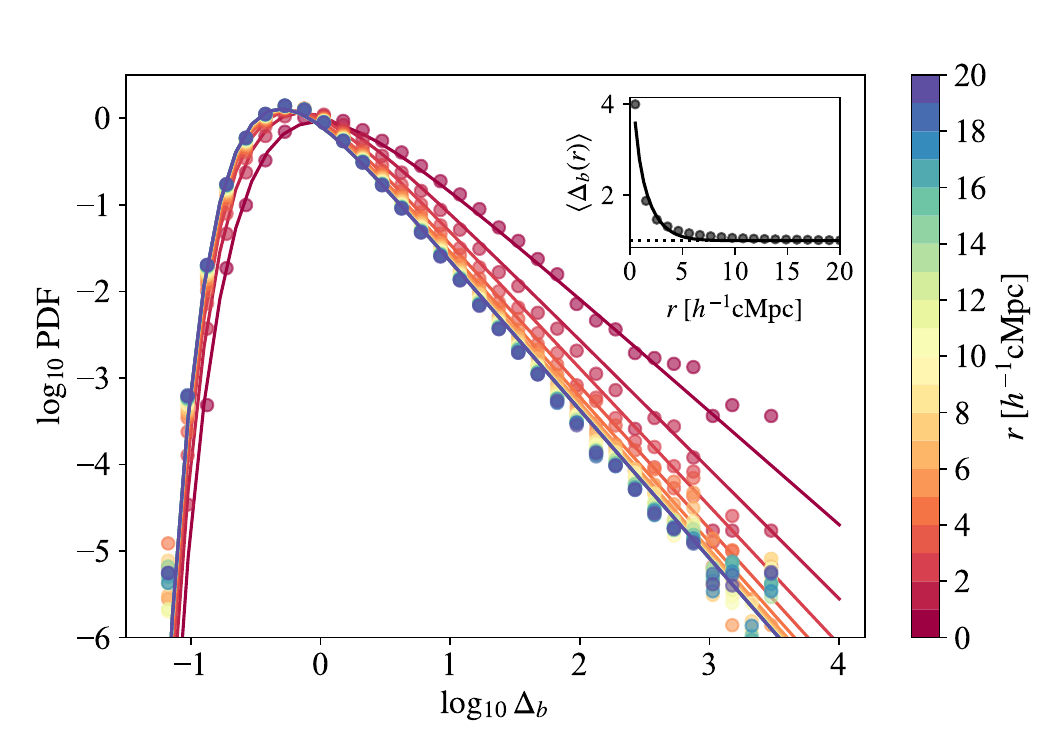}}
        \vspace{-0.3cm}
        \caption{({\bf Left}): Comparison of the probability distribution function of the gas overdensities at $z=6$ between the NyX simulation and previous studies (red solid: \citet{MHR2000}; blue dashed: \citet{Pawlik2009}). ({\bf Right}): Volume-weighted PDF $P_V(\Delta_b|r)$ of gas overdensities at various radii from the central haloes of mass $>10^{11}\rm~M_\odot$. Circles indicate the measured PDF from the NyX simulation and solid curves indicate the analytic fit to the PDF. The colours indicate the different radii from the haloes. The in-set plot shows the average gas overdensities as a function of radius (circles: NyX simulation, solid curve: analytic fit).}
        \label{fig:density_PDF}
\end{figure}


\section{Density PDF}\label{app:PDF}

The volume-weighted density PDF is modelled using the NyX cosmological hydrodynamic simulation \citep{Lukic2015}. The the further detail We refer the reader to the original paper. The simulation is numercially converged on Ly$\alpha$ forest statistics at one percent level, making it sutiable to measure the IGM properties. 

Figure \ref{fig:density_PDF} (left) verifies that the density PDF measured from the $z=6$ NyX snapshot agrees well with previous studies at $z=6$ \citep[]{MHR2000,Pawlik2009}. The deviation at $\log_{10}\Delta_b\gtrsim2$ is likely due to the different treatment of the star formation and feedback. Since the NyX does not convert high dense gas to star particles, it naturally creates a tail of high-density gas similar to that of \citet{MHR2000}. \citet{Pawlik2009} noted that their simulation may not yet be fully converged thus the PDF at large overdensities $\log_{10}\Delta_b\gtrsim2$ still remain uncertain.
At lower density regions $\log_{10}\Delta_b\lesssim-0.5$, the NyX simulation predict a slightly larger number of  low density regions. This should be physical because the NyX's high spatial resolution and large box $100h^{-1}\rm cMpc$ allows us to sample lower density regions of the IGM compared to the $10h^{-1}\rm cMpc$ box from \citet{Pawlik2009} and \citet{MHR2000}. Some of the discrepancies may be attributed to the difference between SPH and grid-based hydrodynamic slover in Gadget \citep{Springel2005} and NyX \citep{Almgren2013}. Overall, the density PDF from the NyX simulation agrees very well with the previous work. 

We generalised the analytic fitting formula proposed by \citet{MHR2000} to allow the radial dependence of the density PDF, $P_V(\Delta_b|r)$, around dark matter haloes. Figure \ref{fig:density_PDF} (right) shows the volume-wieghted density PDF $P_V(\Delta_b|r)$ at various radii from the central haloes of mass $M_{\rm h}>10^{11}\rm\,M_\odot$ measured from the $z=6$ NyX snapshot. We find that the numerical result can be well fit with the following analytic formula,
\begin{equation}
    P_V(\Delta_b|r)d\Delta=A(r)\exp\left[-\frac{(\Delta_b^{-2/3}-C(r))^2}{2(2\delta(r)/3)^2}\right]\Delta_b^{-\beta(r)}d\Delta_b,
\end{equation}
where $A(r)=A_0+A_1 e^{-r/r_A}, C(r)=C_0+C_1 e^{-r/r_C}, \delta(r)=\delta_0+\delta_1 e^{-r/r_\delta},$ and $\beta(r)=\beta_0+\beta_1 e^{-r/r_\beta}$. The best-fit parameters are tabulated in Table \ref{tab:PDF_params}. This functional form asymptotically approaches the PDF of the mean IGM at sufficiently large radius. We use this analytic PDF fit to model the density fluctuations.

\begin{table}
    \centering
    \caption{The best-fit parameters for the volume-weighted density PDF as a function of radius from $M_{\rm h}>10^{11}M_\odot$ haloes at $z=6$.}
    \begin{tabular}{llll}
        \hline
        & \multicolumn{3}{c}{The best-fit parameters}\\
        \hline
        $A(r):$      & $A_0 = 0.4958$    & $A_1 = 2.241$     & $r_A = 1.477$ \\
        $C(r):$      & $C_0 = 0.2389$    & $C_1 = -1.292$    & $r_C = 2.092$ \\
        $\delta(r):$ & $\delta_0 = 1.388$ & $\delta_1 = 0.0868$ & $r_{\delta} = 3.061$ \\
        $\beta(r):$  & $\beta_0 = 2.710$  & $\beta_1 = -0.5442$ & $r_{\beta} = 1.619$ \\
        \hline
    \end{tabular}
    \label{tab:PDF_params}
\end{table}

\section{The contribution of neutral islands to galaxy-Ly$\alpha$ forest cross-correlation}\label{app:CCF_test}

To estimate the contribution of neutral islands to the galaxy-Ly$\alpha$ forest cross-correlation, we compare the simulated galaxy-Ly$\alpha$ forest cross-correlation using all Ly$\alpha$ forest pixels along 300 random skewers with the cross-correlation without the contribution from neutral islands. To do this, we mask the Ly$\alpha$ forest pixels where the underlying $\HI$ fraction is $\xHI > 0.10$ to exclude the regions of neutral islands. We then compute the mean Ly$\alpha$ forest transmission around galaxies without these masked pixels,
\begin{equation}
\langle \TIGM^{\rm mask}(r)\rangle = \frac{\displaystyle \sum_{i\in{{\rm pair}(r)}} m_i e^{-\tau_{\alpha,i}}}{\displaystyle \sum_{i\in{{\rm pair}(r)}} m_i},
\end{equation}
where $m_i=1$ for unmasked pixels and $0$ for masked pixels, with $i$ being the index of all pixels having Ly$\alpha$ optical depth $\tau_{\alpha,i}$. The mean Ly$\alpha$ forest transmission $\overline{T}_{\rm IGM}^{\rm mask}$ is also computed without the masked pixels. Thus, the masked cross-correlation is given by 
$\langle T_{\rm IGM}^{\rm mask}(r)\rangle/\overline{T}_{\rm IGM}^{\rm mask}-1$.
In this way, the masked cross-correlation represents the spatial fluctuations of Ly$\alpha$ forest transmission around galaxies only within ionized bubbles.

\begin{figure*}
    \centering
    \includegraphics[width=0.5\textwidth]    {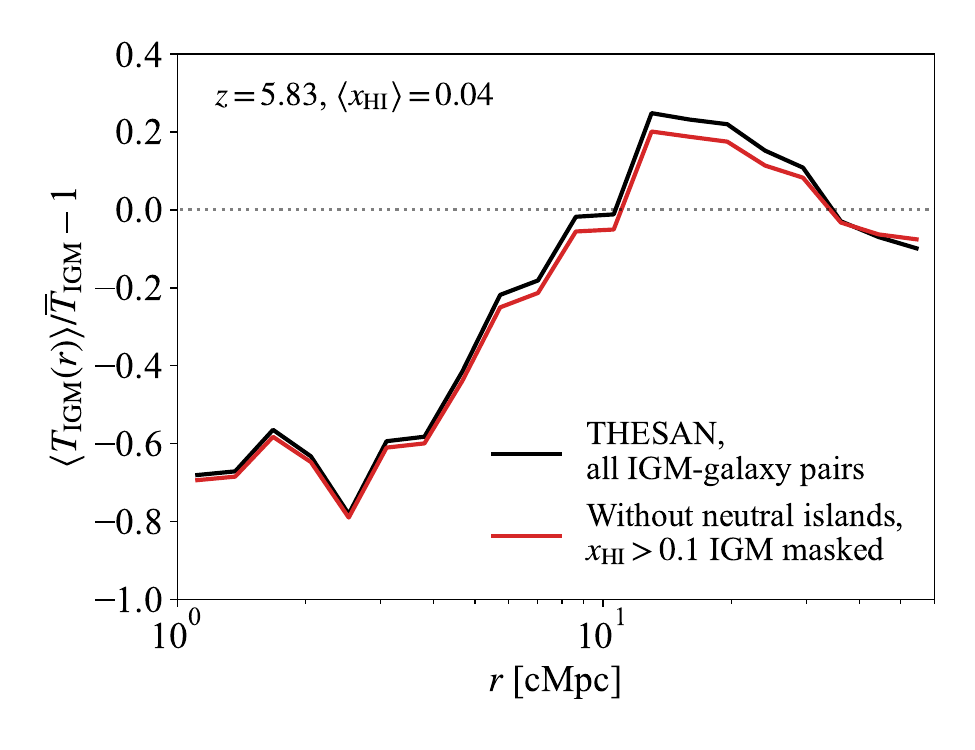}
    \vspace{-0.4cm}
    \caption{Comparison of the simulated galaxy-Ly$\alpha$ forest cross-correlation (black) with the masked cross-correlation excluding the contribution of neutral islands at $z=5.83$ in THESAN. The cross-correlations are computed using galaxies with stellar masses of $>10^{10}\,\rm M_\odot$ and 300 skewers drawn randomly from the simulation box.}\label{fig:CCF_masked}
\end{figure*}

Figure \ref{fig:CCF_masked} shows the comparison between the two cases. We find that the galaxy-Ly$\alpha$ cross-correlation from THESAN with all Ly$\alpha$ forest pixels (black curve) is almost identical to that without the contribution from neutral islands (red curve). We have experimented with different thresholds for $\xHI$ ranging from $0.01$ to $0.5$. In all cases, the masked cross-correlations show a nearly identical shape to the full signal, indicating that excluding the neutral islands has little impact on the galaxy-Ly$\alpha$ forest cross-correlation. The contribution of neutral islands is therefore marginal. The large-scale excess transmission should originate from significant fluctuations of the IGM inside ionized bubbles.

\end{document}